\documentclass{cernrep}

\usepackage{color}
\usepackage[colorlinks=true,citecolor=purple,linkcolor=blue]{hyperref}
\usepackage[normalem]{ulem}
\usepackage{amsmath,amssymb,slashed}
\usepackage[range-units=single,range-phrase=--]{siunitx}
\usepackage{bbm}
\usepackage{epsfig}
\usepackage{grffile,graphicx}
\usepackage{url}
\usepackage{enumerate}
\usepackage{multirow}
\usepackage{placeins}
\usepackage[dvipsnames]{xcolor}
\usepackage{epstopdf}
\usepackage[utf8]{inputenc}
\usepackage[version=4]{mhchem}
\usepackage{tablefootnote}
\usepackage[toc, acronym, nonumberlist]{glossaries}

\newcommand{\newc}{\newcommand*}

\long\def\begincomment#1\endcomment{%
        \begingroup\sf\baselineskip12pt#1\endgroup}

\newcommand{\cl}{\rm {C.L.}}

\newc{\ev}{\ensuremath{\,\mathrm{eV}}}
\newc{\kev}{\ensuremath{\,\mathrm{keV}}}
\newc{\mev}{\ensuremath{\,\mathrm{MeV}}}
\newc{\gev}{\ensuremath{\,\mathrm{GeV}}}
\newc{\tev}{\ensuremath{\,\mathrm{TeV}}}
\newc{\miliev}{\ensuremath{\,\mathrm{meV}}}
\newc{\muev}{\ensuremath{\,\mu\mathrm{eV}}}
\newc{\nev}{\ensuremath{\,\mathrm{neV}}}
\newc{\eV}{\ev}
\newc{\keV}{\kev}
\newc{\MeV}{\mev} 
\newc{\GeV}{\gev}
\newc{\geV}{\gev}
\newc{\TeV}{\tev}
\newc{\milieV}{\miliev}
\newc{\mueV}{\muev}
    \newc{\neV}{\nev}
\newc{\invpb}{\ensuremath{/\text{pb}}}
\newc{\invfb}{\ensuremath{\,\text{fb}^{-1}}}
\newc\nb{\ensuremath{\,\mathrm{nb}}} \newc\pb{\ensuremath{\,\mathrm{pb}}} \newc\fb{\ensuremath{\,\mathrm{fb}}}
\newc\pc{\ensuremath{\,\mathrm{pc}}}
\newc\kpc{\ensuremath{\,\mathrm{kpc}}}
\newc\mpc{\ensuremath{\,\mathrm{Mpc}}}
\newc\ps{\ensuremath{\,\mathrm{ps}}} 
\newc\nmeter{\ensuremath{\,\mathrm{nm}}} 
\newc\mumeter{\ensuremath{\,\mu\mathrm{m}}} 
\newc\mmeter{\ensuremath{\,\mathrm{mm}}} 
\newc\cmeter{\ensuremath{\,\mathrm{cm}}} 
\newc\meter{\ensuremath{\,\mathrm{m}}} 
\newc\kmeter{\ensuremath{\,\mathrm{km}}}

\newc\parsec{\ensuremath{\,\mathrm{pc}}}

\newc\second{\ensuremath{\,\mathrm{s}}}
\newc\msecond{\ensuremath{\,\mathrm{ms}}}
\newc\musecond{\ensuremath{\,\mu\mathrm{s}}}
\newc\nsecond{\ensuremath{\,\mathrm{ns}}}
\newc\psecond{\ensuremath{\,\mathrm{ps}}}
\newc\hours{\ensuremath{\,\mathrm{h}}}
\newc\days{\ensuremath{\,\mathrm{day}}}
\newc\years{\ensuremath{\,\mathrm{year}}}
\newc\kgram{\ensuremath{\,\mathrm{kg}}}
\newc\tonne{\ensuremath{\,\mathrm{t}}}

\newc\ppb{\ensuremath{\,\mathrm{ppb}}}

\newc\events{\ensuremath{\,\mathrm{events}}}

\newc{\sigsi}{\ensuremath{\sigma_{\rm SI}}}
\newc{\sigsd}{\ensuremath{\sigma_{\rm SD}}}
\newc{\Enr}{\ensuremath{E_{\rm nr}}}

\newc{\MHz}{\ensuremath{\,\mathrm{MHz}}}
\newc{\GHz}{\ensuremath{\,\mathrm{GHz}}}
\newc{\THz}{\ensuremath{\,\mathrm{THz}}}
\newc{\tesla}{\ensuremath{\,\mathrm{T}}}
\newc\gram{\ensuremath{\,\mathrm{g}}}
\newc\Bq{\ensuremath{\,\mathrm{Bq}}}
\newc\muBq{\ensuremath{\,\mu\mathrm{Bq}}}
\newc\keVee{\ensuremath{\,\textnormal{keV}_\mathrm{ee}}}
\newc\keVnr{\ensuremath{\,\textnormal{keV}_\mathrm{nr}}}
\newc\eVee{\ensuremath{\,\textnormal{eV}_\mathrm{ee}}}
\newc\eVnr{\ensuremath{\,\textnormal{eV}_\mathrm{nr}}}
\newc\eVr{\eVnr}
\newc{\chisqmin}{\ensuremath{\chi^2_{\mathrm{min}}}}
\newc{\Delchisq}{\ensuremath{\Delta\chi^2}}
\newc{\chisq}{\ensuremath{\chi^2}}
\newc{\like}{\ensuremath{\mathcal{L}}}

\newc\lsim{\ensuremath{\mathrel{\rlap{\lower4pt\hbox{\hskip1pt$\sim$}}\raise1pt\hbox{$<$}}}}
\newc\gsim{\ensuremath{\mathrel{\rlap{\lower4pt\hbox{\hskip1pt$\sim$}}\raise1pt\hbox{$>$}}}}
\newc{\VEV}[1]{\ensuremath{\langle #1 \rangle}}

\newc{\ra}{\rightarrow}

\newcommand{\DSf}{\mbox{DarkSide-50}}
\newcommand{\DSk}{\mbox{DarkSide-20k}}
\newcommand{\DEAP}{\mbox{DEAP-3600}}
\newcommand{\DSl}{\mbox{DarkSide-LM}}
\newcommand{\mCLEAN}{\mbox{MiniCLEAN}}
\newcommand{\ArDM}{\mbox{ArDM}}

\newcommand{\GADM}{\mbox{GADM}}
\newcommand{\GADMC}{\mbox{\GADM C}}
\newcommand{\pDUNE}{\mbox{ProtoDUNE}}
\newcommand{\CEnNS}{\mbox{CE$\nu$NS}}
\newcommand{\LAr}{\ce{LAr}}
\newcommand{\AAr}{\ce{AAr}}
\newcommand{\UAr}{\ce{UAr}}
\newcommand{\grs}{\mbox{$\gamma$-rays}}

\DeclareSIUnit\c{\mbox{$c$}}
\DeclareSIUnit\year{year}
\DeclareSIUnit\ev{events}

\makeglossaries

\newacronym{wimp}{WIMP}{weakly interacting massive particle}
\newacronym{cmb}{CMB}{cosmic microwave background}
\newacronym{dd}{DD}{direct detection}
\newacronym{dm}{DM}{dark matter}
\newacronym{id}{ID}{indirect detection}
\newacronym{alp}{ALPs}{axion-like particles}
\newacronym{bh}{BH}{black hole}
\newacronym{pbh}{PBH}{primordial black hole}
\newacronym{cdm}{CDM}{cold dark matter}
\newacronym{macho}{MACHOs}{MAssive Compact Halo Object}
\newacronym{sm}{SM}{Standard Model of particle physics}
\newacronym{bsm}{BSM}{beyond the Standard Model}
\newacronym{ew}{EW}{electroweak}
\newacronym{hdm}{HDM}{hot dark matter}
\newacronym{wdm}{WDM}{warm dark matter}
\newacronym{susy}{SUSY}{supersymmetry}
\newacronym{mssm}{MSSM}{Minimal Supersymmetric Standard Model}
\newacronym{cmssm}{CMSSM}{Constrained Minimal Supersymmetric Standard Model}
\newacronym{eft}{EFT}{effective field theory}
\newacronym{adm}{ADM}{asymmetric dark matter}
\newacronym{fimp}{FIMP}{feebly interacting massive particle, also called extremely weakly interacting massive particle (EWIMP) or super-WIMP}
\newacronym{cp}{CP}{charge conjugation parity symmetry}
\newacronym{qcd}{QCD}{Quantum Chromodynamics}
\newacronym{edm}{EDM}{electric dipole moment}
\newacronym{nedm}{nEDM}{electric dipole moment of the neutron}
\newacronym{gut}{GUT}{Grand Unified Theory}
\newacronym{pngb}{pNGB}{pseudo Nambu Goldstone boson}
\newacronym{pq}{PQ}{Peccei-Quinn}
\newacronym{ksvz}{KSVZ}{Kim–Shifman–Vainshtein–Zakharov: axion benchmark model}
\newacronym{dfsz}{DFSZ}{Dine–Fischler–Srednicki–Zhitnitsky: axion benchmark model}
\newacronym{mond}{MOND}{MOdified Newtonian Dynamics}
\newacronym{idm}{iDM}{inelastic dark matter}
\newacronym{ccd}{CCD}{charge-coupled device}
\newacronym{hpge}{HPGe}{high purity germanium}
\newacronym{tes}{TES}{transition edge sensor}
\newacronym{tpb}{TPB}{tetraphenyl butadiene}
\newacronym{psd}{PSD}{pulse shape discrimination}
\newacronym{tpc}{TPC}{time projection chamber}
\newacronym{roi}{ROI}{region of interest}
\newacronym{er}{ER}{electronic recoil}
\newacronym{nr}{NR}{nuclear recoil}
\newacronym{lar}{LAr}{liquid argon}
\newacronym{lxe}{LXe}{liquid xenon}
\newacronym{fid}{FID}{fully interdigitised design}
\newacronym{si}{SI}{spin independent}
\newacronym{ntl}{NTL}{Neganov-Trofimov-Luke}
\newacronym{ntd}{NTD}{neutron transmutation doped}
\newacronym{0nu}{0$\nu$DBD}{neutrinoless double beta decay, also 0$\nu\beta\beta$}
\newacronym{lngs}{LNGS}{INFN Laboratori Nazionali del Gran Sasso}
\newacronym{hemt}{HEMT}{high-electron-mobility transistor}
\newacronym{dbd}{DBD}{double beta decay}
\newacronym{sd}{SD}{spin dependent}
\newacronym{snews}{SNEWS}{SuperNova Early Warning System}
\newacronym{pmt}{PMT}{photomultiplier tube}
\newacronym{cdr}{CDR}{conceptual design report}
\newacronym{sipm}{SiPM}{silicon photomultiplier}
\newacronym{simp}{SIMP}{self-interacting massive particle}
\newacronym{ec}{EC}{electron capture}
\newacronym{swot}{SWOT}{Strengths, Weaknesses, Opportunities, Threats}
\newacronym{aar}{AAr}{atmospheric argon}
\newacronym{uar}{UAr}{underground argon}
\newacronym{pdm}{PDM}{photodetector modules}
\newacronym{pmma}{PMMA}{polymethyl methacrylate: acrylic}
\newacronym{cno}{CNO}{carbon-nitrogen-oxygen}
\newacronym{cevnns}{CE$\nu$NS}{coherent elastic neutrino nucleon scattering}
\newacronym{mwpc}{MWPC}{multi-wire proportional chamber}
\newacronym{mpgd}{MPGD}{micro pattern gaseous detectors}
\newacronym{nmr}{NMR}{nuclear magnetic resonance}
\newacronym{lsw}{LSW}{light-shining-through-the-wall}
\newacronym{capp}{CAPP}{Center for Axion and Precision Physics Research in South Korea}
\newacronym{jpa}{JPA}{Josephson parametric amplifiers}
\newacronym{rf}{RF}{radio frequency}
\newacronym{lsc}{LSC}{Laboratorio Subterraneo de Canfranc}
\newacronym{lsm}{LSM}{Laboratoire Subterrain de Modane}
\newacronym{am}{AM}{annual modulation}
\newacronym{snolab}{SNOLAB}{underground laboratory in Canada}
\newacronym{surf}{SURF}{Sanford Underground Research Facility in the United States}
\newacronym{cjpl}{CJPL}{China Jinping Underground Lab}
\newacronym{y2l}{Y2L}{Yangyang underground laboratory in South Korea}
\newacronym{boulby}{Boulby}{underground laboratory in the UK}
\newacronym{hp}{HP}{hidden photon}
\newacronym{bbn}{BBN}{Big Bang Nucleosynthesis}

\RequirePackage[normalem]{ulem}

\begin{document}


\title{\centering
       \mbox{\bf\Large Direct Detection of Dark Matter -- APPEC Committee Report  }
       \footnote{This report has received approval from APPEC
       (1 April 2021; https://www.appec.org/documents).
       } \\[0.2cm]
}
       
\author{\parbox{\textwidth}{\centering
        {\bf Committee Members:} \\
        \rm Julien Billard,$^1$ Mark Boulay,$^2$ Susana Cebri{\'a}n,$^3$
        Laura Covi,$^4$\\
        Giuliana Fiorillo,$^5$ Anne Green,$^6$ Joachim Kopp,$^7$
        B{\'e}la Majorovits,$^8$\\
        Kimberly Palladino,$^{9,12}$ Federica Petricca,$^{8}$
        Leszek Roszkowski (chair),$^{10}$  Marc Schumann$^{11}$ \\[0.2cm]
        }}

\institute{\centering\it
           $^1$ Univ  Lyon,  Universit\'e  Lyon  1,  CNRS/IN2P3,  IP2I-Lyon,  F-69622,  Villeurbanne,  France \\
           $^2$ Department of Physics, Carleton University, Ottawa, Canada \\
           $^3$ Centro de Astropartículas y Física de Altas Energías,\\ Universidad de Zaragoza,  Zaragoza, Spain \\
           $^4$ Institute for Theoretical Physics,
Georg-August University, Goettingen, Germany \\
           $^5$ Physics Department, Università degli Studi “Federico II” di Napoli\\ and INFN Napoli, Naples, Italy  \\
           $^6$ School of Physics and Astronomy, University of Nottingham, Nottingham, UK \\
           $^7$ CERN, Geneva, Switzerland 
                and Johannes Gutenberg University, Mainz, Germany \\
           $^8$ Max-Planck-Institute for Physics, Munich, Germany \\
           $^9$ Department of Physics, University of Wisconsin - Madison, Madison, WI, USA \\
           $^{10}$ Astrocent, Nicolaus Copernicus Astronomical Center PAS and\\ National Centre for Nuclear Research, Warsaw, Poland \\
           $^{11}$ Institute of Physics, University of Freiburg, Freiburg, Germany \\
           $^{12}$ Department of Physics, Oxford University, Oxford, UK \\
           }
           
\date{\today}

\begin{abstract}
  This Report provides an extensive review of the experimental programme of direct detection searches of particle dark matter. It focuses mostly on European efforts, both current and planned, but does it within a broader context of a worldwide activity in the field. It aims at identifying the virtues, opportunities and challenges  associated with the different experimental  approaches and search techniques. It presents  scientific	and technological synergies, both  existing and emerging, with some other areas of particle physics, notably collider and neutrino programmes, and beyond. It addresses the issue of infrastructure in light of the growing needs and challenges of  the different experimental searches. Finally, the Report makes a number of recommendations from the perspective of a long-term future of the field. 
  {They  are introduced, along with some justification, in the opening Overview and Recommendations section and are next summarised at the end of the Report. Overall, we recommend that the direct search for dark matter particle interactions with a detector target should be given top priority in astroparticle physics, and in all particle physics, and beyond, as a positive measurement will provide the most unambiguous confirmation of the particle nature of dark matter in the Universe.}
\end{abstract}

\maketitle
\setcounter{tocdepth}{3}
\tableofcontents

\newpage

\section{Overview and Recommendations}
\label{sec:overview}

The nature of invisible dark matter (DM) that constitutes some $26\%$ of the mass-energy balance of the Universe remains one of the most fundamental puzzles in physics today. The most compelling solution to the DM enigma is provided by postulating some new elementary particle that must be outside of the spectrum of the Standard Model (SM) -- which in fact provides one of the strongest arguments in support of ``new physics'' beyond the SM (BSM).  The hypothetical relic particle is probably cold, i.e.,  non-relativistic, but otherwise in principle not much is known about its basic properties, with the allowed mass range spanning nearly fifty decades, while non-particle DM, for instance in the form of (primordial) black holes, can be even much heavier. Likewise,  so far only gravitational effects of  DM have been observed. Nevertheless, there are good reasons to expect that a particle DM relic exhibits also much less feeble interactions, up to the (electro)weak ones of the SM, 
as is the case of weakly interacting massive particles (WIMPs) in a large variety of BSM models, or intermediate 
ones, in the case of axions and, more generally, pseudoscalar axion-like particles (ALPs).

A decades-long, intense, trans-national and increasingly global worldwide experimental WIMP search is conducted following three main strategies: direct detection (DD) of the scattering of a DM particle off a target in deep underground detectors, indirect detection (ID) of exotic products of DM pair annihilation (or possibly decay) in the Galactic halo and beyond, and their production in accelerators or fixed-target experiments. The underlying principle to search for axions and ALPs in haloscopes (DM axions), helioscopes (solar axions) and laboratory experiments is to primarily make use of axion-photon conversion in the presence of a strong magnetic field, but also other complementary couplings to electrons or nuclei are used.

The prime scientific objectives of DD, in both WIMP and axion/ALP searches, are: (i) to detect a direct interaction of a DM particle with a detector, and (ii) to determine its mass and interaction cross section, or else (iii) to experimentally exclude the broadest accessible ranges of both quantities.  
A detection of a DM particle will clearly constitute a historical landmark in the exploration of the invisible Universe. It will confirm its particle nature and will open a new window on the late, and also very early, Universe. At the same time, it must be emphasised that it will only mark the first step in a long quest to unravel the true nature of the particle constituting DM that will require employing a multi-channel, multimessenger approach combining information also from ID, collider searches, astrophysics, cosmology and astronomy.

{\bf Recommendation~1.\ \ \ 
The search for dark matter with the aim of detecting a direct signal  of DM particle   interactions with a detector should be given top priority in astroparticle physics, and in all particle physics, and beyond, as a positive measurement will provide the most unambiguous confirmation of the particle nature of dark matter in the Universe.}\\[-0.1cm]

\vspace{0.5cm}
Until recently most experimental efforts concentrated on exploring: (i) the mass range from a few $\gev/c^2$ up to the $\tev/c^2$--scale of the commonly called  (thermal) WIMP class of candidates (notwithstanding nuances of theoretical nomenclature), and (ii) the $\sim\!\!\muev/c^2$--mass range of so-called QCD axion. Both classes are predicted  by 
multiple ``new physics'' frameworks addressing other problems of the SM, and, in addition, often have both properties expected of cold DM and are predicted to be detectable. With these two classes of hypothetical particles still remaining by far the most strongly motivated DM candidates, over the last decade or so  the experimental search programme broadened considerably.  In underground WIMP searches, much more effort was put into exploring the previously poorly-probed mass range below a few $\gev/c^2$, down to the sub--$\!\gev/c^2$ regime, and even lower, to the  $\mev/c^2$--scale.  Search for particles interacting like axions   -- not necessarily constituting DM, e.g., solar axions  -- use a variety of methods to probe a very wide mass range from $\sim\!\! 10^{-22}\ev/c^2$ to $\sim\!\!1\ev/c^2$.  

\newpage
{\bf Recommendation~2.\ \ \ 
The diversified approach to probe the broadest experimentally accessible ranges of  particle mass and interactions  is needed to ensure the most conservative and least assumption-dependent exploration of hypothetical candidates for cosmological dark matter or subdominant relics. }\\[-0.1cm]

\vspace{0.5cm}
Over the last thirty years or so a tremendous and sustained increase in detection sensitivity -- on average of nearly three orders of magnitude per decade, possibly the biggest in science and maybe also industry  --   has been achieved by direct detection experiments  searching for WIMPs, due to many technological advances, availability of underground laboratory infrastructure and scientific effort. The biggest advances were made by mature technologies aiming at measuring nuclear recoils  at time projection chambers (TPCs) using  liquid noble gas in probing the  WIMP mass range from $\sim\!\!\gev/c^2$  up to $\sim\!\!\tev/c^2$,   and by bolometers (both scintillators and semiconductors) in exploring the low mass range down to the $\mev/c^2$ regime.

TPC detectors using the liquefied noble gases argon or xenon as target are very successful in exploring 
{WIMPs -- within the ranges specified above -- } via detecting their nuclear recoils and have played the leading role over the last several years by placing the strongest limits. In fact, the liquid xenon experiments XENONnT (mainly Europe-USA-Japan), LZ (mainly USA-UK) and PandaX-4T (mainly China) -- currently all preparing for data taking in 2020 -- are expected to improve the current best limits of their predecessors in WIMP-nucleon elastic scattering cross section by more than a factor of ten in the next years. A similar sensitivity will be reached by the liquid argon TPC DarkSide-20k (mainly Europe-USA-Canada) -- at present under construction. Due to the different masses of the target nuclei and experimental thresholds xenon and argon detectors are more sensitive to lower and higher side of this mass range, respectively. 

The noble-gas TPC detector technology is highly promising as, by the virtue of being scalable to very large target masses, as it allows acquiring high exposures. 
At low mass, below a few $\gev/c^2$, the sensitivity becomes much poorer, although electron recoil and other recently used detection channels allowed one to improve sensitivity to lighter DM candidates, such as low-mass WIMPs or other DM candidates, \eg ALPs or hidden photons.
Complementary technologies with a better sensitivity are thus crucial to explore the few to sub--$\!\gev/c^2$ region and that is where cryogenic detectors (bolometers) are best suited to take a leading position.

European leadership  in highest-sensitivity experiments is  very strong, and providing key contributions, including innovative technologies, extensive research and development (R\&D) effort, and also significant funding for LXe (XENONnT), LAr (DarkSide) and two out of three most sensitive bolometer-type experiments  (CRESST, EDELWEISS).

Liquid noble gas detectors and bolometers will face another challenge -- that will be both an obstacle and opportunity -- of the irreducible background due to the so-called neutrino floor of coherent neutrino interactions with target nuclei. Initial detection is likely to be achieved by XENONnT, LZ and PandaX-4T, around the $6\gev/c^2$ WIMP mass range, just before the solar branch of the neutrino floor  turns abruptly  down into the atmospheric branch with the drop of some five orders in magnitude.  Reaching down to within an order of magnitude from the solar neutrino floor is the main goal of DarkSide-LM on the LAr side, the LXe-based projects XENONnT, LZ and PandaX-4t, and of the bolometer experiments CRESST-III, EDELWEISS-subGeV, as well as SuperCDMS. This is expected to be achieved within the next decade.  

Taking the ten-year perspective and beyond, on the bolometer side, the technology to access low mass DM is already mature enough to start the planning on a tonne-scale experiment to reach down to the solar neutrino floor. Above some $10\gev/c^2$, an improvement of some three orders of magnitude in sensitivity will be needed to reach down to the atmospheric neutrino floor, which would be the ultimate goal of  future liquid noble gas experiments, and for which a substantial amount of innovative R\&D of both LAr and LXe programmes will be needed. The LAr groups from ArDM, DarkSide-50, DEAP-3600 and MiniCLEAN have recently formed the Global Argon DM Collaboration (GADMC) to build \DSk\ and its successor, the $300\tonne$ detector ARGO. On the xenon side, R\&D is ongoing to build the next generation detector DARWIN with a $40\tonne$ active target and a comparable reach, and towards multi-tonne liquid xenon detectors in general.
It is evident that the outcome of these searches will likely have tremendous implications not only for the field of DM itself -- by opening up a new window on the Universe --  but actually on all particle physics, and beyond. For instance, a discovery of a massive, $10\gev/c^2$--scale or heavier, DM particle in underground experiments in the next decade would provide arguably a very strong rationale to boost construction of a new-generation particle collider capable of opening up a new era of discoveries in the dark sector of physics.

{\bf Recommendation~3.\ \ \ 
The experimental underground programmes with the best sensitivity to detect signals induced by dark matter WIMPs scattering off the target should receive enhanced support to continue efforts  to  reach down to the so-called neutrino floor on the shortest possible timescale. }\\[-0.1cm]

\vspace{0.5cm}
The long-standing  claim  from DAMA/LIBRA of detecting a DM signal via annual modulation using a NaI(Tl) target -- while being in severe conflict with results from  several experiments using different targets -- needs to be independently verified using the same target material. This is the main goal of two currently running experiments, ANAIS (Spain) and COSINE (South Korea), as well as some more in preparation. Depending on the outcome 
{in some two years}, the community will likely be in a position to outline their longer-term plans.

Directional detectors aiming at reconstructing the direction and energy of the WIMP-induced nuclear recoil offer an unambiguous way of confirming the Galactic origin of a WIMP signal. Several efforts worldwide, including European-led projects, are underway, and while, due to technological challenges and significantly lower target masses, currently lagging behind conventional WIMP detectors in terms of sensitivity, in the future they may offer some other potential advantages, \eg of reaching down below the neutrino floor. It is  vital to pursue and support this effort as a longer term investment in the field that, after a detection of a DM signal, may be most effective in exploring a new window on the Universe in terms of ``DM astrononomy''. 

Several results on WIMP-electron scattering in the sub--$\!\geV/c^2$ range have already been published.
Extending the search down to the $\mev/c^2$ scale can also be done via DM-electron interactions. This region remained basically unexplored until very recently due not only to the lack of strong theoretical motivation but also because it requires using new targets and/or developing new search techniques. Over the last few years it has been proposed, mostly in the USA, to detect sub--$\!\mev/c^2$ DM.  While these novel approaches may potentially open up new opportunities, it should be noted that this is {\it terra incognita} not only in terms of the potential to discover an ultra-light WIMP particle, but also on the side of facing potentially unexpected challenges due to poorly known, or unexpected, atomic and condensed matter physics effects.

Detector technology R\&D is pursued in order to advance the field 
beyond current limitations and to explore novel ideas.  Current efforts include developing radioactivity reduction programs, 
new sensors and readout techniques, 
new target materials, methods and technologies to scale up target masses as well as novel analysis strategies.  
Potential synergies across the different areas of detector development, in particular also between direct WIMP and axion searches,  
and possible applications in other fields (\eg, quantum technologies, medical physics)
should be exploited by an enhanced  coordination of the R\&D activities and by improving collaboration with industry, in a way similar to what is done collider physics. 

{\bf Recommendation~4.\ \ \ 
European participation in DM search programmes and associated, often novel, R\&D efforts, that currently do not offer the biggest improvement in sensitivity should continue and be encouraged with view of a long-term investment in the field and the promise of potential interdisciplinary benefits. We recommend that coordinated programmes are established for dark matter detector development.  }\\[-0.1cm]

\vspace{0.5cm}
The bedrock of European experimental  leadership  has been the role played by large world-class  laboratories and support facilities, as over the years has been prominently demonstrated by the example of CERN in leading the world's collider physics programme.  In underground science, in contrast, the effort is somewhat divided and scattered among several deep underground laboratories, primarily LNGS in Italy that over the years has become a European hub for many leading neutrino physics and DM search experiments.  

Current and planned DM, and other underground science programmes, with their highly ambitious goals of reaching unprecedented sensitivities face steeply raising needs, not only in terms of larger investment of human, technological and financial resources, but also in terms of an infrastructure for the 21st century. These new challenges call for a much greater international cooperation in sharing existing infrastructure and know-how, as well as for new bold, cross-disciplinary initiatives.  Importantly, these growing demands can to a large extent be met by making a much more efficient use of existing resources.  Recent attempts to form a network of Deep Underground Laboratories in Europe need to be re-invigorated and opened to cooperation with facilities on other continents.

There is  emerging need  to combine the current national underground laboratories into a single legal entity, a {\em European Laboratory of Underground Science} that, while geographically distributed 
across different countries, could have the status of an international organisation. 
The most efficient way to implement this strategic initiative would be to use the platform of a European Research Infrastructure Consortium (ERIC).  Such a 
new European laboratory would provide a coordinated environment 
not only for dark matter direct detection experiments but also for other underground-based areas of astroparticle physics in Europe and beyond, following on the successful example of CERN in growing into the world hub for the particle physics programme. We recommend that a working group is formed to investigate this possibility further.

The creation of the {\em European Laboratory of Underground Science} would help foster in Europe closer and much more efficient cooperation between the European astroparticle physics community,  national laboratories across Europe and CERN in order to  more effectively  exploit the numerous cross-disciplinary synergies  and address common technological challenges facing particle and astroparticle physics for the next decade and beyond. 
We recommend a closer collaboration between these communities to further the progress of both particle and astroparticle physics
that cannot be treated as separate  anymore and in fact are increasingly interdependent. 

{\bf Recommendation~5.\ \ \ The long-term future of underground science in Europe would strongly benefit from creating a distributed but integrated structure of underground laboratories for the needs of the forthcoming generation of new experiments, and beyond. This strategic initiative would be most efficiently implemented by forming the {\em European Laboratory of Underground Science}.} \\[-0.1cm]

\vspace{0.5cm}
Experimental searches for axions, ALPs, and also other hypothetical particles exhibiting similar interactions, 
are performed using a variety of approaches, notably in  cavity and dielectric haloscopes (as dark matter), helioscopes (solar axions), in laboratory experiments, and several others. Because of the primary detection principle of axion/ALP-photon conversion, many of them require dedicated  high-field superconducting magnets with  sufficient aperture and access to appropriate cryogenic and other infrastructure in  well-equipped experimental halls with stable, low electromagnetic background, and necessary know-how, that are available in large laboratories. 
The benefit of support of large European and national labs  was demonstrated, e.g., by the CAST and OSQAR experiments at CERN and ALPS at DESY.

Two prime objectives of the efforts are: (i) to reach sensitivity to detect  the QCD axion  as a highly motivated DM candidate,   and (ii) to explore the widest possible ranges of axion and ALP masses, from $\sim\!\! 10^{-13}\ev/c^2$ (for ALPs even lower) up to $\sim\!\!10\ev/c^2$. 

Over the last years and decades steady progress with increasingly diverse and complementary approaches to explore axion and ALP relics has brought us much closer to this ambitious target, and many upcoming and next-generation of experiments will be reaching sensitivity to a large fraction of QCD axion models within the expected mass range. In the context of the worldwide activity, in recent years European-led teams have obtained some very promising R\&D results that put them in a leading position to set up new and unique experiments directly probing the so far unexplored mass ranges between $20\muev/c^2$ and $10\miliev/c^2$ as well as below $1\nev/c^2$ for dark matter. 
These novel approaches, e.g.,  ALPS II, babyIAXO and MADMAX to be located at DESY Hamburg that is developing to become one of the world hubs for axion/ALP searches, as well as the other promising smaller scale European haloscope projects, for example at CERN or INFN LNL, will complement the presently leading experiments being performed in the USA and South Korea that have been focusing on the mass range $1\muev/c^2 - 20\muev/c^2$.

In parallel,  R\&D efforts towards proving the applicability of some key novel technologies, like sub-quantum limited and single photon detection techniques at the relevant frequencies or the development of low-loss dielectric and meta-materials, as well as experimental demonstration of new conceptual ideas in all mass ranges relevant for axion and ALP dark matter should be supported to further increase experimental sensitivities and range. This would also  prepare the ground  for axion astronomy in case such a dark matter particle is found.

{\bf Recommendation~6.\ \ \
European-led efforts should focus on  axion and ALPs mass ranges that are
complementary to the established cavity approach and this is where European teams have a  unique opportunity to secure the pioneering role in achieving sensitivities in axion/ALP mass ranges not yet explored by experiments conducted elsewhere. In parallel,  R\&D efforts to improve experimental sensitivity and to extend the accessible mass ranges should be supported.} \\[-0.1cm]

\vspace{0.5cm}
Elucidating the nature of dark matter remains one of the most intense research areas in the  theoretical community, both particle and astrophysical, worldwide, and the effort is often stimulated by experimental advances, and vice versa. This activity is essential  since in general a full assessment of the implications of experimental data -- both in terms of limits and eventually positive measurements, and  including all relevant information in the spirit of the multi-channel, multimessenger approach -- for DM particle properties and underlying physical frameworks cannot be achieved in a simple, model-independent way.  Each type of DM and related collider searches for new physics  necessarily provide only  limited amount of information on some specific measurable quantities, along with their inherent uncertainties. Therefore phenomenological studies are key to allow one to infer their combined impact on specific theoretical frameworks while taking into account underlying assumptions and uncertainties coming from particle and nuclear physics, as well as astrophysics, astronomy, and  Big Bang cosmology. Theoretical efforts to interpret various  anomalies and unexpected results of potential DM origin that are sometimes reported in astrophysical observations often help advancing the understanding of many relatively poorly understood astrophysical environment and processes, \eg in the centre of the Milky Way. 

The European theoretical particle and astrophysics community continues to contribute in several key ways to the worldwide effort, including not only ideas but also notably  practical tools, \eg some widely popular and publicly available numerical packages. The enigma of DM particle, strongly overlapping with the subject of new physics beyond the Standard Model, has many natural, and often fruitful, links with several other vital areas of particle physics, including neutrino physics, in particular neutrino less double beta decay, flavour, \etc and also astrophysics and Big Bang cosmology.

{\bf Recommendation~7.\ \ \ 
Continuing dedicated
and diverse theoretical activity should be encouraged not only in its own right but also as it provides some highly stimulating, and mutually  beneficial, interdisciplinary environment for DM and new physics searches.
}


\newpage

\section{Introduction}
\label{sec:intro}


The nature of dark matter (DM) in the Universe remains one of the key unresolved questions in particle physics, astrophysics and cosmology today. The existence of dark matter -- as it was   called by F.~Zwicky \cite{Zwicky:1937zza}, and also others before him -- 
is supported by a wealth of observational evidence that has revealed that  some 26\% of the total mass-energy balance in the Universe comes in the form of non-baryonic, non-dissipative and cold DM; for recent reviews, see, \eg\cite{Massey:2010hh,Popolo2014NONBARYONICDM,Salucci2018DarkMI}. It is generally assumed that it is probably made up of some new, and yet undiscovered, elementary particle that cannot be part of the Standard Model (SM) of particle physics – which in fact provides one of the strongest arguments in support of “new physics” beyond the SM. For recent reviews, see, 
\eg\cite{Baer:2014eja,Gelmini:2016emn,battaglieri2017cosmic_temp,Roszkowski:2017nbc,lin2019tasi,Profumo2019AnIT}.

Of the very many particle (and also non-particle) candidates for resolving the DM puzzle, two main classes  
have over the last decades attracted most attention of both the theoretical as as the experimental community.  One is a {\bf weakly interacting massive particle (WIMP)}  produced thermally via freeze-out, and is commonly called {\bf ``thermal WIMP"}, or just {\bf ``WIMP"}.\footnote{The notion of WIMP used in the literature is actually not unique, nor well defined, as will be explained below. Also, it does not stand for one concrete particle but rather for a wide spectrum of particles.} The other class is an {\bf axion}, produced non-thermally, 
and much experimental  effort has been devoted worldwide to detecting both of them. Of course, one cannot exclude the possibility of having more than one relic species of dark matter, as well, although this option appears to be less appealing for the reason of simplicity.

WIMP  searches have been conducted along three main avenues.  {\bf Direct detection (DD)} of WIMP DM provides the most straightforward  way of discovering this type of DM particles on Earth by attempting to measure the scattering of a DM particle off a target in underground detectors. Another strategy, called  {\bf indirect detection (ID)}, encompasses several possible ways of detecting some exotic products of DM pair annihilation (or possibly decay) in the Galactic halo of the Milky Way and beyond. Finally, particles constituting DM could be produced and detected at  {\bf accelerators} or fixed-target experiments.

The underlying principle to search for {\bf axions} and   {\bf axion-like particles (ALPs)} in {\bf haloscopes} (DM axions), {\bf helioscopes} (solar axions) and {\bf laboratory experiments} mostly relies on their coupling to photons in the presence of a strong magnetic field,  although 
also the coupling to electrons and nuclei are used.  

Additional search windows are provided by astronomical and astrophysical observations, including recently discovered gravitational waves. It is worth stressing that such a {\bf multimessenger} approach will most likely be absolutely key to identify the true nature of a DM particle, be it a WIMP or axion, or something else. This is because, in general, when a DM signal is eventually detected, no single experiment, or type of experiments, will be able to provide all necessary information to establish all DM particle properties. 

Other possible explanations for DM have been proposed, for instance in terms of black holes (BHs), most notably primordial black holes (PBHs), other astrophysical objects (e.g., MACHOs), modifications of gravity, etc, and are briefly mentioned below.

This Report is primarily devoted to a {\em review of the experimental programmes of direct detection searches of particle DM. It focuses mostly on European} experiments,\footnote{ For the purpose of this Report  the term ``European experiments" is meant to include experiments with a dominant, or at least strong,  involvement of at least one European country. In a majority of cases such experiments at present are also located in Europe.} current and planned, but takes into account a broader context of a worldwide activity in the field. In light of the trend of experimental collaborations to grow in size, often through merging and consolidation, and looking into the future,  we aim at identifying the virtues and challenges   associated with the different experimental  approaches and search techniques. We point out synergies and technological spin-offs, and discuss infrastructure needed for providing a long term future to the program. While we attempted to provide a rather complete description of the European  programmes of direct detection searches, we only briefly mentioned, or left out altogether, many other activities done elsewhere, as well as many theoretical DM candidates or ideas -- an attempt at summarising them in a more complete way would take us far beyond the scope of the Report. The mandate of the Report, as defined by the Scientific Advisory Committee of APPEC, is provided in the Appendix.

The first part of the Report (Section~\ref{sec:theory}) is devoted to a review of the evidence for the existence of DM and of the theoretical status of DM candidates, their types, mechanisms of production and general properties. The experimental DM search programs have primarily focused on searching for the (thermal) WIMP and axionic states. In reviewing experimental searches for WIMPs and axions in Sections~\ref{sec:experiment} and \ref{sec:axiondetection}, respectively, 
we will, however, try not to be too strongly motivated by specific theoretical ideas or preferences. In Section~\ref{sec:broadercontext} we will paint a broader context of placing them among other search strategies via indirect detection, colliders, and astrophysical probes, in the multimessenger approach, and also present a a brief discussion of laboratory facilities in Europe and elsewhere.
We conclude the Report with a list of recommendations in Section~\ref{sec:finalrecom}, which are also presented, along with an extended overview, in Section~\ref{sec:overview}.

\section{Dark Matter -- Evidence, Properties and Candidates  }\label{sec:theory}

In this Section we first present the main arguments for the existence of dark matter, next we review its main properties and production mechanisms. This is followed by an overview of various types of particle DM candidates, ranging from the thermal WIMP to its several variants and alternatives, to the axion and related particles. 
Alternatives to particle DM are also briefly addressed. In this section we set $c=1$.

\subsection{Evidence for Dark Matter}
\label{sec:evidence}

Diverse astronomical and cosmological observations, on scales ranging from galaxies to the entire Universe, provide powerful evidence that $85\%$ of the matter in the Universe is in the form of cold, non-baryonic dark matter (CDM). 

\subsubsection{Astronomical and Cosmological Evidence for Non-Baryonic Dark Matter}
\label{sec:astroevidence}

The rotation curves of disc galaxies are close to flat at large radii, rather than having a Keplerian, $r^{-1/2}$, decline as expected from the luminous components. If Newton's law of gravitation is correct, this indicates that these galaxies are surrounded by invisible, extended, close to spherical, DM halos (for a historical review see Ref.~\cite{Bertone:2016nfn}). Using measurements of the velocity dispersion of galaxies and the virial theorem, galaxies clusters have been shown to contain a large fraction of DM~\cite{Zwicky:1937zza}. Furthermore the energy spectrum and radial flux of the X-rays emitted by the hot gas in clusters demonstrate that the majority of the matter they contain is non-baryonic (e.g., Ref.~\cite{Gonzalez:2013awy}), while gravitational lensing observations allow the mass distribution to be mapped out (e.g., Ref.~\cite{Tyson:1998vp}). Dark matter is also required for the small initial density perturbations to grow sufficiently to produce the observed large scale structure and also to explain the heights of the acoustic peaks in the Cosmic Microwave Background (CMB) angular power spectrum (e.g., Ref.~\cite{Holtzman:1989ki}).

\subsubsection{Observational Probes of the Nature and Abundance of Dark Matter}
\label{sec:dmabundance}

The anisotropies in the CMB allow precise measurements of the  cosmological parameters. A combined analysis, including large scale structure data (baryon acoustic oscillations), gives $\Omega_{\rm CDM} h^2 = 0.119 \pm 0.001$, where the {\bf CDM relic density parameter}, $\Omega_{\rm CDM} \equiv \rho_{\rm CDM}/\rho_{\rm c}$, is the present day CDM density, relative to the critical density $\rho_{\rm c}$ and $h$ is the Hubble constant in units of $100  \kmeter \second^{-1} \mpc^{-1}$~\cite{Aghanim:2018eyx}. Using the same data the baryon density parameter is $\Omega_{\rm b} h^2 = 0.0223 \pm 0.0001$~\cite{Aghanim:2018eyx}, consistent with the 
measurement from Big Bang Nucleosynthesis (BBN) and the primordial abundances of the light elements~\cite{Tanabashi:2018oca}, so that $\sim 85\%$ of the matter is cold DM. The event rate in WIMP direct detection and axion haloscope experiments is directly proportional to the  local (i.e., at the Solar location) DM density. This can be measured via a variety of methods (for reviews see Ref.~\cite{Read:2014qva,2020arXiv200411688S}) and a standard value of $\rho_{0} = 0.3  \gev  \cmeter^{-3}$ is usually adopted. 
On small scales density fluctuations can be erased by the free-streaming of DM. The observed clustering of galaxies is reproduced by numerical simulations in which the majority of the DM  is cold, i.e.~non-relativistic during structure formation~\cite{Springel:2006vs}. 
Constraints on the free-streaming scale from Lyman-$\alpha$ forest data place a constraint on the mass of generic thermal relic DM, with the same temperature as the SM, of $m> 5.3  \kev$~\cite{Irsic:2017ixq, Viel:2005qj}.

The CMB anisotropies are sensitive to energy injection due to, for instance, DM annihilation into electromagnetically
charged final states. Using Planck data, this places a constraint on the WIMP annihilation cross section into electrons 
where $m_{\chi}$ is the WIMP mass and $\langle  \sigma v\rangle$ the thermally averaged annihilation cross section~\cite{Aghanim:2018eyx}. 
The separation of the baryonic and total mass in merging galaxy clusters, like the Bullet cluster, place an upper limit on the DM self-interaction cross section $\sigma/m_{\chi} < {\cal O}(1) \cmeter^{2} \gram^{-1} $ (see discussion in Ref.~\cite{Tulin:2017ara}).

\subsection{Particle Physics Candidates}
\label{sec:dmcandidates}

There are many frameworks beyond the Standard Model (ranging from complete theories to sketchy ideas) that contain viable DM  candidate particles. Here we first summarise the required properties of the DM particle before we discuss some of the most motivated or popular candidates in more detail.

\subsubsection{General Properties}
\label{sec:wimps}

Some general properties of DM particle candidates can be established already from observational evidence and numerical simulations alone.
Firstly, most of them favour {\bf cold dark matter (CDM)}, as described above, although some arguments favour warm dark matter (WDM).\footnote{
DM relics can be classified as {\bf hot, cold, or warm}, depending on how relativistic they are around the time of matter-radiation equality and how large is their free-streaming length during structure formation. 
Hot dark matter (HDM), in the mass range of up to a few tens of~eV, has a free-streaming length comparable to the scale of
galaxy clusters and therefore can only contribute a small fraction of the total DM density. A familiar (and known to exist) example of possible HDM is neutrinos with a tiny mass. WDM, as a thermal relic in the mass range of a few keV, has 
a free-streaming length of the size of \mpc s and has
been considered as a possible way of ameliorating some apparent problems of CDM -- for which the free-streaming length is negligible -- because it reduces the power spectrum on small scales, thus reducing the missing satellite problem of CDM~\cite{Lovell:2011rd}, although this has been disputed
~\cite{Kim:2017iwr}. 
}

Secondly, CMB properties imply that DM is {\bf non-baryonic}, although this  in itself does not put too severe restrictions on the possible forms of interactions of DM particles.  However, since DM particles do not emit photons (otherwise they would become visible) they must be electrically neutral. In fact, a more correct name for DM is ``invisible matter".
A similar argument applies to strong forces. DM would loose energy and fall into galactic centres which is not observed. 
More generally, at least some 90\% of DM {\bf does not dissipate its energy}~\cite{Fan:2013yva}.

{To conclude the discussion of DM interactions}, other than via gravity, DM particles should interact with ordinary matter preferably only weakly, where weak may stand for the familiar weak force, or instead some other (sub)weak force defined by some non-negligible coupling to the Standard Model (SM) particles. In other words, weak forces with which WIMPs communicate with the SM sector do not need to be of electroweak (EW) nature. 
DM particles could also interact with themselves, and this type of {\bf self-interactions} is in fact rather poorly constrained, as quoted above. DM particles should also be either absolutely {\bf stable},
or extremely long lived (for instance, a recent analysis finds a lower bound of at least 160\,Gyr~\cite{Audren:2014bca}). 

Finally, the {\em DM particle mass} range can span nearly 50 orders of magnitude, from values as tiny as $10^{-21}\ev$  (fuzzy DM)
for bosons with de Broglie wavelength of the order of typical sizes of 
dwarf galaxies~\cite{Safarzadeh:2019sre} up to the (reduced) Planck scale $\overline{M}_P \simeq 2\times 10^{18}\gev$ (above which it is difficult to consider DM particles as elementary).
This is as much as we can be fairly confident about the general properties of DM which, however, is only a first step towards identifying its real nature, since they can be easily satisfied by a wide range of specific particle candidates, or in fact classes of candidates.

Strong restrictions on the properties of DM particle candidates, and typically also specific models or scenarios that they are a part of, arise from requiring that: (i) DM relics are produced in the early Universe and (ii) they exhibit the correct density. Apart from the most well-known mechanism of thermal freeze-out, there are several distinct modes of non-thermal production of DM relics. As discussed below, both lead to quite distinct DM properties, and ensuing prospects for DM searches.

\subsubsection{Thermal WIMP Dark Matter from Freeze-Out {\label{sec:thermalWIMPs}}}

The class of DM candidates that over the last few decades gained most attention is dubbed {\bf thermal WIMP}.\footnote{While it is often called simply ``WIMP", here we deliberately use the name ``thermal WIMP" to stress how it is produced, and also to distinguish it from other notions of 
the term WIMP used in the literature. The notion of a general WIMP adopted here can basically encompass all possible types of particles that can in principle constitute CDM (or possibly WDM) in the Universe -- which actually reflects the original meaning of the term given to it by its authors Gary Steigman and M. Turner\cite{Steigman:1984ac} --  because it refers to the properties of some particle and not to the mechanism of its production in the early Universe. 
In this sense the term ``weakly interacting'' does not need to refer to (electro)weak interactions of the Standard Model. }
Thermal WIMPs were produced during the very early and hot stage of the Universe's evolution via a thermal {\bf freeze-out mechanism} -- arguably the most robust mechanism for generating WIMP DM relics --  when SM species and DM particles were in thermal equilibrium.  As
the Universe expanded and cooled, thermal WIMPs eventually froze out
of equilibrium with the thermal plasma. This decoupling happened when
the WIMP annihilation rate became roughly less than the expansion rate
of the Universe $\Gamma_{\textrm{ann}}\lesssim H\propto
T_f^2/\overline{M}_P$, where $T_f$ stands for the freeze-out
temperature (the index $f$ indicates that quantities are evaluated at the freeze-out time) and $\overline{M}_P$ is the reduced Planck mass. After freeze-out the WIMP comoving relic number density remained mostly constant.

The relic density can be computed with high accuracy by employing the Boltzmann equation. It is inversely proportional to the thermally averaged product of the cross section for WIMP pair-annihilation and their relative velocity $\langle\sigma_{\textrm{ann}}v\rangle $. 
Numerically one finds that, at freeze-out 
\begin{equation}
\langle\sigma_{\textrm{ann}}v\rangle_f \approx 3\times 10^{-26}\,\textrm{cm}^3/\textrm{s},
\label{eq:sigmavvalue}
\end{equation}
for which  the correct value of the thermal WIMP DM relic density can be obtained (see, \eg~\cite{Steigman:2012nb} for a more detailed study).\footnote{Actually, in many realistic models a very wide range of a few orders of magnitude both above and below $0.1$ is usually obtained, and in some cases an effect of comparable size can be produced by the related mechanism of WIMP {\bf coannihilation} with another nearly mass degenerate state~\cite{Griest:1990kh}. }
For typical velocities $v\approx 0.1\,c$ one obtains $\sigma_{\textrm{ann}}$  of
weak strength order $\sim\!\! 10^{-36}\cmeter^2(=1\pb)$ for WIMP mass around the Fermi scale.

This remarkable coincidence, often referred to as  the ``WIMP miracle'',  motivated a large amount of research into the possibility that DM particles,  in the form of thermal WIMPs, may be part of some ``new physics" beyond the Standard Model (BSM)\footnote{None of the SM species have the right properties to be a DM particle. In particular the neutrinos are too light and could possibly contribute to HDM only.}  with a mass scale not far from the EW  scale, for which there is independent strong theoretical motivation. 
Thermal WIMPs also generated strong experimental interest owing to the fact that they exhibit detection rates that fall into the sensitivity range of today's, or planned, detectors. In other words, they are {\em discoverable}.

Actually, thermal WIMP  properties can be substantially different from the EW scale. 
On dimensional grounds one often finds $\sigma_{\rm ann} \propto {g^4}/{m_\chi^2}$,
where~$g$ denotes the WIMP effective coupling to the SM sector and $m_\chi$ denotes its mass. Keeping the ratio fixed the correct relic density can be achieved for a very wide ranges of~$g$, from gravitational to strong, and also $m_\chi$, from $\sim\!\! 1\ev$ to $\sim\!\! 120\tev$,\footnote{The upper limit saturates the unitarity argument that $g^4\lesssim 4\pi$~\cite{Griest:1989wd}. }  consistent with the freeze-out mechanism~\cite{Feng:2008ya,Profumo:2013yn,Baer:2014eja}. 

In contrast, thermal fermionic WIMPs exhibiting EW-strength interactions with the SM sector via $\sigma\propto G_F^2 m_\chi^2$, even if they are suppressed by some orders of magnitude, must be heavier than some $4\gev$, which is the so-called Lee-Weinberg bound following from the requirement of not generating too much relic density, or, in popular terms, overclosing the Universe.\footnote{The limit was actually derived for massive neutrinos by several authors, see \cite{Kolb:1990vq}. The other option is that it is lighter than some $10\ev$, which would, however, make it HDM.} In fact, this type of the thermal WIMP in the mass range between a few GeV and a few TeV has in recent decades been most abundantly studied in the literature, and for this reason has often been  referred to as the {\em standard thermal WIMP}, or the standard WIMP, or  just the WIMP. The rough mass range of this popular class of thermal CDM WIMP candidates, however, assumes a specific, although typical, dependence of $\sigma_{\rm ann}$  on the WIMP mass. 
It can be relaxed down to some $100\mev$ (or even less  if asymmetry is allowed between the WIMP state and its antiparticle) when one assumes that the main messenger between the WIMP sector and the SM  sector is some additional light vector boson, nor does it apply to scalar WIMPs~\cite{Boehm:2003hm}. Finally, all these bounds hinge on the argument of not producing too much relic density within the usually assumed underlying framework of the standard thermal history of the Universe, and can be significantly, or basically completely, relaxed if additional entropy production is generated in less minimal  scenarios. In conclusion,
while in some sense the EW mass scale (within roughly an
order of magnitude, or so) can be regarded as being most naturally consistent with the thermal WIMP (see, \eg Refs.~\cite{Baer:2014eja,Roszkowski:2017nbc} for a detailed discussion), in fact no firm lower limit on the thermal WIMP mass exists, other than some $10\kev$ from the requirement that it constitutes CDM.

An analogous question arises regarding the detection rates of thermal WIMPs.  There is a fairly straightforward answer for the case of indirect detection but, unfortunately, not for direct detection. In the case of ID WIMPs pair annihilate into SM states with rates determined by $\sigma_{\textrm{ann}}v_{\rm today} \sim\! 3\times 10^{-26}\,\textrm{cm}^3/\textrm{s}$. This is of the same order as the value at freeze out, given by Eq.~(\ref{eq:sigmavvalue}), because the processes involved (or, in other words, the corresponding Feynman diagrams) are the same, and only the WIMP velocities are much lower today.
ID searches therefore have a fairly well defined target to reach, as long as the main interaction of the DM particle is with the SM.
If upper limits go well below that value for a particular ID search
mode, thermal WIMP models relying on that  channel can
be ruled out in a robust way.

The situation is very different for direct detection. WIMP scattering processes off SM particles are typically different from the annihilation ones and the corresponding scattering cross sections (see Sec.~\ref{sec:interactions})  may, and often do, differ by many orders of magnitude. They cannot be too large since, by the argument of crossing symmetry, the corresponding Feynman diagrams would reduce the WIMP relic abundance at freeze-out too much. They can however be very much lower than typical electroweak cross sections, or than the current best experimental limits, see Fig.~\ref{fig:si_status}, or in fact even than the neutrino floor.

Various theoretical predictions are usually strongly model and assumption dependent. Improving experimental upper limits have ruled out many regions of previously favoured parameter space, and, in some cases, even some specific models or exchange channels. {\em It would certainly be premature, however, to claim that the paradigm of the thermal WIMP has somehow been ``ruled out" on the basis of the lack of detection signal so far, simply because there exist no specific predictions for robust, model-independent target scattering cross sections to explore experimentally.} 

Studies of WIMP properties and prospects for detection occupy a large volume of papers and have been performed in a huge multitude of models exhibiting a wide range of approaches. In one popular strand well defined models are  employed at the phenomenological level, or derived from some more complete theory at some higher energy scale. A  popular and theoretically well-motivated example of this approach is effective low-energy {\bf supersymmetry (SUSY)} that was initially developed to 
provide a solution to the gauge hierarchy problem of the Standard Model and where an attractive DM candidate came out as a bonus.
A particularly well studied framework is the Minimal Supersymmetric Standard Model (MSSM) in which, when SUSY-breaking parameters are set at the unification scale, becomes highly predictive (e.g., the constrained MSSM, or CMSSM), but in more general allow a large ranges of WIMP mass and interactions. 

In fact, among thermal WIMP candidates the most prominent role has over the years been played by a {\bf neutralino} as the lightest state in models of low-energy SUSY because of promising detection rates; see, \eg \cite{Baer:2014eja,Roszkowski:2017nbc} 
for recent reviews. As a Majorana fermion coupled to the SM via EW-strength interactions (however, suppressed by mixing angles), its mass range as CDM is between a few~GeV and a few~TeV, with the exact depending on a specific SUSY model and its assumed parameter ranges. The neutralino is arguably the most popular example of the standard (thermal) WIMP class of CDM candidates.

Predictive and well-defined frameworks, like effective low-energy SUSY models, can be experimentally tested, and cross-examined in a variety of channels, including, in addition to DM, also collider searches and rare decays, which is often a virtue. On the other hand, experimental results presented as constraints on the parameter space of specific models cannot be easily translated to other models.  In some situations, especially for direct detection, low-energy {\bf effective field theories (EFTs)} are therefore often used as an alternative approach \cite{Fan:2010gt,Fitzpatrick:2012ix}.  An EFT includes only a minimal set of particles (for instance SM nucleons and the DM particles) and interactions. It does not address the question how these interactions arise in some underlying theory, even though for a given fundamental theory the corresponding EFT can be rigorously derived. As many ultraviolet-complete theories can reduce to the same EFT, constraints on the EFT apply to a broader class of models.

As an intermediate approach between complete theories and EFTs, numerous {\bf simplified models} have become popular over the last decade, especially in the context of searches for new physics at the Large Hadron Collider (LHC) but also in DM searches; see, \eg~\cite{Abdallah:2015ter,Abercrombie:2015wmb}. 
Simplified models are defined by a small number of new particles and their interactions, usually focusing on just one of many possible channels of interactions with the SM that is mediated by a single messenger (although less minimal models have also been studied). In more realistic models, like the MSSM, the channels are usually bundled together by a given model's parameters and the simplified-model approach allows one to map them out into more easily manageable one-channel scenarios. In this sense, while simplified models are not complete, nor model independent, they provide a convenient platform for placing experimental constraints on specific quantities, like masses and cross sections. In the limit of large messenger mass one connects with the EFT approach.

In the context of DM searches, simplified models typically contain the SM as one (visible) sector, a DM candidate, often as part of a dark sector, and a messenger sector -- often called {\bf ``portal''} -- containing one or more states that mediate SM--DM interactions. 
An example of a simplified but self-consistent model is the {\bf Higgs portal} where DM particle can be either a scalar or a fermion and DM--SM interactions are mediated by a SM Higgs doublet; see Ref.~\cite{Arcadi:2019lka} for a recent review.  The viable parameter space of the simplest Higgs portal models has been almost fully probed, with the most important constraints arising from direct detection experiments.   The {\bf dark photon portal} is another recently popular class of models in which a light thermal WIMP (either fermion or scalar), in the MeV mass range,  interacts with the SM sector via a dark photon (a new dark sector gauge boson)  that mixes with the usual photon via kinetic mixing.
In more elaborate scenarios an additional (dark) Higgs boson is also present in the dark sector. Those models are primarily testable in fixed-target experiments as typical WIMP direct detection rates are usually strongly suppressed. However, direct detection experiments are also increasing their 
sensitivity to these type of models by exploiting the DM-electron scattering mode \cite{Aprile:2019xxb, sensei2}.

\begin{figure}
     \centering
     \includegraphics[width=0.9\textwidth]{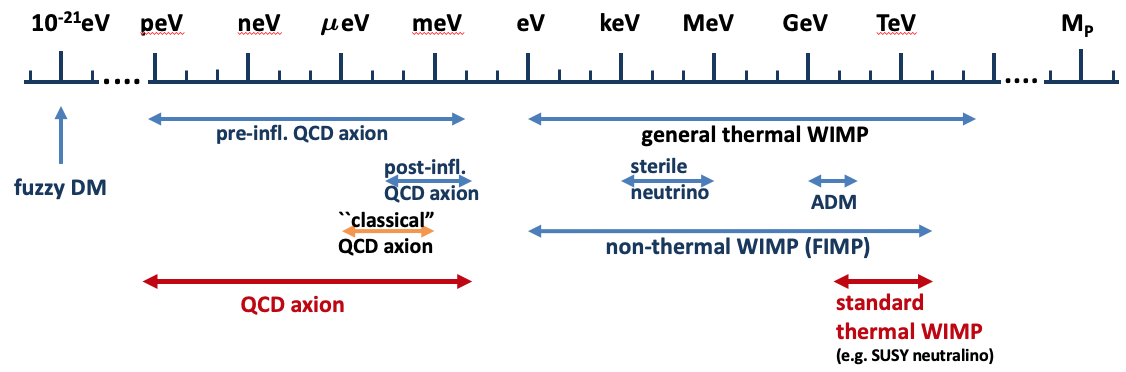}
     \caption{
     A summary of several particle candidates and classes of candidates for DM discussed in the Report. Shown are typical mass ranges, more details can be found in the text. }
     \label{fig:dmmassrange}
 \end{figure}

In another scenario called {\bf asymmetric DM (ADM)}~\cite{Graesser:2011wi,Iminniyaz:2011yp}  an asymmetry between the
DM particle and its antiparticle is generated 
in a way analogous to the mechanism of baryogenesis and modifies their freeze-out. In that case correct relic density can be obtained for DM typically in the mass range from $\sim\!\!1\gev$ to $\sim\!\!15\gev$ with large annihilation
cross section as (partially) asymmetric DM. Since in the ADM scenario the DM is not its own antiparticle and the abundance of $\chi$ and $\bar{\chi}$
particles can be highly asymmetric at present, the expected indirect detection rates from $\chi\bar{\chi}$ annihilations are typically 
suppressed with respect to, \eg~Majorana DM; see, \eg~\cite{Petraki:2013wwa}. On the other hand, elastic
scattering of DM with nuclei can in some model be even
larger than for usual Majorana WIMPs~\cite{Zurek:2013wia}.

To summarise, the standard paradigm of  thermal production of WIMP DM  via the freeze-out mechanism certainly still remains very compelling due to its robustness, simplicity and minimum number of (rather natural) assumptions involved.

\subsubsection{Non-thermal and Other Alternatives to the Thermal WIMP DM Paradigm {\label{sec:nonthermalWIMPs}}}

If  WIMP-SM sector interactions are very much weaker 
than EW ones -- in other words, when $\langle\sigma_{\textrm{ann}} v\rangle$ is too low  --  then $\chi$ particles never reach thermal equilibrium after reheating and actually never freeze out. This category of general WIMPs
is called {\bf feebly interacting massive particles (FIMPs)}~\cite{Hall:2009bx}.\footnote{They are
also called {\bf extremely weakly interacting massive particles (EWIMPs)} or {\bf super-WIMPs}.} 

FIMP DM relics with correct density can still be produced {\bf non-thermally}. The initial population of such DM relics is assumed to be negligible but is generated through a decay of some heavier particles or else via inelastic scatterings
  or decays of some heavier particles in the thermal plasma. This class of processes is now commonly called {\bf freeze-in mechanism} and in recent years has gained much attention.\footnote{The name was introduced in Ref.\cite{Hall:2009bx} in the case of renormalisable WIMP couplings to the SM sector but in fact the mechanism itself had been known much earlier and studied in the context of DM relics in the case of non-renormalisable effective interactions suppressed by some
high energy scale, \eg the Planck mass, $M_\text{Pl} \approx 10^{18}\gev$ for
{\bf gravitinos}~\cite{Ellis:1984eq,Moroi:1993mb,Bolz:2000fu,Feng:2003xh,Feng:2003uy,Pradler:2006qh,Rychkov:2007uq},  the Peccei-Quinn scale, $f_a\approx 10^{11}-10^{12}\gev$ for
{\bf axinos}~\cite{Covi:1999ty,Covi:2001nw,Choi:2011yf} (for a recent review
see~\cite{Choi:2013lwa}), or gauge singlet scalars in \cite{McDonald_2002}. See, \eg \cite{Baer:2014eja,Roszkowski:2017nbc} for a detailed discussion. 
}
Like thermal WIMPs, FIMP DM can be realised in a variety of models, either involving renormalisable interactions or not. In the former case, when DM production involves for instance a light mediator, the low-temperature production dominates over the high-temperature one and freeze-in is largely independent of the reheating temperature $T_\text{reh}$ after inflation~\cite{Hall:2009bx}. The opposite is typically true in models with WIMPs exhibiting non-renormalisable interactions with the SM.

 The detection of FIMPs as DM will be extremely challenging since they  interact  with the SM sector extremely weakly, with usually Yukawa/gauge couplings of the order of $10^{-11}$ required to obtain the right ballpark of the relic density.\footnote{
Note that already the requirement of avoiding DM thermalisation restricts its renormalisable couplings to be below $10^{-7}$~\cite{Bernal:2017kxu}. }
A potentially detectable signal in direct detection is still possible if the very small coupling to the SM is compensated by a very small mediator mass. In that case a scattering cross section is suppressed but could still be within reach of the next generation of detectors. In addition, the scattering with electrons can also play a major role and opens up another opportunity, especially if the FIMP is also very light, in the sub-GeV range, or even much less, down to the eV scale, because, for a fixed local relic mass density, its number density must be large~\cite{Bernal:2017kxu}.

Typically FIMPs whose interactions with the SM are renormalisable, although highly suppressed, and are produced via freeze-in (or some other non-thermal mechanism) tend to be fairly light, with mass in the sub-GeV range, or even much less. This class of FIMPs are in principle experimentally testable via electron scattering interactions. In contrast, FIMPs coupling to the SM with non-renormalisable interactions, e.g., gravitinos or axinos, can be much heavier, up to the TeV scale.
For comparison, thermal WIMPs produced via freeze-out typically feature the mass range from a few GeV up to several TeV, and interactions related to EW ones, although in general they can also be much lighter, as discussed in Sec.\ref{sec:thermalWIMPs}. 
Apart from thermal freeze-out and freeze-in, there are some other mechanisms for producing WIMPs in the early Universe; see~\cite{Bernal:2017kxu} for an exhaustive discussion.

Note also that in  models with a strongly suppressed interaction, FIMP DM does not have to be absolutely stable.
Indeed, especially for Planck scale suppressed couplings, the lifetime of DM can be naturally many of orders of magnitude longer than the age of the Universe.
The DM decay would then produce a signal in indirect detection and possibly a suppressed inelastic signal in direct detection. 

{\bf Sterile neutrinos}  in the keV to MeV mass range form another class of non-thermal relics. They arise for example 
if the SM neutrino sector is augmented by a light right-handed singlet sector. They are called sterile because they do not interact with the electroweak force, apart for the mixing with the active neutrinos.
In general they cannot be in equilibrium with the SM since they would overclose the Universe, therefore their interaction to the SM, as determined by their mixing angle with the active neutrinos, must be suppressed, and are already highly constrained by experiments and X-rays observations~\cite{Boyarsky:2018tvu}. 
This type of DM is not stable as it can decay into an active neutrino and a photon, which may be observed in X-rays observatories. 
Note that sterile neutrinos as DM can only be searched for indirectly since their couplings with direct detection targets are very strongly suppressed~\cite{Adhikari:2016bei}. 

In conclusion, the attractive and robust mechanism of generating thermal WIMP from freeze-out provides only one of several possible ways of
generating DM relics in the early Universe.\footnote{DM particles can be
additionally produced in late-time decays (after DM freeze-out) of
some heavier species. Examples include moduli field
(see~\cite{Moroi:1999zb,Acharya:2008bk} and references therein),
Q-balls (see, \eg~\cite{Fujii:2002kr}), the inflaton field (see,
\eg~\cite{Lyth:1995ka}) or cosmic strings~\cite{Jeannerot:1999yn}. } 
 This fact not only significantly relaxes the standard thermal WIMP paradigm, but also has implications for prospects for DM searches. A partial list of potential DM candidates is shown in Fig.~\ref{fig:dmmassrange}.

Last but not least, a whole class of non-thermal relics that is theoretically highly motivated and experimentally detectable is axions. They, as well as related species exhibiting axion-strength interactions, are reviewed in Section~\ref{sec:axions} below.

\subsubsection{Axions }
\label{sec:axions}

The Standard Model predicts CP violation in the strong interactions proportional to an angle~$\theta$, which is a sum of the QCD vacuum angle and a combination of CP-violating phases of Yukawa couplings~($Y$) of the Higgs to quarks, $\arg\det(YY^{\dagger}$). 
Such CP violation would most prominently appear as a non-vanishing electric dipole moment of the neutron (nEDM) $d_n = 2.4(\pm 1.0) \times \theta \times 10^{-3} \ e \;\rm{fm}$, where $e$ denotes the unit charge~\cite{Pospelov:1999mv}. Experiments set upper limits on the nEDM of $d_n < 1.8\,\times\, 10^{-26} \ e \;\rm{cm}$~\cite{PhysRevLett.124.081803, Afach:2015sja}, 
which in turn limits $\theta$ to be less than $\sim\!\!  10^{-10}$ . To explain this small number in the Standard Model, a large cancellation in the strong sector must be in place between two seemingly unrelated terms contributing to CP violation. This is known as the strong CP problem. 
In this case anthropic arguments have difficulties explaining the smallness of theta, as any value smaller than $\theta \simeq 0.01$ has no noticeable effect on cosmology, astrophysics, nuclear physics, etc. \cite{Ubaldi:2008nf, Dine:2018glh}. 

The arguably most elegant solution to the strong CP problem is the introduction of a $U(1)_{\rm PQ}$ Peccei-Quinn symmetry that is spontaneously broken at an energy scale~$f_a$~\cite{PhysRevLett.38.1440, PhysRevD.16.1791}. 
During the QCD confinement transition the symmetry is explicitly broken, leading to the complex phase of this field to develop a potential with a minimum at $\theta=0$ by non-perturbative QCD effects. The arising particle, a pseudo Nambu Goldstone boson (pNGB) is the \textbf{QCD axion}~\cite{PhysRevLett.40.223, PhysRevLett.40.279}. QCD axions would, hence, automatically be produced in the early Universe by this so called \textbf{misalignment mechanism}, when the initial phase~$\theta_i$ of the $U(1)_{\rm PQ}$ group was dynamically driven to zero during Hubble expansion and the field starts oscillating around its minimum (the amplitude of this oscillation gives the relic density of axions). 
This model is explaining the observations with a minimal set of assumptions, and the existence of the QCD axion as a clear observational consequence.

The QCD axion can be easily accommodated in theories Beyond the Standard Model like supersymmetry and Grand Unified Theories, and it is a firm prediction of string theories~\cite{Svrcek:2006yi}.
Moreover, it has recently been argued that the existence of the QCD axion arises independently from the strong CP problem from QCD as a general consistency requirement~\cite{Dvali:2018dce} due to the necessity of the absence of de Sitter vacua~\cite{Dvali:2017eba}.
The QCD axion mass is directly related to $f_a$ by the relation~\cite{Borsanyi:2016ksw}
\begin{equation}
m_a=(5.7\pm0.7) \mueV \times \frac{10^{12}\GeV}{f_a} \textnormal{.}    
\end{equation}
Their coupling to photons and gluons is naturally suppressed by the energy scale of symmetry breaking $f_a$. There is some model dependence in the calculation of the axion to photon coupling $g_{a\gamma}$. Usually, the KSVZ \cite{PhysRevLett.43.103, SHIFMAN1980493} and DFSZ \cite{Zhitnitsky:1980tq, DINE1981199} benchmark models are used to compare experiments against their sensitivity to DM axions. However, the theoretical uncertainty is larger than the spread between the KSVZ and DFSZ models \cite{DiLuzio:2016sbl, DiLuzio:2020wdo}.

As any spontaneously and explicitly broken $U(1)$ symmetry group leads to pNGBs, this phenomenon is more general. 
In fact, string compactification typically produces a  plenitude of \textbf{ALP} candidates, and many of those could be light, leading to the so-called “axiverse” scenario.
For these, there is no direct relationship between the symmetry breaking scale and the mass. In general these pNGBs are called axion like particles (ALPs).

Additionally, string theory predicts generic "hidden" $U(1)$ gauge factors, leading to ultra light massive particles commonly referred to as hidden (or dark) photons, that would kinetically mix with the standard model photon with an effective kinetic mixing angle $\epsilon$ \cite{Goodsell:2009xc}.
They could be produced with the correct relic abundance in the early universe for a wide mass range, for example by inflationary fluctuations \cite{PhysRevD.93.103520} or by transfer of energy stored in axion/ALP field oscillation to hidden photons via a non vanishing coupling \cite{AGRAWAL2020135136}.
It has also been argued that 
hidden photons could be produced 
by a similar  misalignment mechanism as pNGB axion or ALPs~\cite{Nelson:2011sf}.

Hidden photons, as well as pNGB axion or ALPs, would inevitably contribute to the cold DM  density of the Universe due to their production in the early universe, for example
via a misalignment mechanism or topological defects~\cite{PRESKILL1983127, ABBOTT1983133, DINE1983137}. The overall energy stored in the field oscillations (relic particle density) behaves as $\rho \propto \theta_i^2$, the initial misalignment angle at the QCD phase transition, which is random within each causally connected patch of the Universe; the closer $\theta_i$ is to $\pi$, the more energy is stored. 

For QCD axions, there are two scenarios that differ in their predictive power for the axion relic density, hence for the axion mass, assuming that axions explain the DM problem: If PQ-symmetry breaking occurred before inflation (pre-inflationary scenario), the observable Universe today will constitute of areas that all had the same~$\theta_i$. In this case the axion mass can have any value between $\sim\!\!  10^{-13}\ev$, the mass for  which $f_a$ corresponds to the Planck scale and $\sim\!\!10\miliev$ for which $\theta_i \sim\!\! \pi$, depending on the amplitude of the initial oscillations (relic density).
For the case that PQ symmetry breaking occurred after inflation (post-inflationary scenario), the observable Universe today consists of many patches that were not in causal connection during QCD phase transition. Hence, the average $\theta_i$ of all patches determines the overall energy stored in the mass of axions. The axion mass can hence be constrained. For this scenario, however, a complication arises from additional axion production during the alignment of $\theta$ towards zero by the decay of topological defects: strings and domain walls. 
This can also lead to the formation of gravitationally bound axion clumps~\cite{Kolb:1993hw} called mini-clusters.
Both the relic density and the mini-cluster mass distribution  are notoriously difficult to calculate. Nevertheless, for the post-inflationary scenario the axion mass can be constrained between 25\mueV\ and $\sim\!\!15\miliev$, with the upper bound arising from astrophysical arguments. 
The pre-inflationary and post-inflationary scenarios define the "classical" QCD axion as DM window search range $\sim\!\! 1\muev$ to $\sim\!\!1\miliev$ when one assumes the initial misalignment angle  $\theta_i$ to be $\mathcal{O}(1)$.

For general ALPs (and hidden photons) the constraints in the mass versus coupling constant parameter range are much weaker. In general low mass ALPs could be viable DM  candidates for any coupling constant $g_{a\gamma} \lesssim$ 10$^{-11}$\GeV$^{-1}$ (effective mixing angle of $\epsilon \lesssim 10^{-10}$ for hidden photons) for any mass up to $\sim\!\!\ev$. For a general review see \cite{Arias:2012az}.
Note that such kind of particles are also still possible for a mass $\gtrsim$\MeV.

It is remarkable that a plethora of other cosmological, astrophysical and experimental effects can be explained
by very low mass axions or ALPs. Particularly relevant examples are the accelerated expansion of the Universe~\cite{Kim:2002tq}, transparency of the Universe to gamma rays~\cite{Horns:2012fx, Brun_2013} or altering the evolution of stars, radiation of black hole spin~\cite{Giannotti:2017hny}.  

\subsection{Alternatives to Particle Dark Matter}

There have been attempts at solving the DM puzzle outside of particle physics. In this subsection we briefly review some approaches that have gained more attention.

As discussed in Sect.~\ref{sec:dmabundance}, the vast majority of the DM is non-baryonic. Therefore astrophysical bodies in the form of baryonic {\bf MAssive COmpact Halo Objects (MACHOs)} can only make up a small fraction of the DM. However {\bf primordial black holes (PBHs)}, black holes that may form in the early Universe, remain a viable CDM candidate~\cite{Chapline:1975ojl}. As they form before the time of primordial nucleosynthesis, PBHs are effectively non-baryonic, and if their mass is greater than $5 \times 10^{14} \gram \, (=3 \times 10^{-19} M_{\odot})$ their lifetime is longer than the age of the Universe. The recent discovery of gravitational waves from $\sim\!\! 10 M_{\odot}$  binary black hole (BH) mergers has led to a resurgence of interest in PBHs as a DM  candidate~\cite{Bird:2016dcv,Sasaki:2016jop,Carr:2016drx}. Such massive PBHs are now excluded from making up all of the DM by a combination of lensing, dynamical, accretion and gravitational wave constraints. However, asteroid-mass PBHs, with $ 10^{17} \gram \lesssim M_{\rm PBH} \lesssim 10^{22} \gram $, are challenging to detect and can still make up all of the dark matter. For a recent review of the constraints on PBHs see Ref.~\cite{Carr:2020gox}.

While PBHs are not elementary particles, their production does require physics beyond the Standard Model. The most commonly considered mechanism is the collapse of large density perturbations generated by a period of cosmic inflation. However to form an interesting number of PBHs the perturbations must be several orders of magnitude larger on small scales than measured on cosmological scales, and this can not be achieved generically in single field slow-roll inflation models. For a recent review of PBH formation see Ref.~\cite{Sasaki:2018dmp}.

All of the observational evidence for DM to date comes from its gravitational interactions. Therefore it is in principle possible that the observations could instead be explained by a {\bf modification of the law of gravity}.
Galaxy rotation curves can be explained by a phenomenological modification of Newton's law of gravitation at low accelerations, known as  {\bf Modified Newtonian Dynamics (MOND)}~\cite{Milgrom:1983ca}. To address cosmological observations a relativistic formalism, such as TeVeS~\cite{Bekenstein:2004ne}, is required. These models have difficulties explaining the heights of the higher order peaks in the CMB temperature angular power~\cite{Skordis:2005xk} and are also tightly constrained by the close to simultaneous detection of gravitational waves and electromagnetic signals from a binary neutron star merger~\cite{Boran:2017rdn}. Another challenge is provided by the Bullet cluster, where gravitational weak lensing and X-ray observations show that the dominant mass component is spatially separated from the baryonic mass~\cite{Clowe:2003tk}.
In summary, there is currently no modified gravity model that can explain all of the observational evidence for dark matter.

\newpage

\section{Underground Searches for WIMPs}\label{sec:experiment}

Experiments searching for signals induced by dark matter WIMPs from the Galactic DM halo in terrestrial detectors are called \emph{direct detection} experiments. They require ultra-low background levels to observe the feeble WIMP-matter interactions and are thus conducted in deep underground laboratories for shielding against cosmic rays. Here we first briefly summarise the principles of direct detection (Sect.~\ref{sec:ddbasics}) and the interactions probed by the experiments (Sect.~\ref{sec:interactions}). The different experimental approaches used and the typical backgrounds are discussed in Sects.~\ref{sec:overview_exp} and~\ref{sec:wimpbackgrounds}, respectively. The current status of the field is presented in Sect.~\ref{sec:wimplimits} before we provide a concise summary of the status and prospects of the major technologies employed for the WIMP search in Sect.~\ref{sec:european}. The chapter closes with a comparison of the experimental efforts and an outlook to the projected sensitivity.


\subsection{Direct Detection Principles}
\label{sec:ddbasics}

Direct detection experiments search for signatures of (in)elastic scattering of WIMPs off a target nucleus~\cite{Goodman:1984dc}. The momentum transfer gives rise to a nuclear recoil which might be detectable. (In case of inelastic scattering one also searches for time-coincident de-excitation signals.) The WIMP carries no electric charge and it is not expected (in most scenarios) that interactions with the very light atomic electrons will lead to detectable signals. 
The discussion here largely follows~\cite{Lewin:1995rx} and \cite{Schumann:2019eaa}.

The expected scattering rate is given by 
\begin{equation}\label{eq::rate}
\frac{dR}{d\Enr} = \frac{\rho_0 M}{m_N m_\chi} \int_{v_\textnormal{\tiny min}}^\infty v f(v) \frac{d \sigma}{d\Enr} \ dv \propto \exp \left(- \frac{\Enr}{E_0} \frac{4 m_\chi m_N}{(m_\chi + m_N)^2} \right) F^2(\Enr)\textnormal{,}
\end{equation}
where $m_N$, $m_\chi$ and $M$ are the masses of the target nucleus, the WIMP and the detector, respectively, $\Enr$ \ the nuclear recoil energy, $\sigma$ the scattering cross section and $F(\Enr)$ the form factor (see Sect.~\ref{sec:interactions}). The DM  halo is characterised by the normalised WIMP velocity distribution $f(v)$ and the local DM  density $\rho_{0}$. $E_0$ is the WIMP's most probable kinetic energy. All velocities are defined in the detector's reference frame. $v_\mathrm{min}$ is the minimal WIMP velocity required to induce a nuclear recoil of energy $\Enr$. WIMPs with a velocity above the escape velocity $v_{\rm esc}$ have left the potential well of the Milky Way, i.e., $f(v\,>\,v_{\rm esc})\,=\,0$ in the Galactic rest-frame. The differential rate is eventually described by a featureless falling exponential function (see second part of Eq.~(\ref{eq::rate})), rendering the lower detector threshold $E_{\rm low}$ much more important than the upper boundary $E_{\rm high}$. Integration of Eq.~(\ref{eq::rate}) between these limits yields the expected number of events in an experiment of live time~$T$ and detection efficiency $\epsilon(\Enr)$:
\begin{equation}
N=T \int_{E_{\rm low}}^{E_{\rm high}} d\Enr \ \epsilon(\Enr) \ \frac{dR}{d\Enr} \textnormal{.} 
\end{equation}
The nuclear recoil energies depend on $m_N$ and are typically very small ${\cal O}(10\keVnr)$. 
The energy scale is given in keV$_\textrm{nr}$ (nuclear recoil equivalent) 
{which can differ} from the electronic recoil scale (keV$_\textrm{ee}$) {if the signal is quenched because of the } energy-loss mechanism.

The momentum transfer of very light WIMPs with mass in the $\MeV/c^2$-range to the target nucleus might be too small to generate a detectable nuclear recoil signal. The search for such particles thus often concentrates on WIMP-electron scattering, a signature which is otherwise rejected as background. 

An Earth-based detector moves through the DM  halo with a velocity ($\phi$-component)
\begin{equation}
 v_E = v_\odot + v_\oplus \cos(\theta) \cos \left[ \omega (t-t_0) \right] \textnormal{,} \label{eq::velocities}
\end{equation}
where $v_c=220\kmeter/\second$ is the local circular velocity, $v_\odot = v_c + 12\kmeter/\second$ the motion of the Sun with respect to $v_c$ and $v_\oplus=30\kmeter/\second$ describes the velocity of the Earth orbiting around the Sun. The inclination angle between the Earth's orbit and the galactic plane is $\theta \approx 60^\circ$. $\omega = 2 \pi/T$ with a period $T=1\years$; the phase is fixed to $t_0=$\,June 2, when $v_\odot$ and $v_\oplus$ are parallel. This periodic modulation of $v_E$ leads to a harder (softer) recoil spectrum in summer (winter) when $v_\odot$ and $v_\oplus$ are (anti-)parallel and thus an {\bf annually modulating} DM  signal $S(t)$ above a fixed detector threshold $E_{\rm low}$~\cite{Drukier:1986tm,freese1988,freese2013},
\begin{equation}
 S(t)=B(t) + S_0 + S_m \cos \left[ \omega (t-t_0) \right] \textnormal{.}
\end{equation}
However, most of the signal is unmodulated~$S_0$ as the modulated part~$S_m \sim\! {\cal O}(v_\oplus/v_c) \sim\! 5\%$ is small. The (potentially time-dependent) backgrounds $B(t)$ are typically much larger than $S_0$
A detection of DM  based solely on the modulation signature requires ${\cal O}(10^4)$ signal events. Modifications to the simple DM  halo model (streams, dark disk etc.) will significantly change the expected signal. Astrophysical uncertainties have generally a rather small impact for the ``standard'' direct detection discussed above, however, they are more significant for annual modulation searches~\cite{Green:2017odb}. 

Another way to exploit Eq.~(\ref{eq::velocities}) to reduce backgrounds is to search for a WIMP ``wind'' from the direction in the constellation Cygnus, the point in the sky towards which the Sun is moving. The Earth's daily rotation thus constantly changes the signal direction observed in a detector while most backgrounds are expected to be uniformly distributed (or originate from the Sun, e.g., solar neutrinos). A measurement of the {\bf track direction} could distinguish a DM  signal from background events~\cite{Spergel:1987kx}. The experimental challenge is that the track length $r$\,$<$\,1\,mm is very short for keV-scale nuclear recoils and difficult to reconstruct~\cite{Mayet:2016zxu}.

\subsection{Interactions}
\label{sec:interactions}

The WIMP-nucleus scattering cross section in Eq.\,(\ref{eq::rate}) can be expressed as
\begin{equation}
\frac{d\sigma (\Enr)}{d\Enr} = \frac{m_N}{2 v^2 \mu^2} \left[ \sigsi F_\mathrm{SI}^2(\Enr) + \sigsd F_\textrm{SD}^2(\Enr) \right]\textnormal{,}
\end{equation}
where the unknown interaction is described by a spin-independent (SI) and a spin-dependent (SD) component. The former corresponds to a scalar 
or vector 
effective 4-fermion Lagrangian, the latter has an axial-vector 
structure. All partial waves of the nucleons add up at small momentum transfers~$q$ and the WIMP interacts coherently with the entire nucleus. The loss of coherence at higher~$q$ is accounted for by the finite form factors $F_\textrm{SI}$ and $F_\textrm{SD}$. $F_\textrm{SI}$ is only relevant for WIMP targets with high mass numbers $A\gtrsim 100$ and at high recoil energies~\Enr.

The {\bf spin-independent}~(SI) cross section is given by
\begin{equation}\label{eq::si}
\sigsi=\sigma_n \frac{\mu^2}{\mu_n^2}\frac{(f_pZ+f_n(A-Z))^2}{f_n^2} = \sigma_n \frac{\mu^2}{\mu_n^2} A^2 \textnormal{.}
\end{equation}
$\mu$ and $\mu_n$ are the reduced masses of the WIMP-nucleus and the WIMP-nucleon systems, respectively. The WIMP-nucleus cross section~$\sigma$ is converted to a WIMP-nucleon cross section $\sigma_n$ to facilitate the comparison between different target nuclei. $f_p$ and $f_n$ are coupling constants to protons and neutrons, respectively; the second equality in Eq.~(\ref{eq::si}) assumes $f_p=f_n$. The resulting $A^2$ dependence favours heavy target nuclei to search for spin-independent interactions.

{\bf Spin-dependent}~(SD) interactions describe the WIMP coupling to unpaired nuclear spins~$J$~\cite{Engel:1992bf}:
\begin{equation}\label{eq::sd}
\frac{d\sigsd}{d|\vec{q}|^2} = \frac{8 G_F^2}{\pi v^2} \left[ a_p \langle S_p \rangle + a_n \langle S_n \rangle \right]^2 \frac{J+1}{J} \frac{S(|\vec{q}|)}{S(0)} \textnormal{.} 
\end{equation}
$\vec{q}$ is the the momentum transfer, $\langle S_{p} \rangle$ and $\langle S_{n} \rangle$ the expectation values of the total spin operators for protons and neutrons in the the target nucleus; these have to be calculated using nuclear models and thus carry some systematic uncertainty~\cite{Toivanen:2009zza,Menendez:2012tm}. The cross section depends on the spin-structure function $S(|\vec{q}|)$ and the total nuclear spin $J$ of the target. SD-results are usually quoted assuming that WIMPs couple either only to neutrons ($a_p=0$) or to protons ($a_n=0$). Nuclei with an odd number of protons (e.g., $^1$H, $^7$Li, $^{19}$F, $^{23}$Na, $^{127}$I) or neutrons (e.g., $^{17}$O, $^{27}$Al, $^{29}$Si, $^{73}$Ge, $^{129}$Xe, $^{131}$Xe, $^{183}$W) can effectively probe spin-dependent WIMP-proton or  WIMP-neutron interactions, respectively. 

The various possible 4-point interactions can be described by a number of relativistic and non-relativistic operators. These {\bf effective field theories} (EFTs)~\cite{Fan:2010gt,Fitzpatrick:2012ix,Hoferichter:2016nvd} allow for a direct comparison of direct detection results with collider searches (when the kinematic requirements for the WIMP production are taken into account). The SI and SD interactions are mainly described by the non-relativistic EFT operators ${\cal O}_1 = 1_\chi 1_N$ and ${\cal O}_4=\vec{S}_\chi \cdot \vec{S}_N$, respectively, however, in general any WIMP search data can be interpreted using a plethora of operators~\cite{Aprile:2017aas,Angloher:2018fcs}. In more complex {\bf simplified models}~\cite{Abdallah:2015ter,Abercrombie:2015wmb} the simple 4-point-interactions are replaced by $s$- or $t$-channel exchange of a mediator.

The expected signature of {\bf inelastic dark matter (iDM) interactions} is a nuclear recoil followed by an electronic de-excitation of either the WIMP or the target nucleus. The former process (WIMP gets excited) was originally proposed to reconcile the DAMA/LIBRA claim with the other (null) results~\cite{TuckerSmith:2001hy}, however, the concept can be adapted to other models, e.g., for light thermal DM~\cite{Berlin:2018jbm}. The latter (nucleus gets excited) is expected to occur in WIMP-nucleus interactions~\cite{Goodman:1984dc,Baudis:2013bba}, however, at reduced rates compared to ordinary elastic scattering.  Inelastic interactions can be described within the EFT operator framework as well~\cite{Arcadi:2019hrw}. In both cases, the distinct delayed coincidence signatures can be used to effectively suppress backgrounds but the detector must be able to observe electronic recoils.

Using the {\bf Migdal effect}~\cite{Dolan:2017xbu,Ibe:2017yqa} has recently been proposed to extend the reach of direct detection experiments searching for nuclear recoils further into the $\mev/c^2$-regime: in this hypothetical scenario, the WIMP-nucleus interaction can lead to atomic excitation and ionisation as the electron cloud does not follow the recoiling nucleus instantaneously. For low-mass DM the additional electronic recoil excitation/ionisation signal is above threshold, unlike the nuclear recoil alone, and facilitates the detection of the scattering process. 
A similar strategy to overcome the conventional threshold relies on the detection of {\bf bremsstrahlung} photons following the undetectable nuclear recoil~\cite{Kouvaris:2016afs}. This process, however, leads to weaker results than the Migdal effect. One should also note that the existence of both effects has not yet been proven experimentally.

WIMPs with very low mass, in the MeV/$c^2$-range and below, do not transfer sufficient momentum to the target nucleus to generate NRs of detectable size. The coupling of WIMPs to atomic electrons is thus used to search for such particles~\cite{Essig:2011nj,Essig:2015cda}. {\bf WIMP-electron scattering} will create very small ionisation signals of ER type, typically only one up to a few electrons, which can be seen in detectors with single-electron sensitivity. (The detection with scintillating targets has been studied in~\cite{Derenzo:2016fse}.) The coupling of WIMPs to electrons can be parametrised by a cross section~$\sigma_e$ and a DM  form factor $F_\textrm{DM}(q)$ which generally depends on the momentum transfer~$q$~\cite{Essig:2011nj}. Results are typically quoted for the cases where the interaction is mediated by a heavy ($F_\textrm{DM}=1$) or a very light scalar or vector mediator particle ($F_\textrm{DM}(q)= \alpha m_e^2 /q^2$).  
A recent discussion of WIMP--electron interactions in the context of effective field theory can be found in~\cite{Catena:2019gfa}.

A similar ER signal can be generated by very light axions and ALPs via the {\bf axio-electric effect}~\cite{Derevianko:2010kz,Arisaka:2012pb}. Similar to the photo-electric effect, the absorption of an axion leads to the ionisation of the atom which can be detected. The energy of the ER is given by the sum of the axion's rest mass and its kinetic energy reduced by the electron's binding energy. See also Section~\ref{subsubsec:low_background}.

\subsection{Experimental Approaches}
\label{sec:overview_exp}

\begin{figure}[tb]
    \centering
    \includegraphics[width=0.99\textwidth]{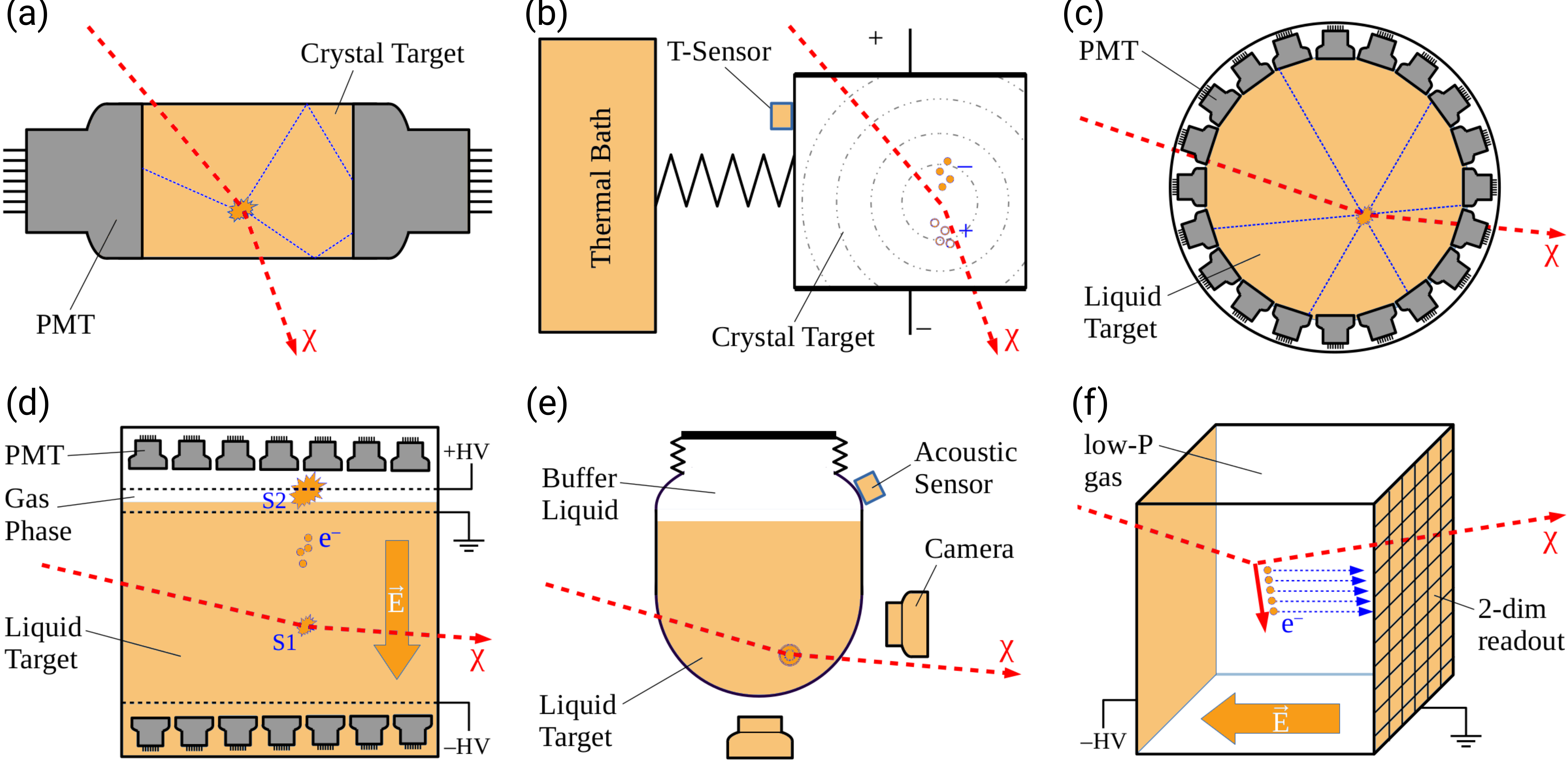}
    \caption{Working principle of common detector types for the direct WIMP search: (a) scintillating crystal, (b) bolometer (here with additional charge-readout), (c) single-phase and (d) dual-phase liquid noble gas detectors, (e) bubble chamber, (e) directional detector. Images adapted from~\cite{Schumann:2019eaa}.}
    \label{fig::dets}
\end{figure}

Several detector designs with various target materials are being used to search for WIMP dark matter. Here we provide a brief overview (the description closely follows a recent review~\cite{Schumann:2019eaa}):

Large target masses can be realised by using arrays of high-purity {\bf scintillator crystals}, mainly NaI(Tl) but also CsI(Tl). They feature a rather simple detector design (see Fig.~\ref{fig::dets}\,a) and can be operated stably for long periods of time. The high mass numbers of~I ($A$\,=\,127) and Cs ($A$\,=\,133) lead to a high sensitivity to spin-independent interactions. The shortcomings of these detectors are a comparatively high intrinsic background level -- the world-record for large-scale crystals used for the DM search is $\sim\!\!1\events/\kgram/\days/\keV$~\cite{Bernabei:2018yyw} -- and the absence of fiducialisation and electronic recoil rejection. The experiments thus concentrate on exploiting the annual modulation signature (above a much larger non-modulated signal and background fraction) to identify a DM  signal; the individual detection of DM  candidate events is not possible. Typical thresholds are $1-2\keVee$ ($\approx8\keVnr$ in Na, $12\keVnr$ in Cs, $22\keVnr$ in I). 

Germanium and silicon semiconductor {\bf ionisation detectors} are used to search for DM-induced charge signals. Only a very small amount of energy is needed to create an electron-hole pair (Ge: 2.9\,eV, Si: 3.6\,eV) which leads to an excellent energy resolution. On the other hand, the signals exhibit a rather slow time constant $\tau \sim\! 1\musecond$ 
and the capacitance of the diodes, leading to high electronic noise, does not allow building detectors beyond the few-kg scale. The state-of-the-art experiments use p-type point contact HPGe crystals at the kg-scale and achieved very low thresholds down to $\sim\! 160\eVee$~\cite{Jiang:2018pic}. 
Background events from the large n$^+$ surface can be distinguished from bulk events based on their longer rise time~\cite{Aalseth:2010vx}. Thanks to their smaller mass number~$A$ silicon detectors have a better sensitivity to low-mass WIMPs than germanium which is exploited, e.g., by using charge-coupled devices (CCDs).

Crystalline {\bf cryogenic detectors (bolometers)} measure either heat or athermal phonon signals by measuring the tiny particle interaction-induced temperature increase~$\Delta$T. Detector operation at cryogenic temperatures~$T$ (typically $\le$50\,mK) and a low heat capacity~$C$ is required to achieve a good sensitivity. Dielectric crystals with good phonon-transport property are particularly well-suited for cryogenic operation. Several methods to measure $\Delta$T are available, frequently used are transition edge sensors (TES) for athermal phonons and neutron transmutation doped (NTD) germanium thermistors for heat. In both cases, the resistivity of the sensors strongly depends on the temperature. A simultaneous measurement of a second observable (ionisation, Fig.~\ref{fig::dets}\,b, or scintillation) allows for signal/background discrimination as the partition of the signal into the two channels depends on the recoil type~\cite{Shutt:1992qg}. Cryogenic detectors feature a precise energy measurement with almost no quenching in the heat channel, excellent energy resolution and background rejection down to energies of $\mathcal{O}(1\keVnr)$), where the distributions start to overlap. The operation at mK-temperatures is challenging and expensive and the requirement of a low-energy threshold constrains the mass of the individual detectors, limiting the reachable exposure. The threshold of cryogenic solid state detectors can be further reduced by operating the crystals with a high bias voltage, which effectively converts the charge signal into heat by the production of additional Neganov-Trofimov-Luke phonons~\cite{Luke:1990ir}, hence boosting the total heat signal. However, ER rejection capabilities are lost in this method.

The {\bf noble gases} argon and xenon are excellent scintillators and can be ionized easily. In liquid state they are used to build massive, dense and compact DM  targets which already exceeded the ton-scale. Interactions produce heat (undetected) and excite (X$^*$) and ionise (X$^+$) the atoms. The X$^*$ form excimer states X$^*_2$ with  with neutral atoms~X. These decay under the emission of ultraviolet light at wavelengths of $128\nmeter$ and $178\nmeter$ 
for argon and xenon, respectively. Photocathode materials with sensitivity to the xenon scintillation light exist but wavelength shifters such as tetraphenyl butadiene (TPB) have to be used for argon. The X$^+$ ions form singly ionised molecules X$^+_2$ with neutral atoms and recombine with electrons forming X$^*$ states, which will de-excite again via the emission of scintillation light. The de-excitation happens with different decay constants, depending on the spin state of the excimer $X^*_2$. The relative population of singlet and triplet states, and thus the scintillation pulse shape, differs for electronic and nuclear recoils. The two constants are very similar for xenon ($4\nsecond$ and $22\nsecond$) 
but differ very significantly for argon ($7\nsecond$ and $1.6\musecond$)  
which allows for very efficient pulse shape discrimination~\cite{Ajaj:2019jk} (see Sect.~\ref{sec:lar}). The high rejection level allows Ar-based searches to mitigate backgrounds from the radioactive $\beta$-decaying isotope $^{39}$Ar ($T_{1/2}=269\years$).  

{\bf Single-phase liquid noble gas detector} measure only the scintillation signal: to optimise light collection efficiency, a spherical target is surrounded by photomultipliers, see Fig.~\ref{fig::dets}\,c. The interaction position can be reconstructed via photon timing and the signal distribution with few cm~resolution. Single-phase argon detectors additionally employ PSD for background reduction. 
{\bf Dual-phase time projection chambers (TPC)} measure the primary scintillation signal (S1) as well as the ionisation electrons. These are swept away from the interaction site by means of a strong electric drift field $|\vec{E}_d| =0.1-1.0$\,kV/cm which is created across the cylindrical target by wire or mesh electrodes, see Fig~\ref{fig::dets}\,d. A stronger extraction field ($\sim\!\! 10$\,kV/cm) across the liquid-gas interface pulls the electrons into the gas phase where they collide with gas atoms producing a secondary scintillation signal (S2) which is proportional to the number of electrons. Both signals are recorded using photosensors installed above and below the target. The simultaneous measurement of the two signals allows the rejection of multiple scatters, reconstruction of the interaction position with mm-precision and improving of the energy resolution as light and charge signals are anti-correlated~\cite{Aprile:2006kx}. The partition into excitation and ionisation depends on the ionisation density and the ratio S2/S1 is used to distinguish electronic from nuclear recoils. The scintillation signal governs the threshold of dual-phase TPCs; a charge-only (``S2-only'') analysis can thus lower the threshold significantly~\cite{Angle:2011th}, however, the background increases.

Superheated liquids are used as target material in {\bf bubble chambers}. The liquids are kept at a temperature just above their boiling point such that a local phase transition will create a bubble if energy above some threshold is deposited into a certain micro-volume (threshold detector; no direct energy reconstruction possible). The position and number of bubbles is recorded with mm-precision via stereoscopic camera readout, see Fig.~\ref{fig::dets}\,e. The probability for bubble formation depends on various operational parameters and can be tuned such that only nuclear recoil events (from $\alpha$-particles, neutrons or WIMPs) will create bubbles. The detector is then almost immune to electronic recoils. $\alpha$-particles can be rejected based on the acoustics of the bubble's explosion~\cite{Aubin:2008qx}. The long time scales to establish the superheated state lead to large dead times and complicated detector calibration.
{\bf Superheated droplet detectors} employ the same detection principle. Their bubbles are trapped in a water-based polymer matrix~\cite{BarnabeHeider:2005ri} which reduces dead time. 

Bubble chambers can be operated with various target fluids of different composition (typically refrigerants such as CF$_3$I, C$_3$F$_8$, C$_4$F$_{10}$, C$_2$ClF$_5$, C$_3$ClF$_8$). Most targets contain the isotope $^{19}$F which has the highest sensitivity to spin-dependent WIMP-proton couplings. A good sensitivity for spin-independent couplings is obtained with targets containing iodine ($A\approx127$).

{\bf Directional detectors} aim at reconstructing the direction of the WIMP-induced nuclear recoil~\cite{Mayet:2016zxu,ohare2021}. As the track length depends on the target density and longer tracks facilitate the reconstruction of the directionality, most directional detectors feature low-pressure gas targets ($\sim\!\!40-100$\,mbar) with  photographic or fine-granularity track readout in a TPC geometry, see Fig.~\ref{fig::dets}\,f. The target masses realised so far are not competitive to other technologies and the backgrounds are higher since self-shielding is not efficient. A large number of channels is required to reconstruct the short tracks. Electronic recoil background can usually be rejected to high levels thanks to their lower ionisation density and longer range. The most common target gas used is CF$_4$, which provides sensitivity to spin-dependent WIMP-nucleon interactions; sometimes it is used in mixtures with other gases (e.g., CS$_2$, CHF$_3$). 

\subsection{Backgrounds and Background Mitigation Strategies}
\label{sec:wimpbackgrounds}

The design of direct detection experiments aims at minimising the background in the region of interest (ROI), such that a few signal events observed during the experiment's exposure yield a high statistical significance. A background-free exposure is important to enhance the discovery potential, and to avoid misinterpretation of positive signals if they are observed. Here we briefly summarise typical background sources and strategies how to reduce them (the discussion is based on~\cite{Schumann:2019eaa}).

{\bf Electronic recoil (ER) backgrounds} are generated by $\beta$- and $\gamma$-particles interacting electromagnetically with the target's atomic electrons. These backgrounds are most abundant in every detector, however, most experimental techniques have means to reduce it to negligible levels. Bubble chambers can even be operated in a mode where ERs do not create a signal. The backgrounds come from long-lived natural radioisotopes ($^{238}$U, $^{232}$Th chains and their daughters, e.g., $^{214}$Pb; $^{40}$K), cosmogenic activation (e.g., $^3$H, $^{39}$Ar) and anthropogenic isotopes (e.g., $^{60}$Co, $^{85}$Kr, $^{110m}$Ag, $^{137}$Cs). Only isotopes with a half-life exceeding $\gtrsim0.5\years$ are relevant. The ultimate ER background will come from elastic collisions of low-energy solar neutrinos with atomic electrons~\cite{Baudis:2013qla}.

{\bf $\alpha$ backgrounds} in the target or on detector surfaces usually deposit too much energy to fall into the ROI, however, they can become relevant if a large part of the $\alpha$-energy is lost in insensitive detector regions. Due to their operation as threshold detectors, $\alpha$-particles are the main sources of background in bubble chambers.

Neutron-induced {\bf nuclear recoil (NR) backgrounds} are most critical for WIMP searches as they can mimic a standard WIMP signal. Radiogenic neutrons arise from ($\alpha,n$) and spontaneous fission reactions, cosmogenic neutrons are induced by cosmic ray muons. Neutrons are harder to shield than $\gamma$-rays since they have a longer mean free path, however, they can be distinguished from WIMPs based on their scatter multiplicity. {\bf Coherent scattering of neutrinos} off target nuclei will produce NRs which are indistinguishable from WIMPs on an event-by-event basis and thus constitute the ultimate background for direct WIMP searches~\cite{Billard:2013qya}, see Sect.~\ref{sec:neutrino}. This so-called {\bf neutrino floor} can be defined in various ways and by no means corresponds to a hard limit.\footnote{ As a matter of fact, DM-nucleus interactions described by operators associated with velocity or momentum dependence do not exhibit a neutrino floor, due to the induced differences in the DM and neutrino nuclear recoil energy spectra~\cite{Dent:2016iht}.} The definition we adopt in this report is a discovery limit which is defined as the cross section~$\sigma_d$ at which a given experiment has a 90\% probability to detect a WIMP with a scattering cross section $\sigma>\sigma_d$ at $\ge$3\,sigma. The methodology and assumptions are described in~\cite{Billard:2011zj,Billard:2013qya}. Only neutrino-induced NRs are taken into account and the flux uncertainties are considered. In this report we extend the neutrino floor to very low WIMP mass ranges by assuming an unrealistic $1\miliev$ threshold below $m_\chi = 0.8\GeV/c^2$ in the calculation.  

Finally, {\bf detector artefacts} (e.g., from incomplete signal collection, incorrect corrections, accidental coincidences) or noise often also lead to background events. 

Backgrounds are reduced using various {\bf background mitigation strategies}. These are often summarised as ``low background techniques''~\cite{Heusser:1995wd}: 

(i) Cosmogenic neutrons are reduced by conducting the experiments in deep-underground laboratories with a significantly reduced muon flux. Active muon vetoes further reduce this contribution.

(ii) ER and NR backgrounds from the environment are reduced by shielding the experiments either with massive, compact shields (e.g., Pb, Cu, PE) or larger shields of lighter isotopes (e.g., water, argon).

(iii) All detector and target materials are selected with various analytical methods for minimal radioactive contamination. Activation at ground level (or at high altitudes during air transport) is minimised. Detailed Monte Carlo simulations are used to determine acceptable activity levels. 

(iv) All detector components are cleaned thoroughly before detector assembly is done in cleanroom environments, sometimes also in low $^{222}$Rn environments to reduce surface plate out. 

(v) The DM  target materials are often purified from trace contaminations: either during production process (e.g., during crystal growth), or at procurement level (e.g., production of low-$^{39}$Ar argon; gas distillation, chromatography) or even online while taking data.

(vi) Detectors are designed to minimise backgrounds, e.g., by choosing appropriate materials, by optimising the material budget or by allowing for background rejection, see (vii) and (viii).

(vii) Backgrounds located at the detector surface are often reduced by fiducialisation, i.e., the selection of clean inner volume. This method requires knowledge of every event's coordinates or a detector design in which surface events generate special signals.

(viii) Active rejection during data analysis makes assumptions on the expected DM  signal (e.g., single scatter NR) and rejects all events which do not fall into this category. Typically the ratio of two out of the three observables: heat, scintillation and ionisation, is used to differentiate between ER and NR events due to their different energy-loss mechanisms. Other methods are scintillation pulse-shape-discrimination (liquid argon), acoustic $\alpha$-rejection (bubble chambers) or the rejection of multiply scattering events. Finite rejection efficiencies might lead to background leaking into the signal region. If the signal assumption is incorrect, the signal might be rejected in the analysis.

\subsection{Current Status}\label{sec:wimplimits}

\begin{figure}
    \centering
    \includegraphics[width=1.0\textwidth]{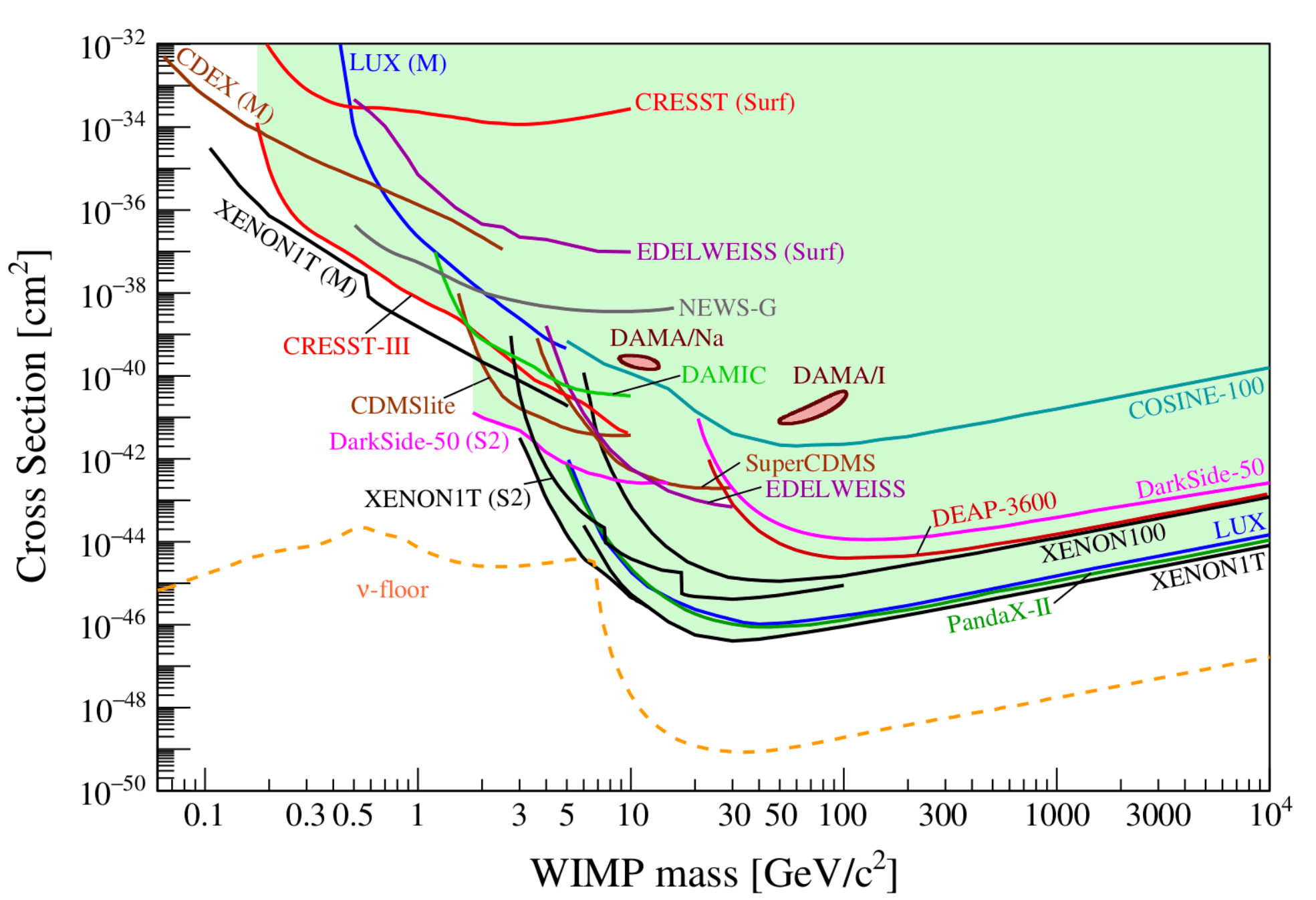}
    \caption{Current status of searches for spin-independent elastic WIMP-nucleus scattering assuming the standard parameters for an isothermal WIMP halo: $\rho_0 = 0.3\GeV/cm^3$, $v_0 = 220\kmeter/\second$, $v_\textrm{esc} = 544\kmeter/\second$. Results labelled "M" were obtained assuming the Migdal effect~\cite{Ibe:2017yqa}. Results labelled "Surf" are from experiments not operated underground. The $\nu$-floor shown here for a Ge~target is a discovery limit defined as the cross section~$\sigma_d$ at which a given experiment has a 90\% probability to detect a WIMP with a scattering cross section $\sigma>\sigma_d$ at $\ge$3\,sigma. It is computed using the assumptions and the methodology described in~\cite{Billard:2011zj,Billard:2013qya}, however, it has been extended to very low DM mass range by assuming an unrealistic $1\milieV$ threshold below $0.8\GeV/c^2$. Shown are results from CDEX~\cite{Liu:2019kzq}, CDMSLite~\cite{Agnese:2018gze}, COSINE-100~\cite{cosinenature}, CRESST~\cite{Angloher:2017sxg,Abdelhameed:2019hmk}, DAMA/LIBRA~\cite{Bernabei:2008yi} (contours from~\cite{Savage:2008er}), DAMIC~\cite{damic3}, DarkSide-50~\cite{Agnes:2018ep,Agnes:2018fg}, DEAP-3600~\cite{Ajaj:2019jk}, EDELWEISS~\cite{Armengaud:2019kfj,Hehn:2016nll}, LUX~\cite{Akerib:2016vxi,Akerib:2018hck}, NEWS-G~\cite{newsg1}, PandaX-II~\cite{Cui:2017nnn}, SuperCDMS~\cite{Agnese:2014aze}, XENON100~\cite{Aprile:2016swn} and XENON1T~\cite{Aprile:2018dbl,Aprile:2019xxb,Aprile:2019jmx,Aprile:2020thb}. 
    }
    \label{fig:si_status}
\end{figure}

The results of DM searches can be interpreted assuming a plethora of different types of WIMP interactions with the target. Here we summarise the status of the field focusing on the most commonly used models introduced in Sect.~\ref{sec:interactions}. Not all published results are mentioned as we concentrate on the projects providing the most stringent constraints (in general and for the technology).

The status of {\bf spin-independent WIMP-nucleon} couplings scattering is summarised in Fig.~\ref{fig:si_status}. Above WIMP mass of $\sim\!\!3\geV/c^2$, the strongest constraints come from the LXe TPCs XENON1T, LUX, and PandaX-II. XENON1T has the best sensitivity to WIMPs in this mass-range thanks to its $2.0\tonne$ target and $1.0\tonne\times\years$ exposure~\cite{Aprile:2018dbl,Aprile:2020thb}. The results from the LAr experiments DarkSide-50~\cite{Agnes:2018ep} (TPC, $46\kgram$ $^{39}$Ar~depleted target) and DEAP-3600~\cite{Ajaj:2019jk} (single-phase, $3.6\tonne$ target) are presently limited due to the requirement for approximately five times higher exposure than xenon (due to favourable enhancement of the cross section on xenon at low threshold), and the currently low acceptance in DEAP-3600. 
In the mass range $1.8-3.0\GeV/c^2$ the most stringent exclusion limits were placed by DarkSide-50~\cite{Agnes:2018fg} using only the ionisation signal to lower the threshold to $0.1\keVee$. A similar analysis was performed by XENON1T~\cite{Aprile:2019xxb}.

Due to their extremely low thresholds well below $1\keVnr$, the cryogenic experiments with ionisation/scintillation and phonon readout are very sensitive to low-mass WIMPs. Ge-based detectors, e.g., CDMS, improve the low-mass sensitivity by exploiting the Neganov-Trofimov-Luke effect~\cite{Luke:1990ir}. Using a $24\gram$~CaWO$_4$ crystal with a threshold of 30.1\,eV, CRESST-III currently places the most stringent constraints from $0.16-1.8\GeV/c^2$~\cite{Abdelhameed:2019hmk}.
Exploiting the Migdal effect~\cite{Dolan:2017xbu,Ibe:2017yqa} extends the reach further into the MeV/$c^2$-regime: if the recoiling atom gets excited and ionised by the WIMP-nucleon interaction this may lead to a detectable signal. Several results using this effect were already published~\cite{Akerib:2018hck,Armengaud:2019kfj} with the strongest ones being from XENON1T~\cite{Aprile:2019jmx} and CDEX~\cite{Liu:2019kzq} above and below $110\MeV/c^2$, respectively. However, calibrating the detector response to this effect is still an open issue which will be addressed by several groups in the near future.  

Bubble chambers filled with targets containing $^{19}$F have the highest sensitivity to {\bf spin-dependent WIMP-proton} couplings. The best limit to date is from PICO-60 using a $52\kgram$ C$_3$F$_8$ target~\cite{Amole:2019fdf}. At lower WIMP mass, between $2\gev/c^2$ and $4\geV/c^2$, the best constraints come from PICASSO ($3.0\kgram$ of C$_4$F$_{10}$~\cite{Behnke:2016lsk}). CRESST used crystals containing lithium to probe spin-dependent DM-proton interactions down to DM mass of $\sim\!\!800\MeV/c^2$~\cite{Abdelhameed:2019szb}.
The strongest constraints on {\bf spin-dependent WIMP-neutron} scattering above $\sim\!\!3\gev/c^2$ are placed by the LXe TPCs with the most sensitive result to-date coming from XENON1T~\cite{Aprile:2019dbj,Aprile:2019xxb}. The results from the cryogenic bolometers (Super)CDMS~\cite{Ahmed:2008eu,Marcos:2015dza} and CRESST give the strongest constraints below $\sim\!\!3\geV/c^2$. CDMSLite~\cite{Agnese:2017jvy} uses the Neganov-Trofimov-Luke effect to constrain spin-dependent WIMP-proton/neutron interactions down to $m_\chi=1.5\GeV/c^2$ and CRESST-III~\cite{Abdelhameed:2019hmk} exploits the presence of the isotope $^{17}$O in the CaWO$_4$ target to constrain spin-dependent WIMP-neutron interactions for DM particle's mass as low as  $160\MeV/c^2$. Exploiting the Migdal effect again significantly enhances the sensitivity of LXe TPCs to low-mass DM  with XENON1T providing the most stringent exclusion limits for both, spin-dependent WIMP-proton and WIMP-neutron couplings between $80\MeV/c^2 -2\GeV/c^2$ and $90\MeV/c^2 -2\GeV/c^2$, respectively~\cite{Aprile:2019jmx}.

The DAMA/LIBRA experiment searches for an {\bf annual modulation} signal with an array of NaI(Tl)-crystals and has reported a 12.9\,$\sigma$-detection of a signal over a total of 20\,annual cycles~\cite{Bernabei:2018yyw} (see Sect.~\ref{sec::nai}). The observed effect shows expected features of a halo DM particle interaction and no other confirmed or viable explanation has been provided. However, the DM nature of this observation is in tension with a large number of results. If interpreted in the standard WIMP scenario, much more sensitive experiments exclude DAMA's claim by many orders of magnitude, see Fig.~\ref{fig:si_status}. Assuming this interpretation, the phase-2 results of DAMA are even inconsistent with the phase~1 results of the same experiment~\cite{Baum:2018ekm}. LXe experiments with a significantly lower background did not find a modulation signal and excluded DAMA's claim with high significance~\cite{Aprile:2017yea,Abe:2015eos,Akerib:2018zoq}. The CDEX experiment also did not find a signal in a $1\kgram$ Ge-crystal with a threshold well below that of DAMA/LIBRA~\cite{cdexmod}. Attempts to solve the discrepancy by so-called ``isospin-violating'' DM  models favouring NaI over Xe targets~\cite{Feng:2011vu} are challenged by COSINE-100~\cite{cosinenature,cosinemod} and ANAIS-112
{\cite{anaismod,anais3y}} which also employ low-background NaI(Tl) crystals. The ANAIS-112 data is consistent with the absence of a modulation signal; COSINE-100 is consistent with both, the null hypothesis and the DAMA/LIBRA best fit, but excludes DAMA if interpreted as being due to standard spin-independent interactions.

Detectors with single-electron sensitivity are required to provide constraints on low-mass DM  interacting via {\bf WIMP-electron scattering}. In models with a heavy mediator, $F_\textrm{DM}=1$, the most stringent limits below $\sim\!\!10\mev/c^2$ come from SENSEI using a Si-CCD target~\cite{sensei2}, reaching down to $500\keV/c^2$. Other competitive results in this mass range are from the Si-detectors of DAMIC~\cite{damic2} and SuperCDMS~\cite{Agnese:2018col} as well as from the Ge-bolometers of  EDELWEISS~\cite{Arnaud:2020svb}. The best limits above $10\mev/c^2$ are from XENON10~\cite{Essig:2017kqs} and XENON1T~\cite{Aprile:2019xxb} and there are also results from DarkSide-50~\cite{Agnes:2018ft} and XENON100~\cite{Essig:2017kqs}. In models with a light mediator where the interaction is described by a DM  form factor $F_\textrm{DM}(q)=\alpha m_e^2/q^2$, SENSEI provides the best limits in the entire mass range above $500\keV/c^2$.

\subsection{Status and Plans of Experimental Activities}\label{sec:european}

The different types of experiments employed for the direct detection of WIMP dark matter were briefly summarised in Sect.~\ref{sec:overview_exp}. Here we provide details on the status, plans, advantages, limitations as well as timelines of the various experimental activities, organised into technology and target nucleus. For each technology, we also list the most important areas of ongoing R\&D. 

\subsubsection{Bolometers}
\label{sec:bolometers}

In recent years, thanks to their sub-100\,eV nuclear recoil energy thresholds, bolometric detectors have consolidated their role as the leading technology in the GeV/$c^2$ and sub--GeV/$c^2$ mass region. The SuperCDMS, CRESST and EDELWEISS collaborations successfully explored DM  mass as low as $1.5\gev/c^2$~\cite{Agnese:2018gze}  and $160\mev/c^2$ \cite{Abdelhameed:2019hmk} for standard DM-nucleus interaction, and down to $45\mev/c^2$ \cite{Armengaud:2019kfj} after including the Migdal effect. 
These results were obtained thanks to the $70\ev_{\rm ee}$, $30\ev$, and $55\ev$ energy thresholds achieved with $\sim\!\!600\gram$ of {\bf Ge} (SuperCDMS~\cite{Agnese:2018gze}), $\sim\!\!24\gram$ of {\bf CaWO$_4$} (CRESST~\cite{Abdelhameed:2019hmk}) and $\sim\!\!33\gram$ of {\bf Ge} (EDELWEISS~\cite{Armengaud:2019kfj}) bolometers, respectively. 

\paragraph{Operating principle}
\label{sec:bol_principle}
Under the category of bolometers, often called cryogenic detectors, there is a variety of approaches where the main characteristic is the conversion of the collected energy into a thermal signal (phonon-mediated). It is important to stress that in this approach almost all the released energy is eventually converted into heat, resulting in an almost non-quenched signal. Therefore uncertainties related to the calibration of the energy scale for nuclear recoils are overcome. To operate these detectors, extremely low operating temperatures are needed, typically in  the 10-20\,mK range where thermal fluctuations are low enough to sense $\mathcal{O}$($\mu$K) temperature rises. Since the energy to produce an elementary excitation (phonon) is extremely small, $\mathcal{O}$(10\,meV), the fluctuation of the number of elementary excitations is not limiting the resolution. Near the energy threshold, the main contributors to the energy resolution are therefore the electronics and thermal fluctuation noises inherent to the technology and bolometer design. Successful approaches in the DM field profited from double readout (heat + light, heat + charge) that allows for particle discrimination down to $\mathcal{O}(1\keVnr)$.

The signal of a cryogenic detector  
cannot be easily fiducialised. This condition makes the detectors vulnerable to background originating from contaminations on the facing surfaces. For this reason, different approaches were developed to solve the problem. In semiconductor bolometers, events near the surface can be rejected by shaping the electric field across the crystals~\cite{Broniatowski:2008yyl}, or by measuring the rise-time of the phonon signal~\cite{Akerib:2007zz}. In scintillating bolometers, such events can be tagged by using a scintillating veto surrounding the target crystals \cite{strauss_detector_2015}. It is worth noting that, due to their high-purity levels, semiconductor-based cryogenic experiments are not limited by their internal radioactive contaminants except from the ones cosmogenically produced~\cite{Armengaud:2016aoz}.
The bolometric approach is aiming to push the exposure into the range of kg$\times$day  
(to be compared with  
tonne$\times$day, or multi--tonne$\times$day,
in noble liquid TPCs). 
Thresholds below 100\eV\ will allow the mass region below $1\gev/c^2$ to be extensively explored. On the other hand, this approach will not be competitive in terms of exposure\footnote{It has to be mentioned that recently bolometric approaches in the tonne scale have been proven very successfully in neutrinoless double beta ($0\nu\beta\beta$)
decay searches, e.g., CUORE~\cite{Adams:2019jhp}, and similar approaches can be considered in the future.} with most sensitive experiments in the mass region ($10-1000\GeV/c^2$). Still such detectors, due to their superior energy resolution and their possibility of using a variety of different target materials, can be of particular interest to study the properties of the interacting particle, should a signal appear in one of the present generation liquid noble gas experiments.

The goal of the next generation of bolometric DM experiments is to explore the solar neutrino floor region, to catch the DM signal on top of the neutrino background, and to probe sub-MeV/$c^2$ DM candidates with their interaction with electrons.
Particularly important in this region are the solar neutrino flux uncertainties (e.g., ranging from 16\% for $^8$B to 1\% for pp) that, in case the solar neutrino background becomes significant, will limit the discovery potential of the next generation dark matter experiments~\cite{Billard:2013qya}. For this reason an improvement in both the theoretical estimations and experimental detection of the solar neutrino flux components is mandatory for an improvement of the discovery reach of low-mass dark matter experiments.

\paragraph{Semiconductor cryogenic detectors}
\label{sec:bol_semiconductors}
The EDELWEISS and SuperCDMS experiments have pioneered the use of cryogenic semiconductor crystals (Ge and Si) to search for DM particles. Following a particle's interaction in the detector medium, the induced recoil will release its energy by creating both phonons (heat) and charge carriers (ionization). 
To first order\footnote{We neglect here the phonon energy loss due to Frenkel defects and to charge trapping.}, the different measurable energy quantities are intertwined as follows:
\begin{equation}
    E_{\rm ion} = Q(E_R) E_R \ , \ \ \
    E_{\rm NTL} = E_{\rm ion}\frac{V}{\epsilon} \ , \ \ \ \text{and} \ \
    E_{\rm heat} = E_R + E_{\rm NTL} = E_R\left[1+Q(E_R)\frac{V}{\epsilon}\right]
    \label{eq:GeNTL}
\end{equation}
where $V$ is the voltage bias and $\epsilon$ is the average energy required for an electron recoil to produce an electron-hole pair. $E_{\rm heat}$ and $E_{\rm ion}$ stand for the heat and ionisation energies, respectively. $E_{\rm NTL}$ is the additional Neganov-Trofimov-Luke heat energy produced by drifting the charge carriers across the crystal~\cite{Luke,Neganov:1985khw}. dupa
The quenching factor $Q(E_R)$ is by definition equal to 1 for ER and between 0 and 0.3 for NR below $20\keV$~\cite{Barker:2013nua}. It is worth highlighting  that, in addition to the event-by-event discrimination, the simultaneous heat and ionisation energy measurements at V $\neq$ 0 also provides a direct measurement of the true nuclear recoil energy, hence avoiding any assumptions on the ionisation yield. Following Eq.~(\ref{eq:GeNTL}), two operating modes can  be considered.

{\bf NR/ER discrimination mode:} By operating the detector at low enough bias voltages, such that $E_{\rm NTL} \ll E_R$ for nuclear recoils, the simultaneous measurement of heat and ionisation provides an event-by-event identification of the recoil type, hence allowing a highly efficient rejection of the dominant gamma backgrounds as well as the majority of beta-backgrounds. Residual gamma- and surface beta-backgrounds are further removed using active surface rejection, based on either veto electrodes or charge asymmetry, of the FID (Fully Interdigitised Design) and iZIP detectors from the EDELWEISS~\cite{Armengaud:2017rzu} and SuperCDMS~\cite{Agnese:2013ixa} experiments, respectively.

{\bf NTL boosted mode:} 
By operating the detector at high voltage biases ($\geq$100\,V), the cryogenic calorimeter is effectively turned into a charge amplifier of mean gain $(1+Q(E_R)\frac{V}{\epsilon})$. As $E_{\rm heat} \simeq E_{\rm ion}$, event-by-event discrimination is no longer possible and an ionization yield model has to be assumed to convert the total heat energy into a nuclear recoil energy equivalent. It should be noted that, thanks to the much higher ionization yield of electron recoils compared to nuclear recoils, this operation mode is highly beneficial to any DM searches looking for interactions with electrons instead of nuclei.  

{\bf EDELWEISS}, in its third phase,  operated an array of 24 germanium cryogenic FID detectors of 870 g each, equipped with two NTD-Ge and aluminium electrodes, in the Modane Underground Laboratory (Laboratoire Souterrain de Modane, LSM) in France~\cite{Armengaud:2017rzu}. The collaboration successfully achieved an average baseline energy resolution (RMS) of $400\eV$ and $200\eVee$ for the heat and ionization channels, respectively. Thanks to the double heat and ionization measurement, EDELWEISS demonstrated gamma and surface-beta rejection factors better than $< 2.5 \times 10^{-6}$ and $< 4\times 10^{-5}$ at 90\%~C.L. respectively, while keeping a nuclear recoil acceptance of about 75\% down to a nuclear recoil energy of $15\keV$~\cite{Armengaud:2017rzu}. Thanks to these performances, the collaboration achieved leading Ge-based exclusion limits on SI interactions for WIMP mass from $5\GeV/c^2$ to $30\GeV/c^2$~\cite{Hehn:2016nll}, provided the first measurement of the cosmogenic activation rate of tritium in Ge above-ground~\cite{Armengaud:2016aoz}, and derived leading limits on ALP dark matter candidates (especially in the $\keV/c^2$ mass range)~\cite{Armengaud:2018cuy}. The next phase of the EDELWEISS experiment, called EDELWEISS-SubGeV, aims at probing DM particle candidates within the eV/$c^2$-to-GeV/$c^2$ mass range by considering both DM interactions on electrons and nuclei.
In this context, the collaboration has recently demonstrated a 55\,eV heat energy threshold on a $33.4\gram$ Ge detector operated above ground -- leading to the most stringent above-ground limit on SI interactions above 600 MeV/$c^2$ (particularly relevant to SIMP dark matter searches)~\cite{Armengaud:2019kfj}, and a $6\eVee$ energy threshold with a $33.6\gram$ Ge detector operated at LSM in NTL boosted mode (78\,V) -- leading to the first Ge-based sub-100 MeV/$c^2$ DM-electron search and one of the most stringent limit on dark photons below $\sim\!\!10\ev/c^2$~\cite{Arnaud:2020svb}.

The {\bf CDMS} (Cryogenic Dark Matter Search) collaboration finished its SuperCDMS-Soudan phase in 2016 during which an array of 15 iZIP Ge detectors of 600 g each, equipped with low-impedance TES and ionization electrodes, were operated in the Soudan underground facility~\cite{Agnese:2013ixa}. The collaboration 
achieved world leading Ge-based WIMP dark matter constraints from $\sim\!\! 1.5\gev/c^2$ to $250\gev/c^2$, combining three different analyses with both high and low voltage operating modes~\cite{Agnese:2018gze, Agnese:2014aze, Agnese:2017njq}. Recently, using a gram-scale Si NTL boosted detector prototype, the SuperCDMS collaboration demonstrated sensitivity to single electron-hole pair~\cite{Romani:2017iwi} for the first time, and subsequently provided the first silicon calorimeter based sub-MeV/$c^2$ DM-electron constraints~\cite{Agnese:2018col}. The CDMS collaboration is now building its next SuperCDMS-SNOLAB phase which will operate 4~detector towers (2~NTL boosted and 2~iZIP), with 6~(Ge/Si) detectors per tower, that will start by mid-2021.

Both CDMS and EDELWEISS are aiming to reach 90\% C.L. excluded cross sections about one order of magnitude above the {\it solar} neutrino floor, for a DM mass between 500 MeV/$c^2$ and 6 GeV/$c^2$. These demonstrators will hence pave the way to their following upgrades that will probe sub-GeV/$c^2$ DM models down to the ultimate solar neutrino background. It should also be mentioned that the technology developed within the CDMS and EDELWEISS experiments is now being exploited in the context of the precise and low-energy measurement of 
coherent elastic neutrino-nucleus scattering process~\cite{Agnolet:2016zir, Billard:2016giu}.

\paragraph{Scintillating cryogenic detectors}
\label{sec:bol_scintillators}
The {\bf CRESST} experiment has pioneered the use of scintillating crystals as cryogenic detectors for DM search and is to date the leading experiment exploiting this approach, simultaneously measuring a phonon/heat and a scintillation light signal. 
The choice of CaWO$_4$ as target material was motivated by its characteristics (good phonon transport properties, high scintillation yield, heavy~$A$ target nucleus in the compound), but an advantage of the technology is the possibility to use any scintillator suitable for low temperature applications. Any particle interaction in the scintillating crystal induces a phonon/heat signal, yielding a precise measure of the energy deposited, and produces scintillation light that is used for particle identification. The scintillation light leaving the crystal is measured by a second cryogenic detector optimised to maximise its absorption (light detector). The discrimination of different types of particles interacting in the scintillating absorber is done on the light yield parameter, defined as the ratio of the energy measured in the light detector to the energy measured in the scintillating crystal. Electron recoils have the highest light yield, all other types of events have a lower light yield, referred as quenching. The combination of the CaWO$_4$ crystalline properties and of the performance of the Transition Edge Sensors (TES) employed, leads to particle identification capabilities down to nuclear recoil energies of $\sim\!\!1\keV$ where the rejection of electron recoil events from the nuclear recoil acceptance region is still at the 90\% level~\cite{Abdelhameed:2019hmk}. 

In recent years, the CRESST technology has been applied to two major spin-offs for the measurement of coherent neutrino-nucleus scattering \cite{Strauss_2017} and DM search \cite{Angloher_2016}. The COSINUS project is now applying the technology to NaI bolometers to verify the explanation of DAMA/LIBRA results in terms of DM scattering off sodium and/or iodine (see Sect.~\ref{sec::nai}).

The third phase of the CRESST experiment, CRESST-III, represents the strengthening of the low-mass DM programme, focusing the experimental search to light DM particles. The detectors in use are optimised for low-mass DM sensitivity by scaling down the crystal mass from $250-300\gram$ for CRESST-II to about $24\gram$ for CRESST-III. The design concept exploits a fully-active enclosure of the target crystals \cite{strauss_prototype_2017}, allowing for a complete vetoing of surface-related background. The first measurement campaign in the cryogenic facility located in the underground laboratory of \textit{INFN Laboratori Nazionali del Gran Sasso} (LNGS) in central Italy, with 10 low-mass detectors lasted from May 2016 until February 2018~\cite{Abdelhameed:2019hmk}.  In total five out of ten detector modules of this type reached or exceeded the design goal of 100\,eV threshold for nuclear recoils, with the best one achieving a threshold as low as 30.1\,eV. The unique combination of this unprecedented threshold with the light element oxygen present in the target yields sensitivity to a DM particle mass as low as $160\MeV/c^2$, allowing the experiment to achieve leading limits on SI interaction for the lowest DM particle mass explored as of today considering standard DM-nucleus interaction\footnote{The isotope $^{17}$O present in the target yields sensitivity for SD neutron-only interactions.}. In the limit of very low-mass DM (sub-GeV/$c^2$) the presence of light nuclei in the absorber becomes increasingly relevant. For this reason, in parallel with the activities aiming to the improvement of detectors based on CaWO$_4$ crystals, an intense R\&D on the development of detectors based on alternative crystals, not necessarily scintillating\footnote{For very small recoil energies, the background discrimination based on the detection of scintillation light, though being of crucial importance for background modelling, becomes progressively less efficient. In this limit it is relevant to have the possibility of neglecting the light channel using crystals intrinsically radio-pure and containing light nuclei.}, is being carried on \cite{Bertoldo2020, Canonica2020}.

The CRESST collaboration is pursuing an upgrade plan to a second phase in which 100~of CRESST-III detectors with thresholds of ${\cal O}${(10\,eV)} and with crystals of significantly improved quality will be used. The goal is to further enhance the sensitivity by several orders of magnitude, improving the capability of detecting light DM. CRESST-III phase-2 will further extend the sensitivity to low-mass DM and reach cross sections at which observation of coherent neutrino-nucleus scattering of solar neutrinos is within reach.

\paragraph{R\&D Programs}
Series of R\&D studies, already ongoing and funded within the CRESST, EDELWEISS and other bolometric experiments, will allow the next-generation of cryogenic experiments to reach the neutrino floor. To achieve such sensitivity, several developments and upgrades are needed and detailed hereafter:

{\bf Background reduction}~~
Scaling a DM experiment is always related to the problem of reducing backgrounds as described in section~\ref{sec:wimpbackgrounds}.  For the next- generation bolometric experiments the limiting backgrounds will be radioactive contaminants and charge leakage (specifically for semiconductors) of internal origin and external neutron and gamma backgrounds. To address the first, there is a wide radioactivity reduction program already ongoing within each cryogenic DM experiments. For instance, the CRESST collaboration is carrying out studies in background reduction from powder purification for crystal growth. The EDELWEISS collaboration, not limited from internal radioactivity levels thanks to the high purity levels of Ge, is investigating ways to further reduce their charge leakage producing single-charge backgrounds. In addition, all cryogenic experiments are also actively investigating  both the origin and the mitigation of the newly observed low-energy excesses affecting their science reach in the lowest DM mass range. Moreover, background reduction can be done actively thanks to particle identification. CRESST for instance has demonstrated the possibility to reject $^{210}$Pb related backgrounds by using active (scintillating) holding structures for the detectors~\cite{angloher_results_2014}, and EDELWEISS is developing a HEMT-based low-noise electronics to extend the  ``heat and ionization'' ER/NR discrimination threshold down to 50 eV$_\textrm{nr}$~\cite{Juillard:2019njs}. 
With increasing exposure and reducing the backgrounds mentioned above, {\it solar} neutrinos will become an irreducible background that will limit the DM sensitivity in the next decade.

{\bf Tonne scale}~~ The exposure needed to deeply explore the solar neutrino floor is  of the order of $\sim\!\!1\tonne\times\years$. In the application of cryogenic/bolometric detectors, reaching one tonne of active material has been for years a far dream. The CUORE project, with its 988 bolometers for a $\sim\!\!750\kgram$ total active mass, was the first  experiment proving that large arrays of bolometric detector can be operated successfully and steadily for years. Many other projects and proposal are consolidating the strategy and the techniques, proving that a tonne$\times$year of active material is a realistic and reachable exposure. It should however be noted that for DM-electron searches, covering the eV/$c^2$-to-MeV/$c^2$ mass range, payloads of only a few kilograms are required to probe most of viable DM models. Eventually, with larger target masses comes the need for instrumenting large arrays of individual detectors. Though multiplexing is not necessarily mandatory to readout about a thousand detectors, several groups over the world are developing multiplexed readouts to ease the scale-up of the cryogenic detector technology. 
    
{\bf Dry dilution cryostat}~~
In recent years the mK~technology evolved in the direction of so-called dry cryostats. In this kind of apparatus the traditional liquid helium (and liquid nitrogen) baths are substitute by cold heads (based on helium gas thermal cycles). These solutions reduced significantly the dead time and maintenance needed in the apparatus but raised the problem of mechanical vibration introduced by the cold heads (typically Pulse Tube cryo-coolers). The CUORE experiment pioneered this technique showing that for double beta decay search (which does not require very low threshold) such solution is suitable. Currently several rare event searches are studying solutions for the utilisation of dry cryostats for low-threshold applications.

{\bf Resolution and threshold}~~
The improvement of resolution and the noise reduction will contribute significantly to improve the sensitivity, especially in the low mass region for DM search. Within all cryogenic DM collaborations, several R\&Ds are carried out, focused on the optimisation of the heat sensors (TES, NTD, or else), the detector design, the magnetic fields control, and the reduction of leakage currents, to ensure improvement of the experimental threshold. In addition to sensor and detector design improvement, several groups are also developing next generation low-noise electronics to further improve the achievable energy resolutions and thresholds.

\subsubsection{Liquid Xenon Detectors}
\label{sec:lxe}

The first multi-ton DM target was realised using the noble gas xenon in liquid form (LXe)~\cite{Aprile:2017aty}. 
Xenon is an excellent scintillator ($\lambda=178\nmeter$)  
and can be ionised easily ($W=15.6\ev$). 
Its high density 
($Z=54$, $\rho=2.8\gram/\cmeter^3$) 
allows building massive but yet  compact detectors with efficient self-shielding capabilities. The high mass number $A$\,=\,131.3 leads to an excellent sensitivity to SI interactions ($\propto A^2$) which is complemented by the world-leading sensitivity to SD WIMP-neutron couplings thanks to a $\sim\!\!50\%$ natural abundance of $^{129}$Xe and $^{131}$Xe with unpaired neutrons. 

The first LXe-based WIMP searches used only the scintillation light ~\cite{Belli:1996sh,Alner:2005pa}. More sensitive detectors also employ the ionization signal: the detector is operated as a TPC (see Sect.~\ref{sec:overview_exp}) drifting the ionization electrons through the LXe target and extracting them into the gas phase~\cite{dolgoshein}. This principle was developed within the ZEPLIN and XENON collaborations. Since the first results from ZEPLIN-II~\cite{Alner:2007ja} and XENON10 in 2007~\cite{Angle:2007uj}, LXe-based dual-phase TPCs are leading the search for WIMPs with mass above a few GeV/$c^2$ (Fig.~\ref{fig:si_status}). The main advantages of the {\bf LXe TPC technology} are:

(i) The very low energy threshold of $\sim\!\!1\keVee$ and $\sim\!\!5\keVnr$ when reading out both light (S1) and charge signals (S2). The threshold can be further reduced by using single scintillation photons in the analysis~\cite{Akerib:2019zrt}. Using only the charge-signal leads to even lower thresholds of a few electrons~\cite{Angle:2011th} and extends the sensitivity down to $\gtrsim 2\GeV/c^2$ (WIMP-nucleon scattering, 1.6\keVnr \ threshold) and to $\gtrsim\,20\MeV/c^2$ (WIMP-electron scattering, 190\,eV$_\textrm{ee}$ threshold)~\cite{Aprile:2019xxb}. Exploiting the Migdal effect~\cite{Ibe:2017yqa} further enhances the WIMP-nucleon sensitivity to $\gtrsim 100\MeV/c^2$~\cite{Akerib:2018hck,Aprile:2019jmx}, see Fig.~\ref{fig:si_status}. The low threshold also provides excellent sensitivity to (all active flavors of) galactic supernova neutrinos~\cite{Chakraborty:2013zua,Lang:2016zhv,Raj:2019wpy}. (The XENON detectors receive alarms from the supernova early warning system (SNEWS) since 2017.)

(ii) The 3D-reconstruction of the interaction position with mm~precision allows for target fiducialisation, precise correction of the detector response and the identification of multiple scatter events.
 
(iii) Rejection of electronic recoil (ER) backgrounds to the $10^{-3}$ level at 50\% nuclear recoil (NR) acceptance based on the  charge-over-light (S2/S1) ratio~\cite{Aprile:2006kx}. The separation of the distributions depends on a number of experimental parameters (e.g., light yield, drift field) and rejection levels of $1\times 10^{-4}$ have been achieved in a small detector operating with a high drift field~\cite{Akimov:2011tj}. Remarkably, the rejection level remains essentially constant down to the low-$E$ threshold and, due to the partially overlapping ER and NR distributions, can be increased by reducing the NR acceptance~\cite{Akerib:2017vbi,Aprile:2019bbb}. Due to the similarity of the singlet and triplet scintillation decay time constants, pulse shape discrimination (PSD) is not powerful for Xe~\cite{Hogenbirk:2018zwf,Akerib:2018kjf}. However, effective PSD would require many detected S1~photons which would be incompatible with the low-threshold goal of LXe detectors.
 
(iv) The target-intrinsic ER backgrounds $^{222}$Rn (problematic are "naked" beta-decays of the daughter $^{214}$Pb) and $^{85}$Kr (present in commercial Xe gas in trace amounts) can be suppressed to extremely low levels by xenon purification~\cite{Abe:2008py,Aprile:2016xhi,Bolozdynya:2007zz,Aprile:2017kop,Akimov:2016yuf}, material selection~\cite{Heusser:1995wd}, detector design as well as S2/S1 discrimination. 
The radiogenic NR background is also minimised by material selection and detector design and is still subdominant at the ton-scale. It can be further reduced using active veto detectors (made of LXe, liquid scintillator or Gd-doped water).  
 
(v) The unprecedented low intrinsic background allows rare event searches using the NR and ER channels. The low background of current LXe detectors directly challenges the DAMA claim~\cite{Aprile:2015ade} and allowed XENON1T to measure the double electron capture of $^{124}$Xe. With $T_{1/2} = 1.8 \times 10^{22}\years$, this is the longest half-life ever measured directly~\cite{XENON:2019dti}. The upcoming generation of LXe detectors will start observing the first solar neutrino signals: irreducible low-energy NRs from $^8$B neutrinos and ERs from pp and $^7$Be neutrinos~\cite{Akerib:2018lyp,Schumann:2015cpa} that can be rejected via S2/S1-based discrimination. Multi-ton scale detectors will measure the solar pp neutrino flux to sub-percent levels~\cite{Baudis:2013qla,Aalbers:2020gsn}. 

(vi) LXe targets are sensitive to many DM models beyond the standard spin-independent and spin-dependent WIMP-nucleon couplings. Examples include (spin-(in)dependent) inelastic WIMP-nucleon interaction models~\cite{Baudis:2013bba,Aprile:2017ngb,Chen:2017cqc} (candidates discussed in Sect.~\ref{sec:nonthermalWIMPs}), WIMP-pion couplings~\cite{Aprile:2018cxk}, and signatures from DM models which predict ER signals~\cite{Akerib:2019diq}. Thanks to the ultra-low ER background, LXe TPCs are also sensitive to solar axions (a recently observed excess of events in XENON1T could be explained by solar axions~\cite{Aprile:2020tmw}), galactic axion-like-particles, bosonic Super-WIMPs~\cite{Arisaka:2012pb,Aprile:2014eoa,Akerib:2017uem,Aalbers:2016jon} as well as light mediators~\cite{Ren:2018gyx} and dark photons~\cite{Aprile:2019xxb} as DM candidates through axio-electric couplings and kinetic mixing, respectively. Moreover, one can search for anomalous magnetic moments of neutrinos~\cite{Harnik:2012ni,Huang:2018nxj} and for signatures of keV-sterile neutrinos~\cite{Campos:2016gjh}. 
 
(vii) Xenon has no long-lived stable isotopes besides the extremely long-lived  $^{136}$Xe ($T_{1/2}=2.2\times10^{21}\years$) and $^{124}$Xe ($T_{1/2}=1.8\times10^{22}\years$)  which decay via 2nd~order weak processes. Their zero-neutrino decay modes are very important science channels, even without isotopic enrichment~\cite{Baudis:2013qla,Ni:2019kms,Akerib:2019dgs,Wittweg:2020fak,Akerib:2019oht,Agostini:2020adk}.

(viii) LUX and XENON demonstrated that LXe TPCs can be operated stably over timescales of years to search for annual modulation signatures (in ER and NR events)~\cite{Aprile:2017yea,Akerib:2018zoq}.

Xenon is present at the $400\ppb$ level (by mass) in the atmosphere. 
As the world-wide annual production rate is around $70\tonne/\years$ (2017) xenon gas is not 
a consumable commodity. Therefore, the gas in the experiments is stored in closed systems with redundant safety features. Apart from the purification to remove trace radioactive contaminants (e.g., $^{85}$Kr) and electronegative impurities (O$_2$, H$_2$O), the xenon gas, in particular its isotopic composition, remains unmodified. The gas thus retains its monetary value and could be sold after the end of an experiment\footnote{Re-selling physics experiment's targets has been done successfully in the past. An example is the vending of $100\tonne$ gallium chloride ($30.3\tonne$ of gallium) of the GALLEX/GNO experiment to Recapture Metals Inc.~(Canada) in 2007.}. The total amount of xenon currently used by DM experiments around the world adds up to $\gtrsim25\tonne$.

%

The LXe target mass was increased by almost 3~orders of magnitude over the last 
$\sim\!\!15\years$ while the ER background (before discrimination) was reduced by a factor of~$10^4$. The upcoming and proposed projects will further follow this trend. 
Since XENON100 introduced the likelihood analysis to direct detection~\cite{Aprile:2011hx}, sensitivity and discovery reach 
of most direct detection experiments and all LXe projects are optimised in the analysis by taking into account the shape of signal and background distributions in a multi-dimensional parameter space (e.g., energy, 3D-position, S2/S1). 

The long-standing competition of several LXe projects has helped tremendously to advance the technology and the field in general. It led to many developments and breakthroughs which were adopted by the community. Some examples: development of calibration sources~\cite{Akerib:2015wdi,Aprile:2016pmc,Akerib:2017hph,Baudis:2020nwe}, novel methods to measure scintillation/ionization efficiencies in situ~\cite{Akerib:2016mzi}, new LXe purification methods~\cite{Bruenner:2016ziq,Aprile:2017kop}, photosensors~\cite{Akerib:2012da,Baudis:2013xva}, materials~\cite{Akerib:2017iwt}, triggerless detector readout~\cite{Aprile:2019cee}, analysis techniques~\cite{Aprile:2011hx,Akerib:2019zrt}, new science channels and many others. Another prominent example is the NEST simulation package~\cite{Szydagis:2011tk} which is by now the standard tool for LXe detectors and used by the LAr community as well.  

In recent years, the field experienced some consolidation: the LUX and ZEPLIN collaborations merged to LZ and the XMASS collaboration, which operated a single-phase LXe detector with a $0.83\tonne$ target in Kamioka~\cite{XMASS:2018bid}, joined XENON. Right now, there are four active collaborations world-wide: 

The {\bf LUX-ZEPLIN (LZ)} dual-phase TPC~\cite{Akerib:2019fml} (mainly USA-UK) is currently being commissioned at the SURF laboratory in the USA. The $7.0\tonne$ active LXe target is monitored by a total of 494~Hamamatsu R11410-22 PMTs (3" diameter); these are operated in the LXe and sensitive to the scintillation light. The TPC electrodes to establish the drift and extraction fields are realised as 90$^\circ$ woven meshes made from 75\,$\mu$m and 100\,$\mu$m stainless steel wire. The $\sim\!\!2\tonne$ of LXe surrounding the TPC are monitored by 131~additional PMTs to veto interactions which happen in the TPC and this optically separated "skin region" simultaneously. The double-wall cryostat is made of low-background titanium and surrounded by an outer detector made of Gd-doped LAB scintillator. This "neutron veto" tags $\sim\!\!90\%$ of the neutrons that generated a single scatter signature inside the TPC. An active water Cherenkov muon veto complements the shielding. The sensitivity of LZ to SI WIMP-nucleon couplings is $1.5 \times 10^{-48}$\,cm$^2$ (90\%\,CL, $40\GeV/c^2$ WIMP) assuming an exposure of $15.3\tonne\times\years$ (see Fig.~\ref{fig::projections}) and an ER background of 
$0.04\events/(\tonne\times\days\times\keVee)$~\cite{Akerib:2019fml}. 
This background is achieved by removing $^{85}$Kr from the Xe by gas chromatography~\cite{Akerib:2016hcd} and by reducing the $^{222}$Rn activity to $1.8\muBq/\kgram$ by material selection/cleanliness requirements as well as active online removal by means of adsorption on charcoal~\cite{Pushkin:2018wdl}. (LUX has achieved a Rn activity of $66\muBq/\kgram$~\cite{Bradley:2015ina}.)

{\bf PandaX-4T} (mainly China) is being built in the Jinping Underground Laboratory (CJPL). The active LXe target of $4.0\tonne$ is monitored by 368~Hamamatsu R11410-23 PMTs. Also PandaX-4T employs a LXe veto by recording scintillation light from interactions in the $\sim\!\!2\tonne$ of LXe surrounding the TPC by 126 PMTs (1"$\times$1") arranged in two rings. The low-background stainless steel cryostat is immersed in an active water shield of 10\,m diameter and 13\,m height. The goal of PandaX-4T is to reach an ER background of $0.05\events/(\tonne\times\days\times\keVee$)~\cite{Zhang:2018xdp} 
by reducing the $^{222}$Rn concentration to $1\muBq/\kgram$ (PandaX-II has achieved $8\muBq/\kgram$). To achieve this, a cryogenic distillation column is employed~\cite{Wang:2014ehv} which is capable to purify at a speed of $57\kgram/\hours$~\cite{ref::pandaX4t}. With an exposure of $5.6\tonne\times\years$ PandaX-4T is expected to be able to probe cross sections down to $6  \times 10^{-48}$\,cm$^2$ (90\%\,CL, $40\GeV/c^2$ WIMP), see Fig.~\ref{fig::projections}. For the future the $30\tonne$-scale experiment PandaX-30T is being proposed~\cite{Liu:2017drf}. 

{\bf XENONnT}~\cite{Aprile:2015uzo,Aprile:2020vtw} (mainly Europe-USA-Japan) is the upgrade phase of XENON1T~\cite{Aprile:2018dbl} which re-uses many of the XENON1T subsystems. It is under commissioning at the Italian Gran Sasso Laboratory (LNGS). A total of 494\,Hamamatsu R11410-21 PMTs record the signals from the active $5.9\tonne$ LXe target. The detector is optimised for a minimal mass of the detector components (to reduce NR backgrounds) and for minimal Rn-emanation. The double-wall cryostat made of low-background stainless steel is installed in a 9.6\,m diameter active water shield. The water is doped with gadolinium (0.2\% of Gd$_2$(SO$_4$)$_3$ dissolved in water) to capture and eventually tag thermalised neutrons which produced a single-scatter signal in the TPC. The expected tagging efficiency is around 90\%. As a novel feature the LXe is constantly purified in liquid phase to achieve a higher electron lifetime. The Kr is removed from the LXe target to negligible levels by means of cryogenic distillation~\cite{Aprile:2016xhi}. Following the collaboration's pioneering work~\cite{Aprile:2017kop} the Rn concentration in the LXe is reduced on-line in a high-throughput ($72\kgram/\hours$) cryogenic distillation column. The design goal is to reach a concentration of $1\muBq/\kgram$ (XENON1T has reached $4.5\muBq/\kgram$ at the end of its operation) and an ER background of $0.05\events/(\tonne\times\days\times\keVee)$~\cite{Aprile:2020vtw}. 
The projected sensitivity is $1.4 \times 10^{-48}\cmeter^2$ (90\%\,CL, $50\geV/c^2$ WIMP) after acquiring an exposure of $20\tonne\times\years$, see Fig.~\ref{fig::projections}.

LZ and XENONnT will encounter a relevant irreducible NR background from the coherent scattering of atmospheric neutrinos off xenon nuclei. Signals from $^8$B solar neutrinos will be observed in the S1-S2-based search only below $\lesssim$6\keVnr. The XENON1T charge-only search already expected a contribution of $(2.0 \pm 0.3)\events$  from $^8$B in the ROI for a $6\geV/c^2$ WIMP~\cite{Aprile:2019xxb}. 

The goal of a multi-ton detector, e.g., as proposed by the {\bf DARWIN} collaboration, is to explore the entire accessible WIMP parameter space until the background is dominated by irreducible coherent neutrino scattering events~\cite{Aalbers:2016jon}. Spin-independent cross sections down to $2\times10^{-49}$\,cm$^2$  (90\%\,CL, $40\GeV/c^2$ WIMP) can be reached in a $200\tonne\times\years$ exposure of a $40\tonne$ LXe target~\cite{Schumann:2015cpa}. This requires the ER background being dominated by solar pp-neutrinos and thus sub-dominant contributions from the target-intrinsic contaminants Kr and Rn. $^\textrm{nat}$Kr concentrations of $<$0.026\,ppt (90\% CL) have been demonstrated~\cite{Aprile:2016xhi}; this is better than required for DARWIN. The number of Rn atoms in the target has to be reduced by a factor $\sim\!\!5$ compared to the current state-of-the art to reduce the specific 
$^{222}$Rn activity from $\sim\!\!4\muBq/\kgram$ to $0.1\muBq/\kgram$. This will be achieved by combining material selection, detector design, surface treatment as well as online Rn-removal. About 9~NRs from atmospheric neutrinos can be observed in the $200\tonne\times\years$ exposure above a 5\keVnr \  threshold. To reduce the ER background to below the NR level S2/S1-based rejection needs to be improved by a factor 3-4 compared to what has been already demonstrated in large detectors by optimising light yield and electric field. 
DARWIN's $40\tonne$ LXe target corresponds to a very compact scale of $\sim\!\!2.6\meter$. It could be realised at the depth of the Italian LNGS laboratory; a CDR is being prepared. The three multi-ton projects described above all serve as R\&D and demonstration platforms; in addition dedicated R\&D for DARWIN is pursued at various places. Common R\&D between continents and collaborations is being discussed and prepared. 

The primary goal of a detector like DARWIN is the unprecedented sensitivity to WIMP dark matter exploring SI, SD couplings as well as inelastic scattering. However, its low ER and NR backgrounds and low threshold will allow addressing a plethora of questions related to the dark Universe, the particle nature of neutrinos as well as various questions in astro- and nuclear physics~\cite{Baudis:2013qla,Aalbers:2016jon,Agostini:2020adk,Aalbers:2020gsn}. 

\subsubsubsection{R\&D towards the Ultimate LXe Detector}

A dedicated R\&D program has started within DARWIN to pave the way towards an ultimate LXe-based WIMP detector with a sensitivity down to the neutrino floor. Various aspects are also addressed by other research groups, e.g., in the USA or in China. The upcoming multi-ton scale projects PandaX-4T, XENONnT and LZ can all be considered as "demonstration platforms" for a multi-ton/DARWIN detector as the additional increase in size is modest. The key requirement to reach the neutrino floor is that all internal backgrounds are subdominant to the one from coherent neutrino nucleus interactions~\cite{Schumann:2015cpa}. This will be achieved by a combination of various efforts which can be categorised in the following way:

{\bf Liquid Xenon Target}~~ The LXe inventory has to be constantly purified to achieve a high electron lifetime and small corrections (which impact S2/S1 rejection). R\&D is ongoing to  directly purify the cryogenic liquid itself to increase the efficiency. It is crucial that the purification systems do not add new background, e.g., from Rn. Radio-pure components (e.g., pumps~\cite{LePort:2011hy,Brown:2018uya})  are being developed.  In addition, a strategy to procure the required Xe inventory has to be identified given the limited availability of the gas on the market.

{\bf Backgrounds}~~ 
The specific $^{222}$Rn concentration in the LXe has to be reduced by a factor~$\sim\!\!50$ compared to what has already been achieved in LXe detectors (and by a factor of~10 compared to the design goal of the upcoming detectors~\cite{Akerib:2019fml,Aprile:2020vtw}). This will be achieved by a combination of online Rn-removal, surface treatment, material selection and detector design. The latter two items also address all other detector-related ER and NR backgrounds. R\&D is ongoing to develop and optimise neutron veto systems which tag neutrons that have created a single scatter interaction in the LXe TPC. Promising systems under study are based on Gd-loaded water (using the EGADS experience~\cite{Xu:2016cfv}) and Gd-loaded LAB liquid scintillator~\cite{Yeh:2007zz,Akerib:2019fml}.

{\bf Detector}~~ 
The $\sim\!\!2.6\meter$-scale of a $40\tonne$ LXe TPC is only moderately larger than the $\sim\!\!1.5\meter$ of the upcoming detectors currently under commissioning. Nevertheless, R\&D is required especially on large-scale TPC electrodes (as drift field parameters directly affect S2/S1-based ER rejection), understanding of rare detector artefacts, e.g,. the delayed emission of individual electrons~\cite{Bernstein:2020cpc,Akerib:2020jud}, identification of clean materials and optimisation of the light collection efficiency which again improves ER rejection.

{\bf Photosensors}~~ PandaX-II~\cite{Cui:2017nnn} and XENON1T~\cite{Aprile:2017aty} were successfully operated for years using the Hamamatsu R11410 PMTs; the same sensor was thus selected for all three upcoming projects and serves as baseline option for DARWIN~\cite{Aalbers:2016jon,Agostini:2020adk}. Nevertheless R\&D on alternative photosensors is ongoing as PMTs contribute to the background and their quantum efficiency and achievable position resolution is not optimal. Possible solutions currently under study are SiPMs in various variants~\cite{Arneodo:2017ozr,Baudis:2020nwe}, hybrid detectors~\cite{Ferenc:2018ymb} as well as micropattern detectors~\cite{Erdal:2018bjg}.

\subsubsection{Liquid Argon Detectors}
\label{sec:lar}

Liquid argon (\LAr) is favourable for WIMP detection due in part to the extremely powerful suppression of ER backgrounds enabled by pulse shape discrimination, rendering these backgrounds essentially negligible. This will allow very large argon detectors to mount DM searches that are free of instrumental backgrounds, allowing strong discovery potential down to the neutrino floor.


{\bf \DEAP}, a single-phase liquid argon experiment with a total mass of $3200 \kgram$ of argon contained in a spherical acrylic vessel located at SNOLAB in Sudbury, Canada, has exploited {\bf pulse shape discrimination (PSD)} to achieve ER background rejection with a leakage probability of $4.1\times{10}^{-9}$ with $90\%$ acceptance for nuclear recoils, rendering ER backgrounds insignificant. DM search data resulted in the current best limit on the WIMP-nucleon spin-independent cross section on a LAr target, $3.9\times{10}^{-45}\cmeter^2$ ($1.5\times{10}^{-44}\cmeter^2$) for a  $100\GeV/c^2$ ($1\tev/c^2$) WIMP mass~\cite{Ajaj:2019jk}.
The experiment has also achieved extremely low levels of radioactivity, including the lowest-achieved radon background in noble gas of  $0.15{\muBq/\kgram}$, 
already low enough to enable next-generation searches~\cite{Ajaj:2019jk}. 
    The experiment is currently implementing a set of hardware modifications to mitigate the currently-limiting background from $^{210}$Po on a detector surface in the experimental cooling system.  After this upgrade, the experiment is targeting a background-free exposure with an exclusion sensitivity similar to the current results from XENON1T.


In \DEAP, several novel and cutting-edge technological solutions were successfully developed and implemented.  The collaboration developed and qualified ultra-pure acrylic for use as a cryogenic vacuum vessel, controlled the acrylic radioactivity during production, developed a large-scale ``resurfacer'' device to produce radiopure surface and maintained radon-free conditions after resurfacing, and developed a large-area in-situ deposition system for the wavelength shifter, depositing $10 \meter^2$ of  tetraphenylbutadiene (TPB) wavelength shifter over the surface area of the \DEAP\ vessel.

With the very low rate of cosmogenic muon-induced neutrons at SNOLAB, and the \DEAP\ design which includes substantial passive neutron shielding for external-source neutrons, \DEAP\ has achieved the lowest rate of neutron-induced nuclear recoils of any DM experiment~\cite{Ajaj:2019jk}. \DEAP\ has also demonstrated for the first time excellent position reconstruction in a large single-phase argon DM experiment, with a resolution better than $10\mmeter$ for external-source low-energy events~\cite{Ajaj:2019jk}.

\DEAP\ has also set a limit on neutrinoless double EC decay of $^{36}$Ar 
with unique sensitivity to all three available detection channels~\cite{Dunford:2018xar}, and performed precision measurements of $^{36}$Ar and $^{42}$Ar 
activity in atmospheric argon (\AAr)~\cite{PhysRevD.100.072009}.

{\bf \DSf} at LNGS uses additional charge information available with a two-phase time projection chamber to measure both the prompt argon scintillation light and the ionised electrons resulting from a particle interaction in the detector. This technique provides excellent position resolution and efficient detector fiducialisation while maintaining PSD capabilities. 

Operating with \AAr, \DSf\ provided a powerful assessment of PSD in this approach by measuring a rejection factor better than one part in $1.5\times{10}^{7}$~\cite{Agnes:2015gu}, 
limited by available statistics. A supplementary analysis of Monte Carlo simulated data predicts an ultimate rejection factor  
$>3\times{10}^9$~\cite{Aalseth:2018gq}. 
\DSf\ also demonstrated the viability of an underground argon (\UAr) target, which can be obtained with an $^{39}$Ar
content that is suppressed by a factor of more than 1400 with respect to \AAr, drastically reducing the expected number of electron recoil events to be discriminated~\cite{Agnes:2016fz,Agnes:2017iw,Agnes:2018ep}. 
\DSf, finally, demonstrated the use of a comprehensive anti-coincidence veto scheme, based on a water Cherenkov and on an organic liquid scintillator, which suppressed the residual background from \grs\ and neutrons~\cite{Agnes:2016fw}.

\DSf\ has performed a blind DM search and observed no background events over a run period in excess of two years~\cite{Agnes:2018ep}. 
In addition to sensitivity to WIMPs with mass above $30\gev/c^2$, the two-phase \DSf\ detector has extended its reach to WIMP mass below  $10\GeV/c^2$ by detecting single ionization electrons (i.e., S2-only) extracted from the liquid argon.
The extremely low background, high stability, and low  $100\eVee$ $(600\eVnr)$ analysis threshold of \DSf, enabled a study of very low-energy events, characterised by the presence of the sole ionization signal, which resulted in world-leading sensitivity for low-mass DM searches in the mass range 
$1.8-3.5\gev/c^2 $~\cite{Agnes:2018fg} (see Fig.~\ref{fig:si_status}).  The same technique also provided very competitive limits for the scattering of DM from electrons~\cite{Agnes:2018ft}. 
With careful control of ER backgrounds from local radioactivity and a reduction of the  $10^{39}$Ar background, a $1\tonne$ \LAr\ detector has the potential to reach the solar neutrino floor in this low-mass parameter space.


Given the potential reach of an argon-based detector, a team of over 350~scientists from all of the major groups currently using \LAr\ to search for DM, including \ArDM, \DSf, \DEAP, and \mCLEAN, have joined to form the {\bf Global Argon Dark Matter Collaboration (GADMC)} with a goal of building a series of future experiments that maximally exploit the advantages of \LAr\ as a detector target. The argon experiments (\DEAP, \DSk\ and ARGO) will also provide valuable information to the  SuperNova Early Warning System (SNEWS) network, allowing the full exploitation of supernova signal towards the understanding of the explosion process and neutralisation burst, and implications on neutrino masses and oscillations~\cite{RossiTorres:2015wh,Scholberg:2017id}.

The enabling technologies of the GADMC program are~\cite{Aalseth:2018gq}: the argon target obtained from the high-throughput extraction of low-radioactivity argon naturally depleted in $^{39}$Ar 
from an underground source (\UAr) via the Urania plant; the target high-throughput purification and active isotopic separation via the Aria cryogenic distillation column; light detection via large-area cryogenic photodetector modules (PDMs) made of custom-designed silicon photomultipliers (SiPMs) assembled in a custom-built factory; operation of the WIMP detector within an active veto with liquefied atmospheric argon (\AAr) as scintillator, hosted inside a membrane cryostat built with the technology developed at CERN for \pDUNE~\cite{Abi:2017wp,Acciarri:2016wz}.

The immediate objective of the \GADMC\ is the construction of the {\bf \DSk} two-phase \LAr\ detector, which will operate in Hall-C of the Gran Sasso National Laboratory (LNGS).  The \DSk\ detector consists of two nested detectors housed within a \pDUNE-style membrane cryostat.  
The inner detector is contained within a vessel made from ultra-pure  acrylic (PMMA) and filled with \UAr.  The central active volume of the TPC is defined by eight vertical reflector panels and the top and bottom windows of the acrylic vessel. 
All the TPC surfaces in contact with the active argon volume will be coated with  wavelength shifter TPB to convert \LAr\ scintillation light to a wavelength detectable by SiPMs.  8280 SiPM-based $5\times5\cmeter $ 
PhotoDetector Modules (PDMs) view the argon volume through the top and bottom windows of the acrylic vessel. The detector will operate like the \DSf\ detector, where the ionization in the liquid is drifted to a high voltage gaseous region where an electro-luminescence signal is produced. The height of the TPC is $350\cmeter$. 
The total mass of \LAr\ in the active volume is $49.7\tonne $. 


The outer veto detector is made of a passive Gd-loaded 
PMMA shell surrounding the inner detector and between two active \AAr\ layers.  The Gd-loaded 
PMMA shell moderates neutrons emitted from the \LAr\ TPC until they capture on Gd, 
resulting in the emission of multiple \grs.  The \grs\ interact in the \AAr\ layers and cause scintillation light that is detected by photodetectors, thereby providing an efficient veto of radiogenic neutrons that could result in a NR in the TPC.  The \pDUNE-like cryostat will be surrounded by layers of plastic to moderate cosmogenic and radiogenic neutrons from the rocks surrounding Hall C in the underground LNGS.

The \DSk\ detector will have ultra-low backgrounds and the ability to measure its backgrounds {\it in situ}, resulting in an expected sensitivity to WIMP-nucleon cross sections of $1.2\times{10}^{-47}\cmeter^2 $ ($1.1\times{10}^{-46}\cmeter^2 $) at $1\tev/c^2$ ($10\tev/c^2$) WIMPs with a $100\tonne\times\years$ exposure.  
This projected sensitivity is a factor of~>50 better than currently-published results above  $1\tev/c^2$ and covers a large fraction of the parameter space currently preferred by supersymmetric models.
The sensitivity of \DSk\ would further improve to $7.4\times{10}^{-48}\cmeter^2$ ($6.9\times{10}^{-47}\cmeter^2$) at $1\tev/c^2$  
($10\tev/c^2$) WIMPs with a  $200\tonne\times\years$ exposure.  


\DSk\ is designed to operate with all sources of instrumental background reduced to less than $0.1\events$  
over a $200\tonne\times\years$ 
exposure.  
All background from minimum-ionising radiation sources will be completely removed by the combined action of PSD of the primary scintillation pulse and comparison of the primary and secondary scintillation.  
The expected radiogenic neutron background contributions of the various detector components following all TPC and veto cuts for the full \DSk\ exposure are negligible.  
The ER signals from solar neutrinos and radiogenic sources will be removed by the pulse shape discrimination capabilities of argon. 
The only remaining background for WIMP searches will be the signal from the coherent scattering of atmospheric neutrinos on argon nuclei, with an expected $3.2\events$ over the $200\tonne\times\years$ 
exposure. The outstanding sensitivity to coherent nuclear recoils will enable \DSk\ to detect a supernova neutrino burst coming from anywhere in the Milky Way Galaxy and, for a majority of the galaxy, clearly identify the neutralisation burst. \DSk\ would perform a flavor-blind measurement of the total neutrino flux and average energy, setting an overall normalisation that is not affected by neutrino oscillations. 

In parallel to \DSk\ detector, a second and important element for this program is the development of an approximately 1 tonne detector specifically optimised for the detection of low-mass DM, {\bf \DSl}.  \DSl\ will achieve a lower energy threshold than \DSk\ by triggering on the electroluminescence signal from ionization electrons, thereby adding sensitivity to WIMP mass below $10\GeV/c^2$ at the expense of the PSD power afforded by argon's prompt scintillation light.  Without PSD, contributors to the ER background in \DSl\ must be reduced beyond the requirements of \DSk\ through careful detector design and material selection. 
Among the technological advancements required to enable \DSl\ and the goal of reaching the neutrino floor for WIMP mass between $ 1\geV/c^2$ and $ 10\GeV/c^2$
are the development of low-background PDMs and the construction of the Aria cryogenic distillation column, which will completely remove $^{85}$Kr and reduce $^{39}$Ar levels to the level of $ 1\muBq/\kgram$.  


The ultimate objective of the \GADMC\ is the construction towards the end of this decade of the {\bf{ARGO}} detector, which will have a $300\tonne $ 
fiducial mass and will push the experimental sensitivity to the point at which the coherent scattering of atmospheric neutrinos becomes a limiting background. The excellent ER rejection possible in argon will eliminate ER backgrounds from solar neutrinos, which will extend the sensitivity of ARGO to high WIMP mass above $\gtrsim100\gev/c^2$, beyond that of technologies with more limited ER discrimination.  
It remains to be decided whether ARGO will be a single- or double-phase detector. The throughput  of the Urania plant and Aria facility will enable $400\tonne $ 
of \UAr\ to be extracted and purified over a period of about $4\years$.  In addition to DM detection, such a large detector would also have excellent sensitivity to a neutrino burst associated with a galactic supernova.  If located at SNOLAB or at similar depth, ARGO\ will also have the potential to provide a very accurate measurement of the flux of CNO neutrinos to solve the Solar Metallicity Problem~\cite{Franco:2016ex};  see section~\ref{sec:neutrino}.

The long-term program including \DSk, \DSl, and ARGO, will completely cover the spin-independent WIMP hypothesis parameter space down to the neutrino floor for WIMP masses from $1\gev/c^2$ 
to several hundreds $\tev/c^2 $. 

It is worth noting that preliminary studies have provided exciting hints of a direction sensitive effect on nuclear recoils in a \LAr-TPC~\cite{Cao:2015ks}. 
Columnar recombination models~\cite{Jaffe,Cataudella_2017} suggest that the magnitude of the recombination should, under some circumstances, vary with the angle between the electric field and the track direction. A difference in the electron-ion recombination effect is, in fact, expected when the ionising track is either parallel or perpendicular to the electric field. 
\GADMC\ is further investigating the possibility to exploit the process of columnar recombination in liquid argon to assess the
directionality of a 2-phase TPC with the {\bf ReD} program \cite{Aalseth:2018gq}. 

The prospects for a \LAr\ directional detector are specially promising at the neutrino floor, to confirm the galactic origin of an observed signal~\cite{Cadeddu_2019}; in addition, at even larger sensitivities, measuring the direction of a nuclear recoil would give the possibility to distinguish the atmospheric neutrino background from a WIMP signal. 

\paragraph{R\&D Programs in view of the \GADMC\ \LAr\ Program}
\label{sec:LArTechnologies}
The following four key technologies will enable the next generation of experiments and the long term scientific goals of the \GADMC. Their development will also have potentially wide-reaching effects within the physics community.

{\bf Low-Radioactive underground argon with Urania~\cite{Aalseth:2018gq}:}
The \DSf\ experiment established that \UAr\ is depleted of $^{39}$Ar 
by a factor of approximately 1400, a sufficiently low rate to be deployed in a detector the size of \DSk. However, constructing \DSk\ will require that large amounts of \UAr\ be procured in a timely fashion. This will be accomplished by Urania, an argon extraction and purification plant capable of extracting $330\kgram/\days $ 
of \UAr.  The Urania plant is being built and will be installed in Colorado. It is projected to collect approximately 60 tonnes of argon for use in the \DSk\ detector by 2022 and could continue to produce underground argon for ARGO and other interested particle physics experiments that require \UAr\ to achieve their scientific objectives.  

{\bf Purification and Active Depletion with Aria~\cite{Aalseth:2018gq}:} 
The Aria plant is a $350\meter $ 
tall cryogenic distillation column that was designed to explore the possibility of chemically separating argon isotopes. It is under construction in Sardinia, Italy. The plant is estimated to be able to process \UAr\ at a rate of $10\kgram/\days$, 
obtaining a $^{39}$Ar 
depletion factor of 10 per pass.  
Preliminary measurements with part of the column indicate that the Aria plant could potentially exceed significantly the design criterion of a factor of 10 depletion per pass, with multiple passes.
At much larger rate  this plant will perform chemical purification of the \UAr\  for \DSk\  to make it detector grade.

{\bf ARGUS Large-scale storage facility for underground argon:}
A facility for storing 400 tonnes of underground argon is being developed at SNOLAB in Sudbury, Canada.  Plans are being made to extend the operation of the Urania extraction plant and to transfer and store the full target required for ARGO at SNOLAB. Long-term underground storage is required to prevent activation of the low-radioactivity argon.

{\bf SiPM-based Cryogenic Photosensors~\cite{Aalseth:2018gq}:} 
The development of low-background, large-area, cryogenic silicon photomultiplier (SiPM) detectors capable of replacing conventional photomultiplier tubes is critically important for achieving the desired sensitivity of \DSk\ and other large-scale \LAr-based experiments, including DUNE, and LXe-based detectors, such as nEXO, NEXT and possibly DARWIN.  The \DSk\ photodetector modules will be assembled at the Nuova Officina Assergi (NOA), a dedicated cleanroom packaging facility that will have future utility for any experiment needing large volume silicon detector production.

{\bf \pDUNE\ Liquid Argon Cryostat~\cite{Abi:2017wp,Acciarri:2016wz}:} 
The \DSk\ detector will operate within a membrane cryostat filled with liquefied atmospheric argon, a technology initially developed at CERN for \pDUNE. Eliminating the organic liquid scintillator veto used in \DSf\ for the \AAr\ veto has several advantages. With the \DSk\ \LAr-TPC directly immersed in \AAr, the massive stainless steel vacuum cryostat that would be needed and its correspondingly large contribution of background events, can be replaced with a transparent, radio-pure PMMA vessel. Photodetector modules can then be mounted outside of the PMMA vessel, reducing their contribution to the background rate and simplifying their assembly strategy. The \pDUNE\ cryostat has the added advantage that it is scalable, making it a technology appropriate for ARGO.
%

\subsubsection{Scintillating Crystals, Ionisation Detectors, Bubble Chambers}
\label{sec:otherdetectors}

Together with experiments based on noble liquids and solid state cryogenic detectors, other projects based on different  technologies, in many cases at  smaller scales, can help to explore the DM landscape. The identification of distinctive signatures of the DM interaction, like the annual modulation in the rates (see Sect.~\ref{sec:ddbasics}), is being pursued using NaI scintillators. There are also projects particularly suited to explore low mass WIMPs, achieving extremely low-energy thresholds down to tens of $\eVee$
and/or searching for different interaction channels, thanks to the development of novel technologies in advanced, ultra-sensitive detectors and sensors 
\cite{battaglieri2017cosmic_temp}.
Bubble chambers using target fluids containing $^{19}$F have shown leading sensitivity to SD interactions.

\paragraph{NaI(Tl) scintillators}
\label{sec::nai}

As pointed out in Sect.~\ref{sec:wimplimits}, the strong tension with other results when interpreting the DAMA/LIBRA annual modulation signal as DM in different halo and interaction models has made a model-independent test with the same NaI target mandatory. This is the goal of COSINE-100 and ANAIS-112, now in data-taking phase, as well as of other projects in preparation.

The {\bf DAMA/LIBRA} experiment operates in the Laboratori Nazionali del Gran Sasso (LNGS) in Italy~\cite{bernabeippnp2020}. Detectors produced by Saint Gobain company with a mass of $9.7\kgram$ are being used, firstly 9 units and since 2003, 25 detectors. In 2011 all PMTs were replaced, allowing to reduce the software energy threshold from 2 to $1\keVee$ in the second phase of the DAMA/Libra experiment. The background level in the region of interest is from 0.5 to $1\events/(\keV\times\kgram\times\days)$~\cite{Bernabei:2018yyw}. The results from the phase-1~\cite{damaphase1} were confirmed by those of phase-2~\cite{Bernabei:2018yyw}, favouring the presence of a modulation with all the proper features expected from the standard halo model at 12.9\,$\sigma$~\cl with an exposure of $2.46\tonne\times\years$ over 20 annual cycles. The deduced modulation amplitude for the 2-6\keVee\ region is $S_{m}=(0.0103\pm0.0008)\events/(\keV\times\kgram\times\days)$; 
compatible values were found for different fitting procedures, periods of time, energy regions and detector units. Improved model-dependent corollary analyses after DAMA/LIBRA phase-2 have been presented \cite{damacorollary,bernabeippnp2020}, applying a maximum likelihood procedure to derive allowed regions in the parameters' space of many different considered scenarios by comparing the measured annual modulation amplitude with the theoretical expectation. Data collection is expected to go on until the end of 2024, while work is underway for phase~3, updating hardware to lower the software energy threshold below 1\keVee.

{\bf COSINE} is a joint effort between the KIMS collaboration in Korea and the DM-Ice experiment carried out at the South Pole. Eight NaI(Tl) detectors from Alpha Spectra company are operated ($106\kgram$ in total, COSINE-100) immersed in 2200~l of liquid scintillator at the Yangyang underground Laboratory (Y2L) in South Korea \cite{cosineperformance,cosinebkg}. The liquid scintillator system is mainly intended to veto the $^{40}$K events from the NaI(Tl) crystals, producing a peak at 3.2\kev. The Physics run started in September 2016 with a threshold at 2\kev$_{ee}$. From the first $59.5\days$ of data  the DAMA/LIBRA signal was excluded as due to Spin-Independent (SI) WIMPs with a standard halo model \cite{cosinenature}. The first annual modulation analysis using $1.7\years$ of data has been presented \cite{cosinemod}; total exposure analysed is $97.79\kgram\times\years$ as three large crystals were excluded due to low light yield and excessive PMT noise. The COSINE-100 event rate for 2-6\keVee\ in the crystals used is $2.7\events/(\keV\times\kgram\times\days)$~\cite{cosinemod}.
The best fit modulation amplitude derived for the 2-6\keVee\ region is $S_{m}=(0.0083\pm0.0068)\events/(\keV\times\kgram\times\days)$. 
Data taking is going on and COSINE-100 expects to attain 3$\sigma$ coverage of the DAMA region with five years of data exposure. The development of crystals to improve radiopurity is underway in Korea for a second phase with $\sim 200\kgram$ of NaI(Tl) crystals (COSINE-200); first results from small ($\sim 0.7\kgram$) crystals have been presented showing a reduction of $^{210}$Pb \cite{cosinecrystals}. 

{\bf ANAIS} (Annual modulation with NAI Scintillators) is operating nine NaI(Tl) modules also built by Alpha Spectra ($112.5\kgram$ in total, ANAIS-112) at the Canfranc Underground Laboratory in Spain \cite{anaisperformance,anaisbkg}. The DM run is underway since August 2017, with an outstanding light collection of $\sim 15$~photons per keV for all modules allowing an energy threshold of 1\keVee. 
In the region from 1 to 6\keVee, the measured, efficiency corrected background level is $3.6\events/(\keV\times\kgram\times\days)$~\cite{anaisbkg}. The first results for $1.5\years$ of data and exposure of $157.55\kgram\times\years$, focused on a model-independent analysis of annual modulation, were published \cite{anaismod}; updated results for 
{2 years \cite{anaistaup} and 3 years of data \cite{anais3y} were presented. From the analysis of $313.95\kgram\times\years$ exposure, the best fit modulation amplitude derived for the 2-6\keVee\ region is $S_{m}=(0.0003\pm0.0037)\events/(\keV\times\kgram\times\days)$ while for the 1-6\keVee\ region it is $S_{m}=(-0.0034\pm0.0042)\events/(\keV\times\kgram\times\days)$. Both are incompatible, respectively, at 2.6\,$\sigma$ and 3.3\,$\sigma$ with the DAMA/LIBRA results~\cite{anais3y}}. The evaluated sensitivity from the measured background in 1-6\keVee\ (corroborated by the 
{3}-year results) confirms the possibility to detect the annual modulation in the 3\,$\sigma$ region compatible with the DAMA/LIBRA result for five years of measurement in the present conditions \cite{anaissensitivity}. Data taking is progressing smoothly and it is expected to go on to reach this exposure time.

{\bf SABRE} (Sodium-iodide with Active Background REjection) is under preparation at LNGS \cite{sabre}. It is focused on the development of ultra-high purity NaI(Tl) crystals; a potassium content of $(4.3\pm 0.2)\ppb$ has been quantified by ICPMS for a new crystal \cite{sabreK}. Tests with one detector (SABRE Proof of Principle, PoP) are underway with one $3.5\kgram$ crystal shipped from the USA to Gran Sasso \cite{sabrenai33}. SABRE  plans to operate $\sim\!\!50\kgram$ and use passive and active (liquid scintillator veto) shielding. The goal is to reach a background level in the region of interest around one order of magnitude lower than that of DAMA/LIBRA.
After three years of exposure, the experiment is expected to be sensitive to WIMP-nucleon scattering cross sections down to $2\times10^{-42}\cmeter^{2}$ for a WIMP mass of $40-50\gev/c^{2}$. The main asset of the project is that twin detectors in northern and southern hemispheres (at LNGS and in the Stawell Laboratory, being built in Australia) will be implemented to investigate any seasonal effect due to backgrounds, which should show opposite phase. The plan is to operate first the SABRE PoP set-up when authorised, before producing all the required crystals and starting the full experiment at LNGS.

A different approach can be pursued using NaI. After the first study of NaI(Tl) crystals at low temperatures for bolometric applications \cite{coron2013}, {\bf COSINUS} (Cryogenic Observatory for SIgnatures seen in Next-generation Underground Searches) is developing at LNGS NaI scintillating bolometers based on the CRESST technology~\cite{Angloher_2016}. As the phonon signal is independent of the particle type but the scintillation light is not, such a detector has the potential to discriminate nuclear recoil events from electronic background on an event-by-event basis, which has been proven with crystals with a mass of tens of grams. The unquenched phonon channel is used and an energy threshold lower than in conventional NaI(Tl) scintillators is expected. It has been evaluated that if COSINUS excludes a DM scattering rate of about $0.01\events/(\kgram\times\days)$, with an energy threshold of 1.8\kev, it will rule out the explanations of DAMA/LIBRA in terms of DM scattering off sodium and/or iodine \cite{cosinusjcap}. After successful R\&D, the construction of the experiment has started; 
first DM results from $100\kgram\times\days$ could be obtained in 2023. 

The {\bf PICOLON} project (Pure Inorganic Crystal Observatory for LOw-energy Neutr(al)ino) is working in Kamioka also in the development of highly radio-pure NaI(Tl) scintillators after several re-crystallisation processes \cite{picolon2021}. In the longer term, PICOLON plans to install hundreds of kg of NaI(Tl) inside the KamLAND liquid scintillator detector in $\sim$2030.

In summary, important results (even if still with low significance) have been recently released by NaI(Tl) experiments (COSINE-100 and ANAIS-112), whose aim is to solve the long-standing conundrum of the DAMA/LIBRA annual modulation signal. Other NaI(Tl) projects, with interesting features, are ongoing, too.

\paragraph{Ionisation detectors}
\label{sec:iondet}

To probe low-mass DM candidates there are some specific requirements: the use of lighter targets (to keep the kinematic matching with the DM particle), to lower the energy threshold (to detect smaller signals) and even to change the search channel (see Sect.~\ref{sec:ddbasics}). Different ideas have been proposed  using semiconductor devices and noble gas detectors.

Silicon-based sensors offer a high sensitivity to single-electron signals and very low ionization energy. {\bf DAMIC} (DArk Matter In CCDs) is using Si charge-coupled devices where the charge produced in the interaction drifts towards the pixel gates, until readout. In this way, 3D position reconstruction and effective particle identification for background rejection are possible. At SNOLAB in Canada, 7~CCDs with a total mass of $40\gram$  are operated since 2017, achieving a leakage current of 2~e$^{-}/\mmeter^{2}/\days$ and a threshold of 50\eVee. Precise measurements of the quenching factor in Si have been made. Limits for low-mass DM including interaction not only with nucleons but also with electrons and hidden photon DM have been presented \cite{damic0,damic1,damic2}. Results from an exposure of $11\kgram\times\days$ have been released, showing an excess of ionization events above the analysis threshold requiring further investigation; updated limits on SI WIMP-nucleon cross sections have been derived~\cite{damic3}. For DAMIC-M~\cite{damicm}, to be operated at the Modane underground laboratory in France, more massive CCDs ($13.5\gram$ each) will be used, based on the Skipper readout~\cite{skipper}: the multiple, non-destructive measurement of the pixel charge allows to reduce noise and achieve single electron counting with high resolution, as already proved with readout noise equivalent to 0.07~e$^{-}$. The low readout noise and low leakage current will allow DAMIC-M to observe physics processes with collision energies as low as $1\ev$. The {\bf DAMIC-M} experiment will consist of an array of 50 CCDs with more than 36 million pixels in each CCD. The goal is to reach a reduction of the background level in the region of interest of a factor $\sim$50 in comparison to the set-up at SNOLAB (down to $\sim\!\! 0.1\events/\kev/\kgram/\days$), mitigating for instance surface backgrounds from $^{210}$Pb and controlling cosmogenic $^{3}$H. DAMIC-M is now in preparation; a proof-of-concept prototype will be installed soon and commissioning of the final apparatus could start in 2023. Assuming an exposure of $1\kgram\times\years$, DAMIC-M could reach unprecedented sensitivity in the low-mass DM searches, including the GeV-scale WIMPs through nucleon spin-independent scattering and DM particles interacting with electrons with a mass from $1\mev/c^{2}$ to $1\gev/c^{2}$ \cite{damicm}, and also explore a broad range of hidden-sector DM candidates.

The innovative Skipper readout~\cite{sensei} is also used by {\bf SENSEI} (Sub-Electron-Noise Skipper CCD Experimental Instrument), working with a new generation of CCDs. Results from operating CCD detectors of $0.0947\gram$ and $2\gram$ at the shallow MINOS cavern of Fermilab in the USA yield world-leading constraints on DM-electron scattering for mass below $1\mev/c^{2}$~\cite{sensei,sensei2}. The collaboration plans to install a $100\gram$ detector at SNOLAB. 

{\bf CDEX} (China Dark matter EXperiment) is using Point-Contact Ge detectors, allowing to reach sub-keV thresholds thanks to a very small capacitance in combination with a rather large detector mass. This was also the approach of the CoGENT detector at Soudan in USA~\cite{cogent}. Operating at the Jinping underground laboratory in China, CDEX-1 used two detectors of $\sim\!\!1\kgram$ each reaching an energy threshold of $160\keVee$. Limits from an annual modulation analysis \cite{cdexmod} as well as on SI nucleus scattering with sub-GeV WIMPs were derived~\cite{Liu:2019kzq}. In CDEX-10, a $10\kgram$ detector array immersed in liquid N$_{2}$ is being operated and constraints on both  WIMP-nucleon SI and SD couplings have been presented \cite{Jiang:2018pic}. Work is underway for future phases with significantly larger masses, CDEX-100 and CDEX-1T, searching for DM and $^{76}$Ge neutrinoless double beta decay process as well \cite{cdexfuture}.

{\bf NEWS-G} (New Experiments With Spheres-Gas) uses a spherical proportional counter, able to achieve 
single-electron thresholds thanks to a very low capacitance ($<$1~pF) for a large volume~\cite{giomataris}. 
First results were obtained with the SEDINE detector, consisting of a copper sphere, $60\cmeter$ in diameter, filled with Ne-CH$_{4}$ at 3.1~bar ($310\gram$ active mass) 
operating at Modane. 
Exclusion limits for sub-GeV WIMPs were derived from a $42\days$ run
\cite{newsg1}. A new, larger copper sphere ($140\cmeter$ in diameter) to be filled with gases with low atomic masses has been built; great effort has been devoted to mitigate the contribution from $^{210}$Pb in bulk copper from the sphere by electroplating a layer of ultra-pure copper onto the inner detector surface~\cite{Balogh:2020nmo}. 
After a first installation at Modane it was moved to SNOLAB and is currently under commissioning. A significant improvement in sensitivity to low-mass WIMPs from 0.05~to $10\gev/c^{2}$ is expected
thanks to background reduction, better sensor performance and improved
analysis methods~\cite{giroux}. The design of ECUME, a $140\cmeter$ in diameter fully electroformed underground spherical proportional counter, is ongoing and construction will begin at SNOLAB in Fall 2021. For the future, the construction of DarkSPHERE, a $3\meter$-diameter detector, is being investigated, with a projected sensitivity reaching the neutrino floor in the sub-GeV mass range.

{\bf TREX-DM} (TPCs for Rare Event eXperiments-Dark Matter) is based on a gas TPC holding a pressurised gas at 10~bar inside a copper vessel, equipped with the largest microbulk Micromegas readouts ever built \cite{trexdmigor,trexdmbkg}. It operates at the Canfranc Underground Laboratory, being presently in the commissioning phase. Runs with mixtures of Neon with isobutane (Ne+2\% iC$_{4}$H$_{10}$) and, if possible, with underground Argon (Ar+1\% C$_{4}$H$_{10}$) are foreseen. The prospects for the energy threshold range from 0.4\keVee\ down to 0.1\keVee. Even at a prototype scale (with hundreds of grams of gas), competitive sensitivity at the level of $10^{-38}\cmeter^{2}$ could be reached in the direct detection of low mass WIMPs below $1\gev/c^{2}$ if a background at the level of $0.1\events/\kev/\kgram/\days$ can be achieved \cite{trexdmbkg}.

\paragraph{Bubble chambers}
\label{sec::bubcham}
As presented in Sect.~\ref{sec:overview_exp}, bubble chambers offer some very interesting features for WIMP detection. The {\bf PICO} experiment was formed by a merger of the PICASSO and COUPP collaborations in 2012. It uses this technology and has developed a series of bubble chambers operated at SNOLAB. In PICO-60, operating a $52\kgram$ of C$_3$F$_8$ target, a $2.45\keVnr$ threshold was achieved and the best SD WIMP-proton limit from direct detection has been derived~\cite{Amole:2019fdf} (see also Sect.~\ref{sec:wimplimits}). Some changes in the design, like the buffer-free concept, have been implemented in PICO-40L, already starting the data taking. PICO-500 is a fully funded tonne-scale chamber which is now in design phase; it could reach sensitivity to $10^{-42}\cmeter^2$ for the proton interaction cross section for WIMPs with a mass of tens of $\gev/c^2$.

\paragraph{R\&D programs}
To face the technology challenges posed in the context of these DM detection approaches, several R\&D programs are underway involving  international collaboration:

{\bf Development of ultra-pure NaI(Tl) crystals}: special detectors with radiopurity levels (for $^{40}$K or $^{210}$Pb, for instance) much lower than those offered by commercial low-background NaI(Tl) scintillators have been or are being developed in the framework of different experiments in collaboration with different companies. Potassium contents even below that of DAMA/LIBRA crystals, i.e., lower than $\sim\!\!20\ppb$, seems to be now in reach~\cite{zhu19}. Cosmogenic activation of NaI produced during the manufacturing and transport of detectors has been identified as one limiting background~\cite{cosinebkg,anaisbkg}; a proposal to implement underground crystal growth and detector development to avoid cosmogenics has been put forward  by several European groups.

{\bf Determination of the Relative Efficiency Factor (quenching factor) for Na and I recoils}: different measurements of this factor, related to the detection mechanism, are available for NaI(Tl) crystals (see for instance~\cite{joo2019} and the references therein), but there are important discrepancies among them which are not understood. As this significantly affects the interpretation of the DAMA/LIBRA result and the comparison between DAMA/LIBRA and other NaI(Tl) experiments~\cite{ko2019}, a more precise characterisation of quenching factors in the relevant energy ranges is being addressed by different groups considering the possible dependence on different factors like impurities, the content of~Tl as dopant or other crystal properties.

{\bf Development of NaI scintillating bolometers}: these are intended to exploit the capability of discriminating nuclear and electronic recoils on event-by-event basis, in contrast to conventional NaI(Tl) scintillators, profiting from experience on cryogenic detectors. The use of pure (undoped) NaI is being considered following the intrinsic scintillation process at cryogenic temperature. Tests performed within the COSINUS experiment are very promising.

{\bf Development of sensors and readout schemes for ionization detectors}: CCDs have been used for years in digital cameras and in astronomical telescopes for digital imaging of astrophysical objects but unconventional, thicker CCDs have been designed and proved in the search for DM. The novel concept of Skipper-CCDs is being developed involving European and American institutions. Being sensitive to extremely small energy transfers, this will open new windows for the exploration of DM candidates giving very weak signals. Also for gas chambers an active work is ongoing for sensors and readout techniques to improve energy threshold or pattern signal~\cite{Katsioulas:2018pyh,Giomataris:2020rna,giroux}.

{\bf Bubble chamber development}: the PICO collaboration is improving the design of their chambers, implementing for instance the buffer-free design (with the chamber constructed ``right-side-up'' and no water inside the inner vessel) to avoid background events induced by surface tension~\cite{picobufferfree}. Moreover, scintillating bubble chambers, combining the advantages of a bubble chamber with the event-by-event resolution of the liquid noble gas scintillators xenon \cite{scintbubcham} or argon are being developed. The SBC (Scintillating Bubble Chamber) collaboration plans to realise such a detector at SNOLAB.

\subsubsection{Directional Detectors}
\label{sec:directionaldetectors}

Directional detectors, which measure the directions of nuclear recoils as well as their energies, in  principle offer a powerful way of confirming the Galactic origin of a WIMP signal~\cite{Spergel:1987kx}. As mentioned in Sect.~\ref{sec:ddbasics}, due to the Sun's motion with respect to the Galactic rest frame, the directional recoil rate peaks around the opposite of the direction of Solar motion, while nuclear recoils from neutron backgrounds will not have this property.

An ideal directional detector could discriminate a WIMP signal from isotropic backgrounds with only of order $10\events$. In practice, the number of events and exposure, required depends strongly on the capabilities of the detector. Aspects such as whether recoil tracks are reconstructed in 1D, 2D or 3D, angular resolution, energy threshold, whether or not there is an asymmetry in the tracks that allows their sense ($+ \vec{r}$ versus $-\vec{r}$). 
For a review see~\cite{Mayet:2016zxu}.

Directional experiments can in principle probe cross sections below the neutrino floor with smaller exposures than conventional direct detection experiments due to differences in the angular distributions of WIMP and neutrino induced recoils~\cite{Grothaus:2014hja,OHare:2015utx}. Recoils from solar neutrinos peak in the direction opposite to the position of the Sun, which varies over the year and is separated from the direction of Solar motion by between $60^{\circ}$ and $120^{\circ}$~\cite{Mayet:2016zxu}, while the directional fluxes from atmospheric and diffuse supernova neutrinos are approximately isotropic. 
If a WIMP signal is detected, directional experiments could reconstruct the WIMP velocity distribution and do `WIMP astronomy'~\cite{Mayet:2016zxu}. They could also probe the WIMP particle physics more effectively than non-directional experiments, in particular inelastic DM and non-relativistic operators~\cite{Mayet:2016zxu}.

There are two approaches being pursued for directional detectors (see Sect.~\ref{sec:overview_exp}): the use of nuclear emulsions and the operation of low pressure ($\sim$0.1~atm) gas targets in TPCs with different electron amplification devices and track readouts, like Multi-Wire Proportional Chambers (MWPC), Micro Pattern Gaseous Detectors (MPGDs) and optical readouts~\cite{battat}. Many of the gas mixtures used contain $^{19}$F, which provides sensitivity to SD WIMP-nucleon interactions. The reconstruction of tracks is not easy, as they are very short for keV~scale nuclear recoils: $\sim\!\!1\mmeter$ in gas, $\sim\!\!0.1\mumeter$ in solids. In order to improve sensitivity, it is also desirable to register direction (axis, sense), or at least a head-tail asymmetry, by measuring the relative energy loss along the track.

There are several directional detection projects world-wide. However, due to the technological challenges, none of them have yet reached sensitivities comparable to "conventional" detectors.  Prototypes of medium-size (with volumes from 0.1 to $1\meter^{3}$) have already been built and significant progress on basic requirements (like radiopurity, homogeneity, stability and scalability) is being made.

{\bf DRIFT} (Directional Recoil Identification From Tracks) was the pioneer of directional detectors, using MWPCs attached to a TPC with a large conversion volume ($1\meter^{3}$, corresponding to a target mass of 140~g) filled with electronegative gas; in this way, the formed ions (not electrons) are drifted to the readout, to reduce diffusion and optimise track resolution. Electronic recoil background can be rejected to high levels based on their longer range and lower ionization density but alpha background is still problematic. It operated at Boulby, UK, over more than a decade, using a CS$_{2}$+CF$_{4}$+O$_{2}$ mixture. Directional nuclear recoils (from $^{252}$Cf neutrons) quantifying the head-tail asymmetry parameter have been measured~\cite{battat2} and the best limits for SD WIMP-proton interaction from directional detectors ($\sigma < 2.8 \times 10^{-37}\cmeter^2$ at $m_{\chi} \approx 100 \gev/c^2$) were derived from 54.7~live-days~\cite{battat3}. 

{\bf MIMAC} (MIcro-tpc MAtrix of Chambers) also operates a dual TPC with a common cathode, but equipped with pixelized bulk Micromegas (micromesh gas structures), at the Modane Underground Laboratory in France since 2012. MIMAC works with CHF$_{3}$+CF$_{4}$+C$_{4}$H$_{10}$ and 3D tracks of radon progeny nuclear recoils have been registered \cite{riffard}.
A competitive low threshold of 2\keVee\ has been achieved in prototypes, lower than typical thresholds in other directional detectors. First observation of $^{19}$F ion tracks at ion beam facilities with angular resolution at 10-20$^{o}$ has been reported\cite{Tao:2019wfh} 
and quenching factors of He and F with an ion source in Grenoble have been measured. A $1\meter^{3}$ detector is in preparation and could be installed in Modane in the future.

{\bf NEWAGE} (NEw generation WIMP search with an Advanced Gaseous tracker Experiment) uses a simplified system with amplification structure and readout in a monolithic detector with a TPC and a micro-pixel chamber. After first operation in surface, long runs at Kamioka in Japan have been made using CF$_{4}$. The head/tail effect above $100\kev$ has also been confirmed. After the release of first results for SD proton interaction \cite{nakamura}, a low background detector is running since 2018 and very first new limits have been presented from 108 days and an exposure of $1.1\kgram\times\days$ \cite{newage2021}.  

{\bf DMTPC} (Dark Matter Time-Projection Chamber) is based on a TPC equipped with external optical (CCD, PMTs) and charge readouts. Several prototypes have been developed since 2007, operated first at MIT and then underground at WIPP in the USA, working now also for the $1\meter^{3}$ scale (corresponding to $\sim\!\!150\gram$ at 30~Torr). First limits on SD WIMP-proton cross section were obtained from a 10-litre detector \cite{dmtpcresults}. The measurement of the direction of recoils has been reported and the sensitivity to directionality was estimated for the first time~\cite{deaconu}.  

An emulsion film made of silver halide crystals dispersed in a polymer can act as target and tracking detector. Nuclear recoils produce nm-sized silver clusters and 3D tracks are reconstructed with an optical microscope. This is the approach followed by {\bf NEWSdm} (Nuclear Emulsions for WIMP Search with directional measurement)~\cite{newsdm}. New generation nuclear emulsions with nanometric grains (NIT (Nano Imaging Tracker) emulsions) have been developed and new fully automated scanning systems overcoming diffraction limits are being prepared; a spatial resolution of 10~nm has been achieved. Tests at LNGS with a $10\gram$ target are underway to assess backgrounds. A Physics run with $10\kgram\times\years$ using a detector placed on an equatorial telescope (to absorb Earth rotation) to keep orientation towards the Cygnus constellation has been proposed.

\paragraph{R\&D program}
The {\bf CYGNUS} proto-collaboration has been formed, evolving from the workshop series of the same name. It gathers most of the groups working on directional DM detection in the world  
and is carrying out R\&D to determine the optimum configuration for a large target mass directional detector \cite{cygnus}. The objectives include reducing the energy threshold below $1\keVee$; analysing jointly different gas mixtures with varying densities; enlarging the volumes up to 10 to $1000\meter^{3}$ (corresponding to tens of kg target mass, depending on gas and pressure); and considering TPCs with different electron amplification devices and track readouts, with both optical (PMTs, CCDs) and charge readouts (MWPCs, MPGDs).  Conceived as a modular and multi-site observatory, there are proposals for CYGNUS detectors in labs in Australia, Italy, Japan, the UK and the USA. Expectations for SD WIMP-proton interaction sensitivity are very promising,
for instance, with a $1000\meter^{3}$ detector  of He:SF$_6$ and taking data for 6~years, cross sections for SD interaction at the level of $10^{-43}\cmeter^2$ could be reached for $m_{\chi} \sim\!\! (10-100) \gev/c^{2}$ \cite{cygnus}. CYGNO, working at LNGS, has already operated some prototypes with He/CF$_{4}$ using GEMs, CMOS cameras and PMTs~\cite{cygno} en route to building a detector of $1\meter^{3}$, and nuclear recoils from a neutron gun with measurable direction and sense have been registered in the LEMOn prototype.

\subsection{Future Prospects of WIMP Dark Matter Searches }\label{sec:exptcomp}

\begin{table}
\begin{center}
\small

\begin{tabular}[t]{| m{2cm}| m{1.25cm}| m{1.5cm}| m{0.85cm}| m{0.5cm}| m{2.5cm} | m{1.3cm}| m{1.3cm}| m{0.85cm}|}
\hline\hline
Experiment & Lab & Target & Mass [\kgram] & Ch & Sensitivity [$\cmeter^2$@$\gev/c^2$] & Exposure [t$\times\years$] & Timescale & Ref. \\ 
\hline\hline
\multicolumn{9}{|l|}{ {\bf Cryogenic bolometers}~~(Section~\ref{sec:bolometers}) } \\ \hline
EDELWEISS-subGeV & LSM & Ge & 20 & SI & 10$^{-43}$~@~2 & 0.14 & in prep.  & \cite{Arnaud:2017usi}\\  
 SuperCDMS & SNOLAB & Ge,\,Si & 24 & SI & $4\times 10^{-44}$~@~2 & 0.11 &constr. & \cite{Agnese:2016cpb}  \\
CRESST-III & LNGS & CaWO$_4$,\dots 
& {2.5} & SI & $6\times 10^{-43}$~@~1  & $3\times 10^{-3}$ & running &  \cite{Abdelhameed:2019hmk} \\
\hline \hline
\multicolumn{9}{|l|}{ {\bf LXe detectors}~~(Section~ \ref{sec:lxe}) }\\ \hline
LZ & SURF & LXe$^2$ & $7.0\tonne$ & SI & $1.5\times 10^{-48}$~@~40 & 15.3 & comm. &  \cite{Akerib:2019fml} \\
PandaX-4T & CJPL & LXe$^2$ & $4.0\tonne$ & SI & $6\times10^{-48}$~@~40 & 5.6 &constr. & \cite{Zhang:2018xdp} \\
XENONnT & LNGS & LXe$^2$ & $5.9\tonne$ & SI & $1.4\times 10^{-48}$~@~50 & 20 &comm. & \cite{Aprile:2020vtw} \\
DARWIN & LNGS$^*$ & LXe$^2$ & $40\tonne$ & SI & $2\times10^{-49}$~@~40 & 200  & 
$\sim$2026 &  \cite{Aalbers:2016jon}\\
\hline\hline
\multicolumn{9}{|l|}{ {\bf LAr detectors}~~(Section~ \ref{sec:lar}) } \\ \hline
DarkSide-50 & LNGS & LAr$^2$ & 46.4 & SI & $1\times10^{-44}$~@~100& 0.05 & running & \cite{Agnes:2018ep}  \\
 DEAP-3600 & SNOLAB & LAr$^1$ & $3.6\tonne$ &SI & $1\times10^{-46}$~@~100& 3 &running & \cite{Ajaj:2019jk} \\
 DarkSide-20k & LNGS & LAr$^2$ & $40\tonne$ &SI & $2\times 10^{-48}$~@~100& 200 &2023 & \cite{DS_ESPP} \\
 ARGO & SNOLAB & LAr$^\dagger$ & $400\tonne$ &SI & $3\times10^{-49}$~@~100 & 3000 &TBD & \cite{DS_ESPP}  \\
\hline\hline
\multicolumn{9}{|l|}{ {\bf NaI(Tl) scintillators}~~(Section~ \ref{sec::nai}) } \\ \hline
DAMA/LIBRA & LNGS & NaI & 250 & AM & & 2.46 & running & \cite{Bernabei:2018yyw} \\ 
COSINE-100 & Y2L & NaI & 106 & AM & $3\times10^{-42}$ @ 30 & 0.212 &running & \cite{cosineperformance} \\ 
ANAIS-112 & LSC & NaI & 112 & AM & $1.6\times10^{-42}$ @ 40 & 0.560 &running & \cite{anaissensitivity} \\ 
SABRE & LNGS & NaI & 50 & AM & $2\times10^{-42}$ @ 40 & 0.150  &in prep. & \cite{sabre} \\ 
\mbox{COSINUS-1$\pi$} & LNGS & NaI & $\sim$1 & 
SI & $1\times10^{-43}$ @ 40 & $3\times10^{-4}$ &2022 &  \cite{cosinusjcap} \\ 
\hline\hline
\multicolumn{9}{|l|}{ {\bf Ionisation detectors}~~(Section~\ref{sec:iondet}) } \\ \hline
DAMIC & SNOLAB & Si & 0.04 & SI & $2\times10^{-41}$ @ 3-10 & $4\times10^{-5}$ &running & \cite{damictaup2019}\\ 
DAMIC-M & LSM & Si & $\sim$0.7 & SI & $3\times10^{-43}$ @ 3 & 0.001 &2023 & \cite{damicm} \\ 
CDEX & CJPL & Ge & 10 & SI & $2\times10^{-43}$ @ 5 & 0.01 &running & \cite{Jiang:2018pic}\\ 
NEWS-G & SNOLAB & 
Ne:CH$_4$ & 
$\sim 1$ & SI & 
$1.8\times10^{-42}$ @ 2 & 
$6\times10^{-4}$ & comm. & \cite{newsg2} \\
TREX-DM & LSC & Ne & 0.16 & SI & $2\times10^{-39}$ @ 0.7 & 0.01  & comm. & \cite{trexdmbkg}  \\ 
\hline\hline
\multicolumn{9}{|l|}{ {\bf Bubble chambers}~~(Section~ \ref{sec::bubcham}) }\\ \hline
PICO-40L & SNOLAB & C$_3$F$_8$ & 59 & SD &  $5\times10^{-42}$ @ 25 & 0.044 & running & \cite{picotaup2019} \\ 
PICO-500 & SNOLAB & C$_3$F$_8$& $1\tonne$ & SD & $\sim\!\!1\times10^{-42}$ @ 50 & & in prep. &  \\ 
\hline\hline
\multicolumn{9}{|l|}{ {\bf Directional detectors}~~(Section~ \ref{sec:directionaldetectors}) } \\ \hline
CYGNUS & Several & He:SF$_6$ & {10$^3$\meter$^3$} & SD & $3\times10^{-43}$ @ 45 & 6 y & R\&D & \cite{cygnus} \\ 
NEWSdm & LNGS & Ag,Br,C,\dots & & SI & $8\times10^{-43}$ @ 200 & 0.1 & R\&D & \cite{newsdm}  \\ 
\hline\hline
\end{tabular}
\caption{Current, upcoming and proposed experiments for the direct detection of WIMPs. Mass is given in kg unless explicitly specified. The experiments' main detection channel (Ch) is abbreviated as: SI (spin independent WIMP-nucleon interactions), SD (spin dependent), AM (annual modulation). The sensitivity is reported for this channel, assuming the quoted exposure. Note that many projects have several detection channels. "comm." denotes experiment under commissioning.~~~ 
$^1$ Single-phase detector. $^2$ Dual-phase detector. 
$^*$No decision yet. A CDR for LNGS is being prepared. 
$^\dagger$Technology not yet selected. 
\label{WIMP_experiments:table} }
\end{center}
\end{table}

\begin{figure}[h!]
    \centering
    \includegraphics[width=1.0\textwidth]{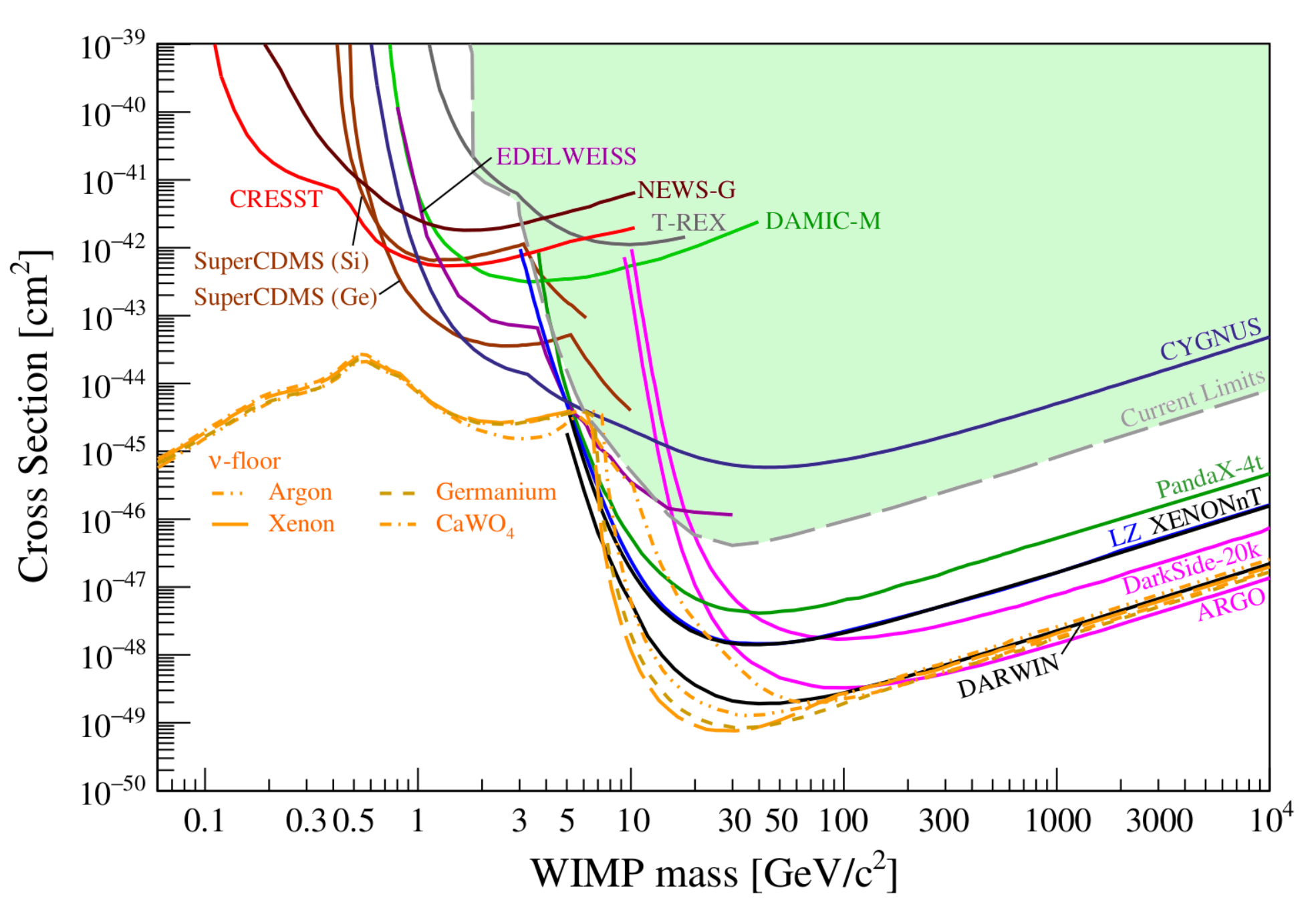}
   \caption{Sensitivity projections (90\% CL) for spin-independent WIMP-nucleon scattering. The neutrino floor is defined as in Fig.~\ref{fig:si_status} and shown for different targets.  
   Shown are projections from ARGO~\cite{DS_ESPP}, CRESST, CYGNUS ($1000\meter^3$)~\cite{cygnus}, DAMIC-M~\cite{damicm}, DarkSide-20k~\cite{DS_ESPP}, DARWIN~\cite{Schumann:2015cpa,Aalbers:2016jon}, EDELWEISS~\cite{Arnaud:2017usi}, LZ~\cite{Akerib:2018lyp}, 
   NEWS-G (ECUME)~\cite{giroux}, 
   PandaX-4t~\cite{Zhang:2018xdp}, SuperCDMS~\cite{Agnese:2016cpb}, T-REX~\cite{trexdmbkg}, XENONnT~\cite{Aprile:2020vtw} along with the envelope of the current results from Fig.~\ref{fig:si_status}. 
    \label{fig::projections}}
\end{figure}

Most of the upcoming and proposed projects presented in the sections above are summarised with some basic information in Table~\ref{WIMP_experiments:table}. Many of the projects are sensitive to a variety of different channels while the table states only the "main" channel. The sensitivity to spin independent WIMP-nucleon interactions of upcoming and proposed projects is shown in Fig.~\ref{fig::projections}; no efforts were made to unify the underlying assumptions.

\begin{figure}[h!]
\begin{center}
\includegraphics[width=0.7\textwidth]{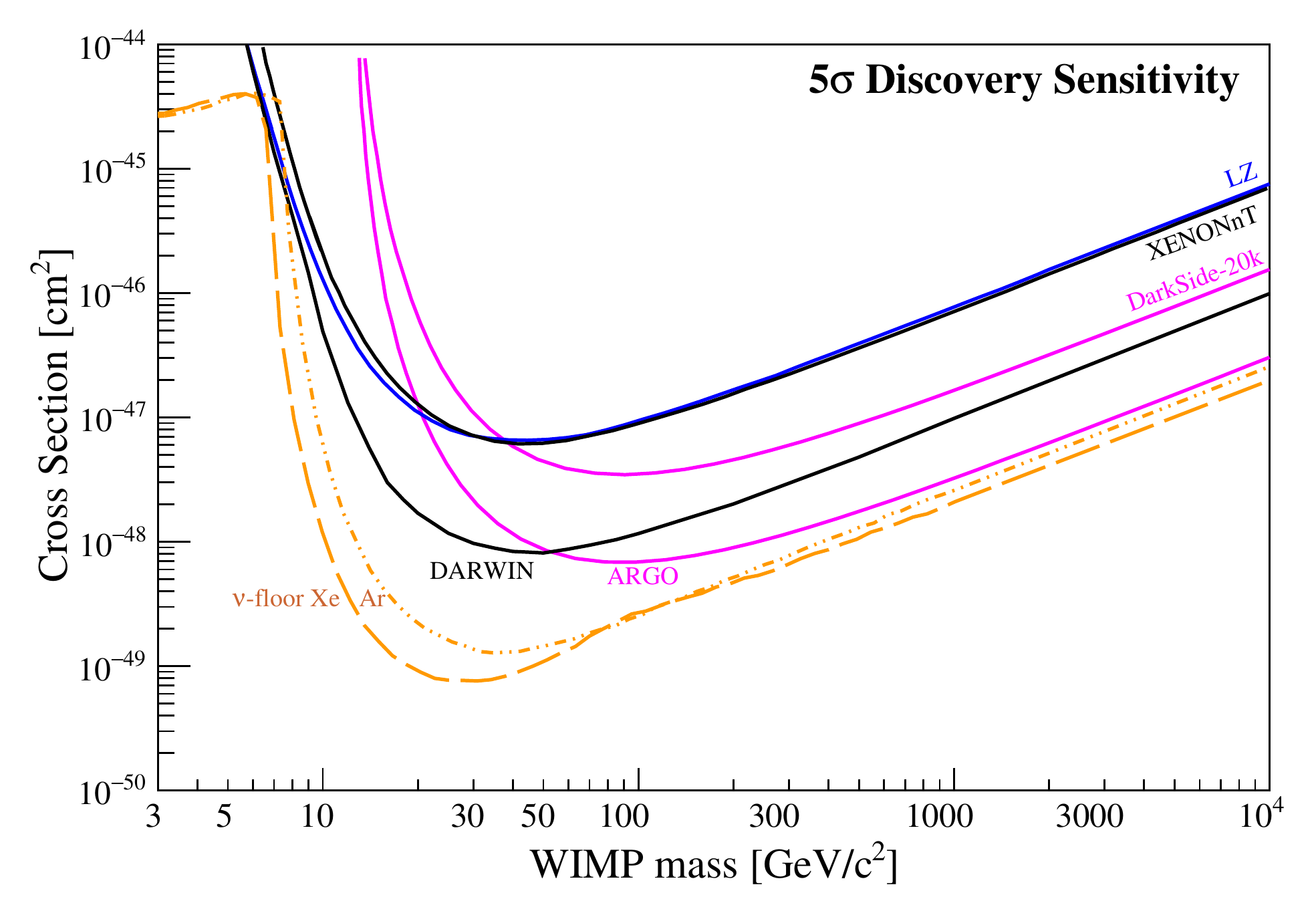}
\caption{ Projected $5\sigma$ discovery potential of some upcoming and proposed liquid noble gas dark matter projects. Shown are projections from ARGO (LAr, $3000\tonne\times\years$ exposure~\cite{DS_ESPP}),  DarkSide-20k (LAr, $200\tonne\times\years$~\cite{DS_ESPP}), DARWIN~(LXe, $200\tonne\times\years$, estimated using~\cite{Schumann:2015cpa}), LZ~(LXe, $15.3\tonne\times\years$~\cite{Akerib:2018lyp}) and XENONnT~(LXe, $20\tonne\times\years$~\cite{Aprile:2015uzo}). }
\label{fig:Physics-NobleDiscoveryComp}
\end{center}
\end{figure}

The low-mass region, from $\sim\!\!100\mev/c^2$ to $\sim\!\!5\gev/c^2$, will be best explored by the cryogenic bolometers (CRESST, SuperCDMS, EDELWEISS) with their extremely low-energy thresholds. Here, despite their reduced SI DM-nucleus cross section, lighter targets (Si, O in CaWO$_4$) are kinematically favoured to probe light DM candidates. Also the CCD-based DAMIC-M and the gas TPC T-REX will be sensitive to new cross section regions in this mass range. The exploration of the medium to high-mass range requires very large exposures and will be dominated by the massive LAr (DarkSide-20k, ARGO) and LXe TPCs (PandaX-4T, XENONnT, LZ, DARWIN).

The discovery potential of DM experiments at their limit of sensitivity is strongly affected by exposure, threshold, uncertainties and the level of background events. Next-generation DM experiments will observe neutrino-induced background events via both $\nu$-$e$ elastic scattering and \CEnNS, generating ER and NR events, respectively. The sensitivity of the largest proposed projects will be limited by these neutrino-induced backgrounds.
The ultimately lower background achievable in argon experiments due to the pulse-shape discrimination of ERs allows a better discovery potential for higher WIMP mass, see Fig.~\ref{fig:Physics-NobleDiscoveryComp}. The discovery potential at lower mass is better in xenon experiments thanks to their much lower experimental energy threshold.
When operated in charge-only mode, the large liquid noble gas TPCs also have a good sensitivity in the low mass region below $\sim\!\!5\gev/c^2$, however, the discovery potential is superior for the dedicated low-mass searches using bolometers and crystals thanks to their lower backgrounds and energy thresholds. 

It is important to emphasise that the whole spectrum of direct WIMP searches with all its complementary approaches, targets and search channels cannot be put into one common figure. Experiments with targets containing $^{19}$F are needed to optimally probe spin-dependent WIMP-proton couplings. Xenon targets ($^{129}$Xe, $^{131}$Xe) are required to test spin-dependent WIMP-neutron couplings with the highest sensitivity, however, there are a number of isotopes which can also provide excellent results in one or/and the other channel (e.g., $^{7}$Li, $^{17}$O, $^{23}$Na, $^{27}$Al, $^{29}$Si, $^{73}$Ge, $^{127}$I, $^{183}$W). The search for signatures of inelastic scattering requires a low background in both, NR and ER (before rejection), channels; an additional excellent energy resolution will allow for an optimal characterisation of the process. Interactions of DM particles in the mass range of ${\cal O}(1-100)\mev/c^2$ are best searched for by detectors with a sensitivity to single electrons, e.g., Si CCDs, Ge bolometers or liquid noble gas TPCs in charge-only mode. Other models introduce different coupling between DM and protons vs.~neutrons to explain the apparent tension between DM claims and limits (e.g., \cite{Feng:2011vu}): in such a "xenophobic" model, parameter space exists where DEAP-3600 has leading exclusion sensitivity~\cite{Yaguna:2019llp}.  In addition there is the long-standing claim of the observation of an annually modulating DM signal in the NaI(Tl) scintillators of DAMA/LIBRA~\cite{Bernabei:2018yyw} which has been investigated and rejected by modulation searches using other targets and detection technologies. However, the ultimate check of the claim can eventually only come from independent projects using the same target material and searching for the same signature.

In general, if a putative DM discovery is made in any experiment, a confirmation using a second target and possibly even a second technology is required, both to rule out potentially mis-identified experimental backgrounds or artefacts, and to start probing the relevant DM particle properties. 
Combining data from different targets can significantly improve the reconstruction of the WIMP mass and cross sections, as well as other WIMP properties such as its spin, self-conjugacy, and coupling structure~\cite{Pato:2010zk,Cerdeno:2013gqa,Peter:2013aha,Catena:2014uqa,Kavanagh:2017hcl}. Moreover, using different targets can even allow for a self-calibration of some astrophysical parameters~\cite{Pato:2010zk,Peter:2013aha}. The comparison of results from different targets/isotopes directly constrains the possible WIMP-matter interaction channel ($A^2$ for SI, target (in)sensitivity for SD, etc.). Ideally, the DM nature of a signal will eventually be confirmed by detecting the directionality of the signal in dedicated detectors, as well as by observations in indirect detection and/or collider searches. 
However, even in optimistic scenarios of detecting a simultaneous signal in both direct and indirect detection experiments, e.g.\ Fermi LAT and/or CTA, a good reconstruction of the WIMP mass, cross sections, and other properties, can typically only be achieved at rather low WIMP mass, below some $200\gev/c^2$, and fairly large values of the cross sections close to current experimental sensitivity~\cite{Roszkowski:2016bhs}.

\subsection{SWOT Tables for WIMP Experiments}\label{sec:wimpswot}

In order to facilitate the comparison between the different experimental approaches, the technology-intrinsic advantages (strengths~{\bf S}) and limitations (weaknesses~{\bf W}) as well as opportunities~{\bf O} and general risks (threats~{\bf T}) of each experimental approach described in Sect.~\ref{sec:european} are summarised in dedicated  {\bf SWOT} tables on the following pages for: cryogenic experiments, 
LXe TPC experiments, LAr detectors, 
NaI(Tl) and ionization detectors and 
directional detectors. 
The different targets and technologies also provide different options for physics beyond WIMP dark matter (see Sect.~\ref{sec:beyonddm}).

\begin{table}
  \begin{tabular}{ |p{7.6cm}|p{7.6cm}|}
    \hline\hline
    \multicolumn{2}{|c|}{\bf SWOT Analysis: Cryogenic Experiments}    \\ \hline \hline
    Strengths & Weaknesses  \\ \hline
\begin{itemize}
            \item unique in reaching simultaneously \eVnr- and \eVee-scale energy resolutions and thresholds with massive target materials (up to $\sim\!\!100$s gram single crystals)
            \item very well controlled nuclear recoil energy scale in pure calorimetric mode, i.e., without Neganov-Trofimov-Luke (NTL) amplification for semiconductors
            \item particle identification based on heat/light or heat/ionisation measurements
           \item low excitation energies avoid quantisation effects of energy information
        \end{itemize}
        &   
        \begin{itemize}
            \item thousands of $\mathcal{O}(100)\gram$ individual detectors required to reach tonne-scale exposure
            \item no or poor self-shielding capacity on an individual detector basis
            \item dispersion of performance among individual detectors in an array
            \item usually slow detectors ($10\musecond$-to-ms scale time response) 
            with respect to other technologies
        \end{itemize}
     \\
    \hline\hline
    Opportunities & Threats \\ 
    \hline
      \begin{itemize}
          \item possibility of using a large number of different target materials providing complementary sensitivities to various DM models
          \item possibility of using target materials with sensitivity to spin-dependent coupling (e.g. $^7$Li, $^{19}$F, $^{23}$Na, $^{127}$I, $^{17}$O, $^{27}$Al, $^{29}$Si, $^{73}$Ge, $^{183}$W)
          \item excellent sensitivity for the exploration of the {\it solar} neutrino floor below $\sim\!\!6\gev/c^2$ with kg-scale exposures thanks to ultra-low-energy threshold 
          \item leading sensitivities of semiconductors to DM-electron interactions thanks to low band gaps (few eV)
          \item NTL amplification in semiconducting materials can provide NR and ER discrimination down to few tens of eV recoiling energies
          \item uniquely well suited in the search for any non-standard DM interaction inducing spectral distortions thanks to their vastly superior energy resolution
          \item well suited to explore various science channels beyond standard WIMP (e.g. axions, ALPs, dark photons, etc.)
      \end{itemize}
           &       
      \begin{itemize}
            \item small exposure of current cryogenic DM experiments 
            \item DM not in the optimal search region of cryogenic experiments, \eg $m_\chi>10\gev/c^2$, where other technologies are much better suited
            \item low-energy excesses (sub--$200\ev$) observed by ongoing experiment yet to be explained that could potentially limit the science reach of the technology
            \item cosmogenic activation, e.g., $^3$H in Ge, has to be properly mitigated
    \end{itemize}
    \\
    \hline\hline
  \end{tabular}
 \caption{SWOT analysis for cryogenic DM experiments.}\label{cryogenic_swot:table}
 \end{table}

\begin{table}
  \begin{tabular}{ |p{7.6cm}|p{7.6cm}|}
    \hline\hline
    \multicolumn{2}{|c|}{\bf SWOT Analysis: Liquid Xenon Experiments}    \\ \hline
        \hline
    Strengths & Weaknesses  \\ \hline
\begin{itemize}
\item massive detectors of moderate dimensions with excellent self-shielding
\item 3D position information of events
\item no long-lived radioactive Xe isotopes; no isotopic depletion necessary
\item $\sim$50\% natural abundance of odd isotopes leads to high sensitivity to spin dependent interactions 
\item well-established purification techniques
\item very low NR and ER backgrounds
\item very low threshold in S1-S2 and S2-only mode (down to single electrons)
\item scintillation wavelength for which photocathodes and transmission windows exist
\item stable operation over years demonstrated
\end{itemize}     
& 
\begin{itemize}
\item only moderate ER rejection (but approximately constant down to threshold)
\item current TPC design limits photon collection
\item high cathode voltage required to establish drift field
\end{itemize}     
\\ \hline\hline
    Opportunities & Threats \\ \hline
\begin{itemize}
\item large community: fruitful competition and chance for coalescence
\item more than $20\tonne$ of Xe gas already in the hand of DM researchers
\item xenon inventory is an investment that can be capitalised after final experiment
\item competitive $0\nu\beta\beta$ search without enrichment possible
\item sensitivity to pp and $^8$B solar neutrinos, atmospheric and supernova neutrinos
\item BSM science, also beyond DM, thanks to low ER background and low threshold
\item after a discovery, isotopic separation of Xe (a proven technology) 
allows separation of odd and even isotopes to study WIMP interactions or removal of isotopes such as $^{136}$Xe if interfering with other science channels
\item potential synergy with $0\nu\beta\beta$ experiments 
\end{itemize}     
&       
\begin{itemize}
\item xenon gas market is finite (production $\sim\!\!70\tonne/\years$); price dictated by bigger players
\item Rn concentration must be reduced by factor $\sim$50 compared to current detectors to reach the ultimate WIMP sensitivity
\item accidental coincidence background may impact final sensitivity
\end{itemize}     
      \\
    \hline\hline
  \end{tabular}
 \caption{SWOT analysis for liquid xenon DM experiments.}\label{lxeswot:table}
\end{table}

\begin{table}
  \begin{tabular}{ |p{7.6cm}|p{7.6cm}|}
    \hline\hline
    \multicolumn{2}{|c|}{\bf SWOT Analysis: Liquid Argon Experiments}    \\ \hline \hline
    Strengths & Weaknesses  \\ \hline
\begin{itemize}
\item both single-phase and two-phase options proven technologies
\item large background-free exposure possible
\item limiting instrumental backgrounds from surfaces become easier to mitigate with increasing detector size using 3D position reconstruction
\item argon easy to purify; already-achieved internal background levels sufficient for all planned future searches
\item allow to span a wide mass range for DM searches: S2-only at low mass, either S1 only or S1-S2 at high mass
\item excellent Pulse Shape Discrimination (PSD) allows suppression of ER events, limiting backgrounds are from coherent neutrino scattering, not from radioactivity or pp ER events
\item very low threshold in S2-only mode

\end{itemize}     
& 
\begin{itemize}
\item{require five times larger target mass than xenon for similar sensitivity at high WIMP mass}
\item{require very large target of underground argon}
\item PSD in S1 and S1-S2 mode implies relatively high thresholds in argon, but this allows complementarity to xenon searches which are primarily sensitive at low energies
\end{itemize}     
\\ \hline\hline
    Opportunities & Threats \\ \hline
\begin{itemize}
\item significant coalescence into single international collaboration allowing for a phased experimental program with progressively increasing sensitivity
\item background-free operation provides excellent discovery potential
\item adding directional detection capabilities to the readout would further improve the discovery potential 
\item complementarity with xenon-based searches allows exploration of model dependence
\item interesting non-DM physics include solar neutrinos and supernova search
\item large inventory of underground argon potential beneficial to other experimental programs
\item synergy with DUNE argon technology
\end{itemize}     
&       
\begin{itemize}
\item require collecting and storing large target masses of underground argon
\item low-mass search requires further isotopic purification of argon 
\end{itemize}     
      \\
    \hline\hline
  \end{tabular}
 \caption{SWOT analysis for liquid argon DM experiments.}\label{larswot:table}
 \end{table}

\begin{table}
  \begin{tabular}{ |p{7.6cm}|p{7.6cm}|}
    \hline\hline
    \multicolumn{2}{|c|}{\bf SWOT Analysis: NaI(Tl) Scintillators, Ionisation Experiments and Bubble Chambers}    \\ \hline \hline 
    Strengths & Weaknesses  \\ \hline
\begin{itemize}
\item NaI(Tl) scintillators can operate in very stable conditions for a long time, accumulating a large target mass (essential requirements to identify a rate modulation as a distinctive signature of DM)
\item some targets containing nuclei with non-zero spin ($^{127}$I, $^{73}$Ge, $^{29}$Si, $^{23}$Na, $^{19}$F, \dots) offer sensitivity to SD interactions
\item ionization detectors have achieved very low-energy thresholds ($\leq 0.1\keVee$) thanks to very low ionization energy and/or low capacitance
\item Si CCDs offer 3D position reconstruction and effective particle identification for background rejection
\item bubble chambers are insensitive to electronic backgrounds
\end{itemize}
&
\begin{itemize}
\item intrinsic background in NaI(Tl) detectors is higher than in other detectors, with absence of fiducialisation or electronic recoil rejection
\item energy thresholds in NaI(Tl) detectors are quite high, presently at 1\keVee
\item accumulation of large target mass is difficult for ionization detectors
\item bubble chambers give no direct measurement of recoil energy
\end{itemize}
\\ \hline\hline
    Opportunities & Threats \\ \hline
\begin{itemize}
\item the ultimate test of the DAMA/LIBRA claim requires using the same target 
\item targets with low mass number are particularly suited to explore low mass WIMPs, which can be accomplished in Si CCDs and in gas detectors with Ar, Ne or He
\item searching for different interaction channels in some ionization detectors allows to explore sub-GeV DM particles interacting with electrons or from the hidden-sector
\end{itemize}
&
\begin{itemize}
\item growth of NaI(Tl) crystals and detector production with required low background is still in development
\item complete development of novel technologies in ionization detectors and related sensors is still underway
\end{itemize}
      \\
    \hline\hline
  \end{tabular}
 \caption{SWOT analysis for NaI(Tl) scintillators,  ionization detectors and bubble chambers DM experiments.}\label{Others_swot:table}
\end{table}

\begin{table}[h!]
  \begin{tabular}{|p{7.6cm}|p{7.6cm}|}
    \hline\hline
    \multicolumn{2}{|c|}{\bf SWOT Analysis: Directional Experiments} \\ \hline \hline
    Strengths & Weaknesses  \\ \hline
\begin{itemize}
\item recoil direction reconstruction offers powerful method for power background discrimination and signal identification


\item in gaseous detectors, many mixtures possible, some with $^{19}$F which gives sensitivity to SD WIMP-nucleon interactions 

\item new generation nuclear emulsions with automated scanning systems provide very good spatial resolution

\end{itemize}
&
\begin{itemize}

\item energy thresholds are typically tens of \keVee

\item achieving large exposures hard, as low density gas is typically used 
or nanometric grains in emulsions

\item  detector readout costly, since large number of electronic channels  required 

\end{itemize}
\\ \hline\hline
    Opportunities & Threats \\ \hline
\begin{itemize}

\item confirm the Galactic origin of a WIMP signal

\item probe cross sections below the neutrino floor with smaller exposures than non-directional experiments 


\item reconstruct WIMP velocity distribution and probe WIMP particle physics 
\end{itemize}
&
\begin{itemize}

\item track reconstruction is very challenging, especially at lowest energies

\item scalability of present prototypes has yet to be demonstrated

\end{itemize}
      \\
    \hline\hline
  \end{tabular}
 \caption{SWOT analysis for directional DM experiments.}\label{Directional_swot:table}
\end{table}




\section{Searches for Axions and ALPs}   
\label{sec:axiondetection}

\subsection{Principles of Detection}\label{sec:axionprinciples}

\begin{figure}
    \centering
    \includegraphics[width=0.9\textwidth]{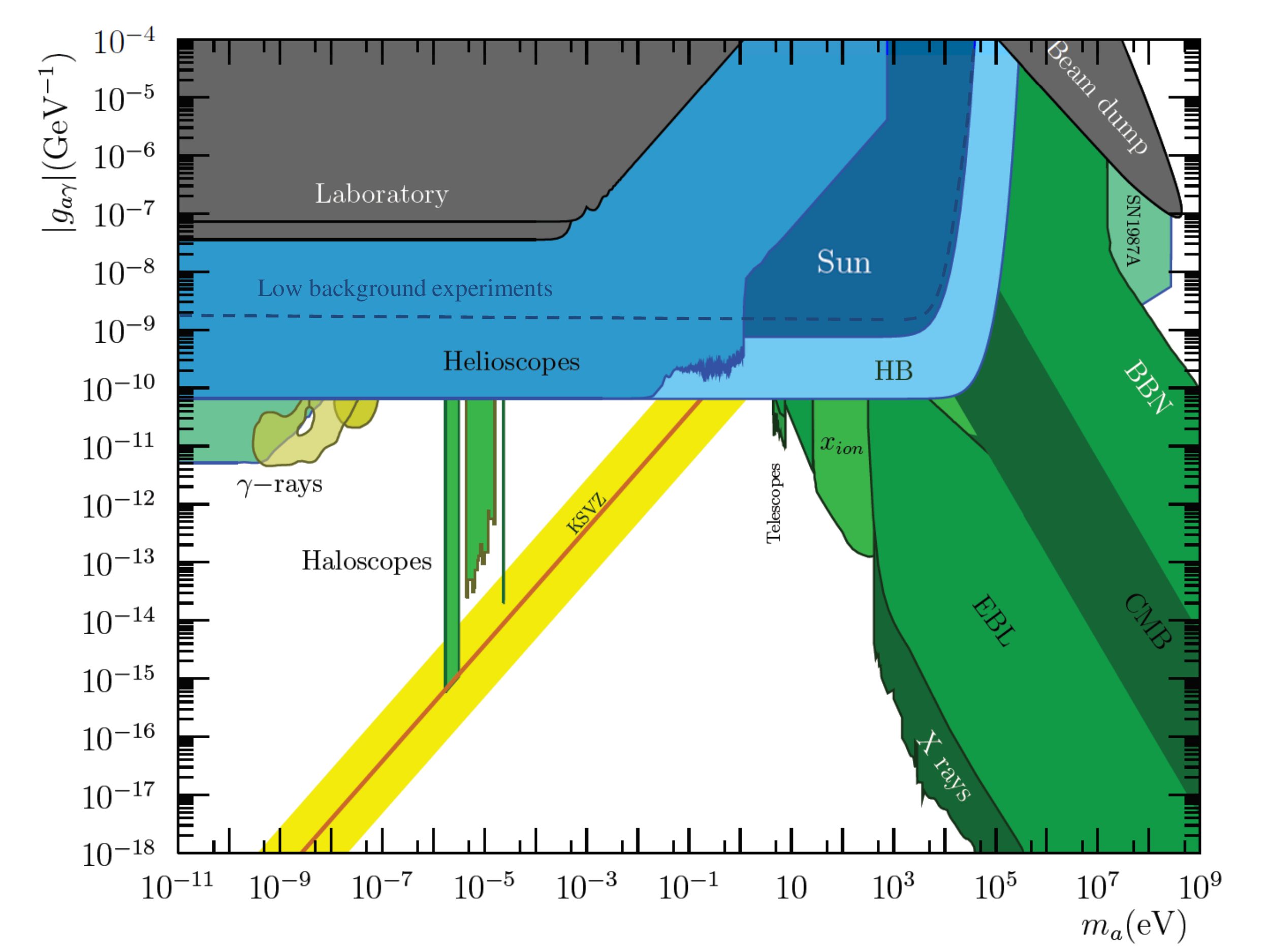}
    \caption{Current exclusion limits in the axion/ALPs to photon coupling constant $g_{a\gamma}$ vs.~mass. Greenish colours denote limits that use cosmological arguments or assumptions. Bluish colours use astrophysical arguments and limits from laboratory experiments are shown with grey colours. The QCD axion models  lie on the yellow band (the KSVZ model is shown explicitly as a red line, the DFSZ model lies at the lower end of the band). The width of the band is determined by theoretical uncertainty of axion models. Note that all of the yellow band as well as most of the white unexplored region is compatible with axions or ALPs as dark matter, respectively. It should also be mentioned that the uncertainty on the exclusion regions from $\gamma$-rays 
    {is} controversial. Figure taken from \cite{Irastorza:2018dyq} which also includes details and references for the individual limits. The dashed blue line for low background limits from Primakoff-Bragg effect has been added.}
    \label{fig:axion_limits}
\end{figure}

The allowed range for ALP and axion mass and ALP couplings spans many orders of magnitudes, resulting in a wide variety of different detection technologies. An overview of the existing limits from different considerations and experiments in the mass vs.~coupling to photon parameter range is given in Fig.~\ref{fig:axion_limits} (adapted from~\cite{Irastorza:2018dyq}). Remarkably, most of the allowed parameter range shown is consistent with some ALP models constituting all of DM~\cite{Arias:2012az}.

The principle of axion and ALP detection of most experiments to cover the very prominent mass range for DM axions between\footnote{In this Section we set $c=1$.} $\sim\!\!1\mueV$, and $\sim\!\! 1\milieV$, is based on the coupling of axions to photons $g_{a\gamma}$.\footnote{Note the adapted explicit distinction between axions and ALPs: Whenever experiments or proposals do not or are not projected to have the sensitivity to probe benchmark QCD axion models we refer to ALPs. If the sensitivity is or could be reached we refer to axions. In the latter case, for a positive detection also ALPs could explain the signal.}
To reach sensitivity to lower axion mass (here defined as axions with mass $\lesssim 1\mueV$) also the coupling to electrons $g_{ae}$ or nuclei $g_{a{\rm N}}$ are used.
The conversion probability is suppressed by $f_a^2$ (which  for axions is proportional to $1/m_a^2$) that for the coupling $g_{a\gamma}$ scales with~$B^2$. Thus experiments exploiting the coupling $g_{a\gamma}$ require a strong static magnetic field~$B$, which is usually provided by an external magnet. For some applications also the electric field inside crystals can be exploited.
Also the axio-electric effect can be used. It is the analogue of the photo-electric effect but with absorption of an ALP or axion instead of a photon and arises due to coupling of ALPs or axions to electrons $g_{{\rm ALP}e}$~\cite{Donnelly:1978ty}.

In general axion and ALP searches can be categorised into three classes, depending on the source of the axions/ALPs searched for:
\begin{itemize}
\item \textbf{Haloscope experiments} utilise the coupling of the DM axion or ALP field with a static magnetic field.
\item \textbf{Helioscope experiments} try to detect axions or ALPs produced in the Sun.
\item In \textbf{lab experiments} ALPs are produced on the site of the experiment.
\end{itemize}

One should remember that only the haloscopes  are direct axion/ALP dark matter detection experiments. However, also the helioscope and lab experiments give important constraints on the couplings of axions and especially ALPs as dark matter candidates. Hence, for completeness, these will also be briefly discussed in the following.

\subsubsection{Haloscopes}

\begin{figure}[b!]
    \centering
    \includegraphics[width=0.45\textwidth]{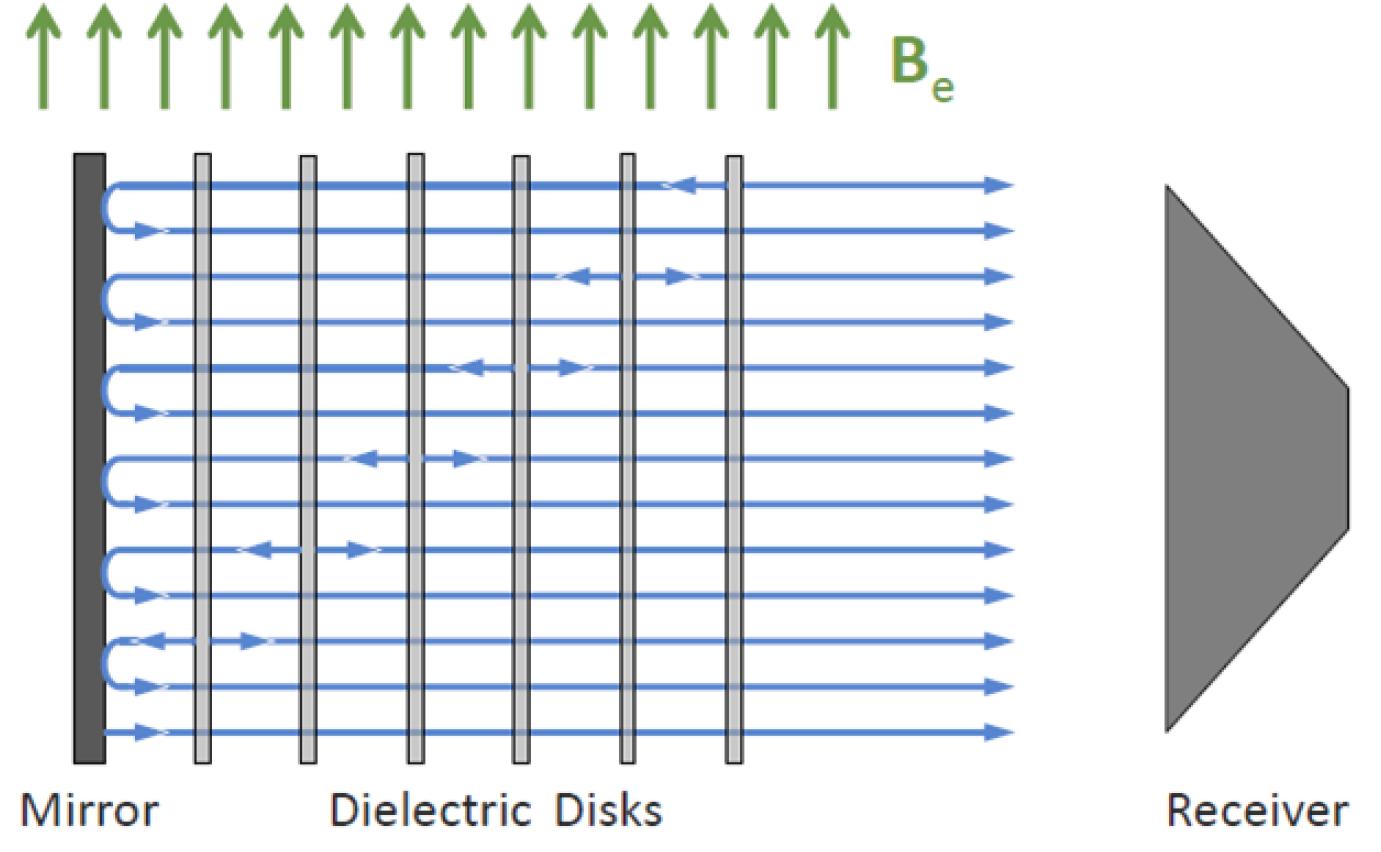}
    \includegraphics[width=0.45\textwidth]{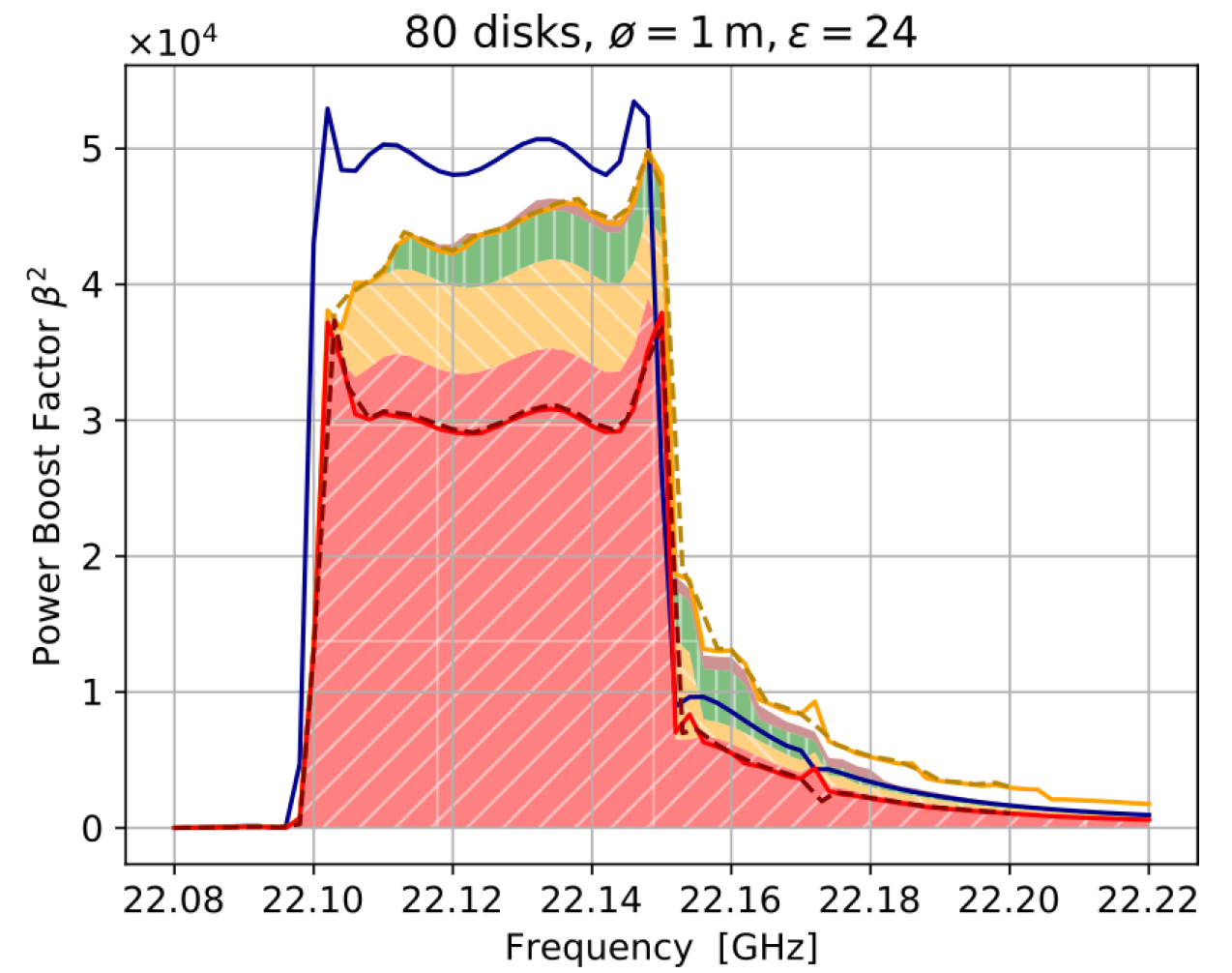}
    \caption{Left: Sketch of the concept of a dielectric haloscope. A number of discs with high dielectric constant is stacked in front of a perfect mirror inside a magnet guaranteeing a $B$-field parallel to the disc surfaces. The signal emitted from this system is collected by an antenna and detected by a receiver (Figure taken from \cite{TheMADMAXWorkingGroup:2016hpc}). Right: Achievable power amplification of a dielectric haloscope with 80 discs and a dielectric constant $\epsilon=24$ (LaAlO$_3$) as a function of frequency with respect to emission from a single mirror. The results a simplified 1D simulation is shown with the blue line. The shaded areas correspond to results of a 3-D simulation of the setup, whereas the different colours denote the contributions to the amplification in different modes of the system. The red dashed line corresponds to the power enhancement of the signal with respect to a single mirror surface after antenna coupling to the first mode only \cite{Beurthey:2020yuq}.}
    \label{fig:dielectric_haloscope}
\end{figure}

Most \textbf{cavity} and \textbf{dielectric haloscope} experiments make use of the fact that the coupling $g_{a\gamma}$ induces an additional source term in Maxwell's equations. Hence, in a static magnetic field any oscillation of the axion field induces a tiny oscillating $E$-field with the oscillating frequency corresponding to the axion mass. This method, originally proposed by P.~Sikivie \cite{PhysRevLett.51.1415, PhysRevLett.52.695.2}, is also known as inverse Primakoff-effect.
Thus, in an external $B$-field the primordial oscillation of the axion field around its minimum induces $E$-field oscillations that could in principle be detected. 

An oscillating $E$-field can be amplified by a {\bf cavity}, if it is in resonance with the respective frequency  \cite{PhysRevLett.51.1415, PhysRevLett.52.695.2}. 
A cavity with high enough Q-factor (of the order of $\gtrsim 10^5$) that can be tuned in frequency can contain high enough power to be detected by quantum 
detectors. Tuning of the cavity resonance frequency can, for example be done using a movable sapphire rod.

Alternatively, the surface between media with different dielectric constants can be used. 
As discontinuities of the $E$-field appearing at the transition between two media are not allowed, 
emission of photons (compensating for the discontinuity) perpendicular to the surface is induced by the oscillation of the axion field    \cite{Horns:2012jf}.
In the {\bf dish antenna} approach a spherical mirror (dielectric constant $\epsilon \rightarrow \infty$) surface focuses all radiation to the 
focal point of the mirror, where it can be detected, if the product of the surface and the static $B$-field perpendicular to the mirror surface are big enough. 
Also, the {\bf dielectric haloscope} concept \cite{Jaeckel:2013eha, TheMADMAXWorkingGroup:2016hpc} utilises the photon emission forced by the discontinuity. In this case a number of transparent discs with high $\epsilon$ and low dielectric loss are stacked in front a plane mirror. Like this, the coherent emission is enhanced by the additional surfaces. Additionally, the discs can be placed in a manner that the coherent emission constructively interferes such to additionally boost the axion to photon conversion probability. By adjusting the distances between the individual discs a considerable boost (of the order of $\gtrsim 10^4$ for 80 discs) of the emitted power can be achieved for a rather broad frequency range (50\MHz). 
The basic principle is sketched in Fig.~\ref{fig:dielectric_haloscope}, left. The power boost of the axion signal for a configuration with 80 discs made from LaAlO$_3$ expected from 1D and 3D simulations is depicted in Fig.~\ref{fig:dielectric_haloscope}, right \cite{Beurthey:2020yuq}.
The system, if disc distances can be tuned to the $10\mumeter$ level,
could in principle cover the mass range between 40\muev and 400\muev  \cite{TheMADMAXWorkingGroup:2016hpc}. Also, it has been recently investigated that, once a DM axion signal has been found, this concept could be used to obtain directional information on the in-falling DM axions, hence could allow to perform DM axion astronomy. \cite{Knirck:2018knd}

The same approach can in principle also be utilised at optical wavelength using stacked lenses with varying disc thicknesses and dielectric constants to access a different mass range between $\sim\!\! 200\miliev$ and $\sim\!\! 5\eV$ \cite{Baryakhtar:2018doz}.
Lately another interesting concept, a \textbf{plasma haloscope} has been proposed: it relies on resonant axion to photon conversion by tuning the effective photon mass in a meta-material induced "plasma" to the axion mass \cite{Lawson:2019brd}. 
This might have the potential to reach a sensitivity to QCD DM axions in a similar mass range as the dielectric haloscope.

Another interesting and complementary concept is based on \textbf{topological insulators}. For these materials axion like quasi-particles are predicted that would, couple to the axion field and lead to resonant conversion to polaritons that are detectable \cite{Marsh:2018dlj}. This method, if suitable materials can be produced, may be sensitive to DM axions in a mass range between $\sim\!\! 0.5\milieV$ and $5\miliev$. 


Also the axion couplings to electrons $g_{ae}$ or nuclei $g_{a{\rm N}}$ are proposed for experimental searches. These couplings can lead to spin interaction, hence to spin precession of electrons or nuclei with frequency corresponding to $m_a$ which can be detected. Furthermore, an oscillating axion field would also induce an oscillation of the electric dipole moment of nuclei, which could be detected using \textbf{NMR techniques}.
In low background experiments the axio-electric effect can be utilised to search for high mass DM  ALPs (see~\ref{subsubsec:low_background}).


\subsubsection{Helioscopes}
Solar axions would be thermally produced with energies corresponding to the temperature in the interior of the Sun, i.e. a few keV. \textbf{Helioscopes} make use of dipole magnets that are pointed towards the Sun. The axions produced in the Sun via different couplings $g_{ae}$, $g_{a\gamma}$ and $g_{aN}$ with energies up to $\sim\!\!10\kev$ can be converted to photons and can then be focused and detected using low background X-ray detectors. Due to axion production in the Sun via different processes, helioscope experiments are sensitive to different couplings, most notably to $g_{ae}$ and $g_{a\gamma}$.
 
Some low background experiments designed to search for WIMP DM or neutrinoless double beta decay are sensitive to ALP interactions via the $g_{ae}$  and $g_{a\gamma}$ couplings to  solar (and DM) ALPs in the mass range $\sim\!\!\eV$ to $\MeV$ via the axio-electric effect \cite{Donnelly:1978ty}. For axions this mass range is strongly disfavoured by limits on $f_a$ by astrophysical arguments \cite{Raffelt:1990yz}.

\begin{figure}[t!]
    \centering
    \includegraphics[width=0.7\textwidth]{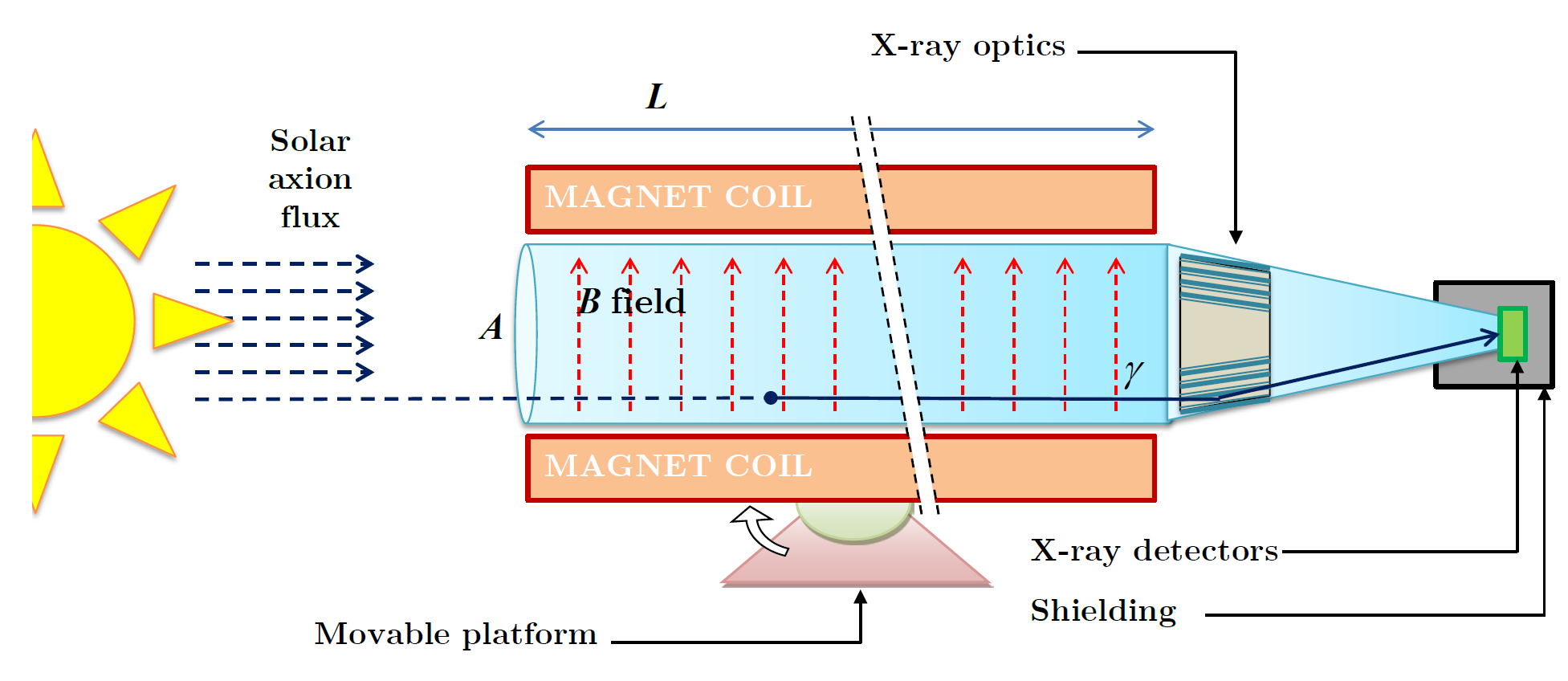}
    \caption{Sketch of concept of the IAXO helioscope: The aperture of a dipole magnet is pointed towards the Sun. Axions from the Sun can be converted into photons (X-rays) inside the $B$-field of the magnet. These can be focused by X-ray optics onto specially developed low background X-ray detectors (taken from \cite{Armengaud:2014gea}). }
    \label{fig:helioscope}
\end{figure}

\subsubsection{Laboratory experiments}
Experiments relying on production of axions in the lab have the advantage that they do not rely on cosmological or astrophysical assumptions and can probe the particle properties model independently.

ALPs converted from laser photons in the laboratory inside a $B$-field can be re-converted to photons behind a wall that blocks the laser light of the initial laser beam. As photons first have to be converted to axions and then back to photons again, the probability for this process scales with $g_{a\gamma}^4$, thus is additionally suppressed. However, by using a strong laser source and by placing optical cavities inside the conversion areas within the magnet, the path lengths, hence the probability for this process to occur can be increased by the product of the finesse of the two cavities \cite{Mueller:2009wt}. Nevertheless, these \textbf{light-shining-through-the-wall (LSW)} experiments yield results that do not rely on astrophysical or cosmological modelling.

An axion field oscillation can be induced by the nuclei of rotating non-circular bodies for the case that there is extra CP violation in axion/ALP-nucleon coupling $g_{a{\rm N}}$. Due its macroscopic de Broglie wavelength, this leads to \textbf{long range forces} that in principle could be detectable from outside the rotating body with the very well known frequency of the rotating body \cite{Crescini:2016lwj, Geraci:2017bmq}. 

\begin{figure}
    \centering
    \includegraphics[width=0.8\textwidth]{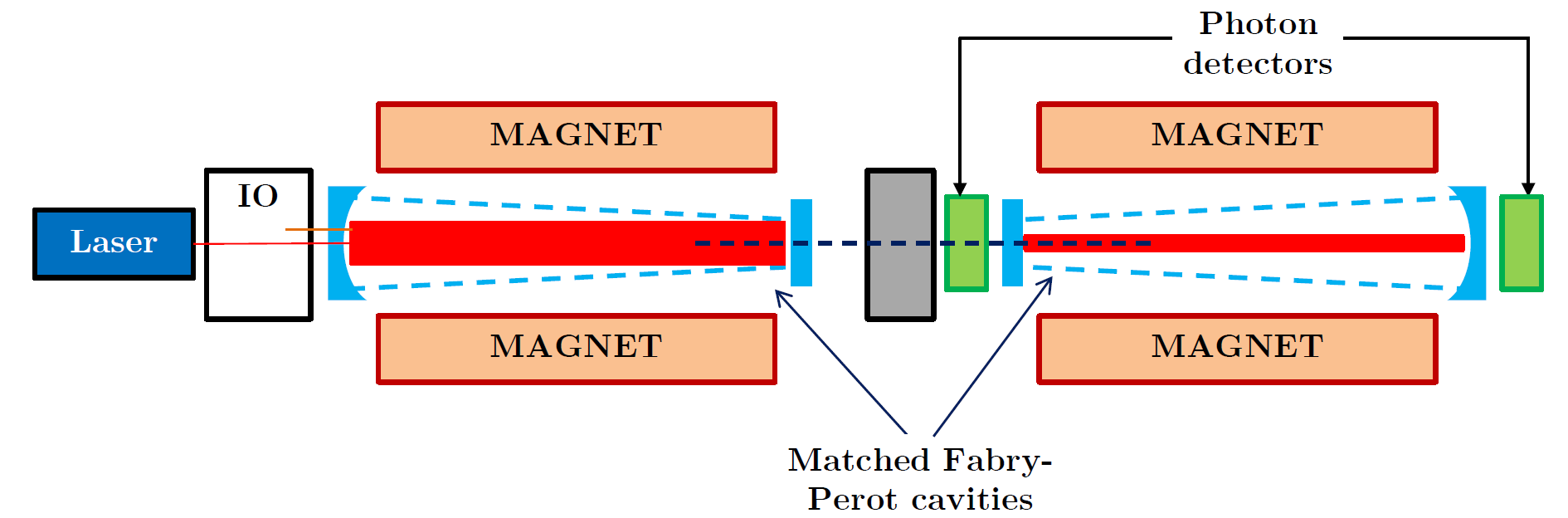}
    \caption{Sketch of the concept of LSW experiments. Laser 
    light is introduced inside a optical cavity that is placed inside a dipole magnet. Photons can be converted into ALPs and propagate unhindered through a wall into a separated optical cavity inside a magnet on the opposite side of the wall. Here, the ALPs can be re-converted to photons and be detected. (Taken from \cite{Irastorza:2018dyq}.) }
    \label{fig:lsw}
\end{figure}


\subsection{Cavity Haloscope Experiments}
\label{sec:cavity}

Presently the cavity-based approach is leading the field of axion searches. The USA-based {\bf ADMX} experiment~\cite{Asztalos:2009yp} located at University of Washington (UW) is presently taking data with a sensitivity that is sufficient for detection of DM axions for the most prominent benchmark models in a mass range around $2\muev - 10\muev$ , assuming that axions make up all DM \cite{PhysRevLett.124.101303, Du:2018uak}.

Presently strong efforts are underway in the USA ({\bf ADMX Gen2} , {\bf HAYSTAC} \cite{Brubaker:2016ktl} located at Yale university), South Korea ({\bf CULTASK} at CAPP \cite{Kwon:2020sav, Lee:2020cfj, Semertzidis:2019gkj}) and Australia ({\bf ORGAN} \cite{McAllister:2018ona, McAllister:2017lkb} at University of Western Australia - UWA) but also in Europe ({\bf QUAX} \cite{Alesini:2019ajt, Alesini:2020vny} at INFN Legnaro) to extend the sensitivity of cavity experiments to mass ranges up to $\sim\!\!70\muev$, maybe even beyond. These are mainly based on the challenging tasks of increasing  magnetic field and volume \cite{Gupta:2019pfm, Battesti:2018bgc}, improving on detection technology  to (sub) quantum limited detection of (quantum squeezed) signals in the 100\MHz \, to 10\GHz \,regime
\cite{Droster:2019fur, Matlashov:2018yid} and by increasing the Q-factor of cavities using superconducting foils \cite{Alesini:2019ajt, Ahn:2020qyq} or using the dielectric cavity approach. Here, low loss dielectric movable cylinders or bars (photonic band-gap cavity) are placed inside a copper cavity, like this increasing the Q-factor by a significant decrease of the losses from the walls of the copper cavity. This technology is being developed by several groups and some very promising results have recently been released by {\bf QUAX} \cite{Alesini:2020vwh}, at CAPP \cite{Kim:2019asb} and by the ORGAN \cite{McAllister:2017ern} groups.
These experiments could reach DM sensitivity to detect KSVZ benchmark model DM axion in a mass range  up to $\sim\!\!70\muev$ or even above, if simultaneously $>14\tesla$ magnets and (sub)-quantum limited receiver (squeezed states) are used.


Single cavity experiments can in principle cover the axion mass range between and $\sim\!\!1\muev$ and maybe $\sim\!\!70\muev$. The sensitivity of these experiments scales with the volume and the Q-factor of the cavity, hence the sensitivity decreases for higher frequencies, as the size of the cavity scales with the radius, quadratically with frequency. Additionally for non-superconducting/dielectric cavities the Q-factor decreases for increasing frequency. Technologies are under development
using multiple mode matched cavities in {\bf RADES} \cite{Melcon:2020xvj, Melcon:2018dba} and at CAPP, South Korea \cite{JEONG2018412}. 

The RADES experiment could utilise the  (baby)IAXO magnet and could explore - so far in a rather narrow bandwidth - both lower and higher mass ranges than conventional cavity experiments. In principle both concepts, the one utilised by  RADES and the one by QUAX, CAPP and ORGAN with sapphire bars could be transformed to confirmation mode experiments once an axion or ALP signal has been found at a well defined frequency.

Also, the QUAX collaboration is exploiting a technology to measure modulations of axion electron coupling induced spin precession due to Earth's movement through the DM halo  \cite{Crescini:2020cvl}.  A first promising R\&D run gives a upper limit on $g_{ae}$ of $\lesssim 2\times 10^{-11}\geV^{-1}$  around $40\mueV$ in a $0.4\mueV$ wide window. 
This concept, sensitive at a rather "high" axion mass range, would be a nice method in case of a positive detection (confirmation mode) if able to reach the sensitivity to probe benchmark axion models: after enough R\&D it could be in shape to quickly tune to the proper frequency and test the axion electron coupling $g_{ae}$.

{\bf KLASH} is another European based proposal, based on the idea to use an existing large aperture superconducting magnet at INFN. The appropriate cavity could be tunable in a frequency range 60\MHz$\,-\,$250\MHz, corresponding to a mass range $0.3\mueV - 1\mueV$ \cite{Alesini:2019nzq, Alesini:2017ifp}) reaching sensitivity to DM axion benchmark models. Presently, the availability of a suitable magnets is being investigated.

\subsection{Dielectric Haloscope and Dish Antenna Experiments}
\label{sec:haloscopes}
The dielectric haloscope concept is presently exploited by the  European based MADMAX collaboration \cite{TheMADMAXWorkingGroup:2016hpc}). 
The main challenges of this endeavour are: the large aperture (1.35\meter\, warm bore) 9\tesla{} dipole magnet, the necessity of (near) quantum noise limited detection in the frequency range 10\GHz$-$\,100\GHz{} and the proper understanding of the calibration of the "booster" as a transducer of axion field oscillations into real photons. 
First proof of principle studies have been finalised \cite{Egge:2020hyo, Brun:2019lyf}. The collaboration is presently planning to build a first down-scaled prototype that could be tested in the existing large aperture MORPURGO dipole magnet at CERN~\cite{MORPURGO1979411}. These prototype measurements could exploit so far uncovered ALP parameter space at a mass range around $m_{\rm ALP}$ around 80\mueV.
The final experiment would be sensitive to DM axions in the mass range $40\mueV- 120\mueV$ in a first phase utilising travelling wave amplification based on JPA technology \cite{Planat:2019}.
In a second phase the experiment could be sensitive to axions with a mass between 120\mueV and 400\mueV. The (sub) quantum limited detection technology for this second phase still needs development, though.
The designated location of the final {\bf MADMAX} experiment is at DESY Hamburg.

The dish antenna concept is being investigated by the European based {\bf SHUKET} at CEA Saclay (CEA)  and {\bf FUNK} project at Karlsruhe institute for Technology (KIT) for hidden (dark) photons without an external $B$-field. First limits on the kinetic mixing of DM hidden photons have been released recently by both experiments, with $\epsilon\lesssim 5\times10^{-12}$ for the mass range between 20\mueV and 30\mueV \cite{PhysRevLett.122.201801} and  $\epsilon\lesssim 10^{-12}$ in the mass range $2\eV - 8\eV$ \cite{Andrianavalomahefa:2020ucg}, respectively. 
The {\bf BRASS}\footnote{https://www.physik.uni-hamburg.de/iexp/gruppe-horns/forschung/brass.html} project is also based on the dish antenna approach, but here the mirror will be magnetised using a Halbach array in order to be sensitive to ALPs. BRASS could, within a decade or so after some more R\&D allow for a very broadband measurement for the mass range between 1\mueV and 10\milieV{} with a sensitivity nearly reaching requirement to detect DM axions.

\subsection{Experiments for Low Mass Dark Matter Axions and ALPs}\label{sec:lowmassaxion}

The USA-based experiments {\bf ABRACADABRA} \cite{Ouellet:2019tlz}, {\bf DM Radio},
\cite{Phipps:2019cqy} 
{\bf ADMX SLIC} and {\bf SHAFT} all  use the fact that due to the axion modified Maxwell equations in the presence of the oscillating axion field via $g_{a\gamma}$ coupling also $B$-field oscillations would be induced along the symmetric axis of a toroidal magnet. Such modulations can in principle be detected by a pick up coil placed in the centre of the magnet.
For a toroidal magnet with few~\tesla{} $B$-field and a volume of many~$\meter^3$. 
This approach in principle has the potential to reach the sensitivity to detect DM axions for the axion mass range 20$\nev\lesssim\,m_a\,\lesssim\,0.8\muev$~\cite{Kahn:2016aff}.
First promising results of proof of principle  setups have already been published~\cite{Ouellet:2018beu, Crisosto:2019fcj, Gramolin:2020ict}. The ABRACADABRA and DM Radio collaborations have recently merged to join forces. 
{Recently, it has also been suggested to use the axion induced $E$-field oscillations to drive power stored in one mode of a superconducting resonant cavity into another specially prepared tunable mode, separated by the axion field-oscillation frequency, where it could then be detected. This heterodyne approach may also  have the potential to reach axion dark matter sensitivity in the mass range between $\sim 1\mueV$ and 10\neV\ \cite{Berlin:2019ahk}.}

The partly European-based Cosmic Axion Spin Precession Experiment {\bf CASPEr} collaboration runs a collection of experiments based on  NMR technologies at the Helmholtz Institute Mainz (HIM) and  at Boston University  that, at least partly, may ultimately  have the possibility to reach sensitivity to detect low mass DM axions with $m_a \lesssim 1\nev$~\cite{Budker:2013hfa}.
Large regions of unexplored parameter space down to a mass of $m_a \sim\!\! 10^{-22}\eV$ for ALPs are being explored and excluded by exploiting the search for DM ALP wind induced frequency shift in nuclear spins \cite{Wu:2019exd, Garcon:2019inh}, or ALP induced oscillations of nuclear electric dipole moment \cite{JacksonKimball:2017elr}. Also, experiments to search for nEDM, like the search at Paul Scherrer Institut at Villigen in Switzerland (PSI) are sensitive to DM ALPs in a mass range $10^{-22}\eV - 10^{-17}\eV$~\cite{PhysRevX.7.041034}.
Note that these techniques are sensitive to different kind of axion couplings to nuclear spin, hence are complementary to most other searches in terms of the coupling constant.

The Global Network of Optical Magnetometers to search for Exotic
physics, {\bf GNOME} collaboration is looking for transition of Earth through topological defects or so called axion stars, for which the encounter rate in some models could be sufficiently high. Such an encounter  could lead to  atomic spins experiencing an oscillating energy shift that may be observable \cite{JacksonKimball:2017qgk}.   

\subsection{Helioscope Experiments}
\label{sec:helioscopes}

The {\bf CAST} experiment has made use of a prototype LHC dipole magnet at CERN. It was mounted onto a auto-tracking structure allowing to point the bore axis of the magnet towards the Sun~\cite{Kuster:2007ue}.
The experiment in its final stage produced the most stringent bound on the ALP-photon coupling $g_{{\rm ALP}\gamma} < 0.66 \times 10^{-10}\gev^{-1}$ for ALPs with m$_{\rm ALP}\lesssim 10\miliev$, and, in particular, probed axions with mass from $\sim\!\!0.5\ev$ to $1.0\ev$~\cite{Anastassopoulos:2017ftl}  or the $g_{a\gamma}$  coupling in some axion benchmark models.

The {\bf IAXO} collaboration aims to use the same concept, but with improved detector technology as well as with longer and stronger larger aperture magnets to achieve the sensitivity for detection of solar axions with a mass between 10\,meV and 1\,eV~\cite{Armengaud:2014gea}. Note that this axion mass range is not or only slightly in conflict with the astrophysical bounds~\cite{Raffelt:1990yz}.
As a first step the {\bf babyIAXO} experiment will be built using a down-scaled magnet based on existing technology and will serve as a test bed for all components of the  IAXO detection technology. Already babyIAXO would surpass the sensitivity of CAST by a factor of 4 in coupling constant and could be sensitive to some benchmark axion models for a mass above 10\,meV~\cite{Armengaud:2019uso}. It will already allow scanning the ALPs parameter range compatible with astrophysical hints (see Sect.~\ref{sec:theory}).  Also it is worthwhile to note that IAXO will be the first experiment with sensitivity to $g_{ae}$ values that are not yet excluded by astrophysical arguments.
While the signal searched for in IAXO
is independent of the mass for $m_a \lesssim 0.02\eV$, some information about the axion mass can be obtained due to the presence of axion-photon oscillations in the higher mass range \cite{Dafni:2018tvj}. This could be relevant to point towards a mass range to detect axion or ALP DM. The designated site for  babyIAXO is DESY Hamburg.

\subsection{Light-Shining-Through-the-Wall Experiments}
\label{sec:alps}

\textbf{Light-shining-through-the-wall (LSW)} initiatives are laboratory experiments. They use strong magnetic fields to convert photons from a laser (or other sources) into axions.  These axions can pass unhindered through a wall, where they can be back-converted to photons inside a second magnet and eventually be detected. As here the production of ALPs happens under controlled conditions, LSW laboratory experiments do not suffer from cosmological or astrophysical uncertainties. 

Two LSW experiments, {\bf ALPS} at DESY~\cite{EHRET2010149} and {\bf OSQAR} at CERN~\cite{Ballou:2015cka} using decommissioned
accelerator dipole magnets have so far have reached sensitivities to exclude ALPs with a mass lower than $\sim\!\!\milieV$ and
$g_{a\gamma}\lesssim 5.8\times{10}^{-8}\gev^{-1}$. 

Presently at DESY the {\bf ALPS II} LSW experiment is under 
construction, aiming for a first data run in 2021 \cite{Bahre:2013ywa}. This experiment utilises 24 modified HERA magnets in combination with
large baseline optical cavities similar to gravitational wave interferometers. The experiment will reach a sensitivity that will allow to detect ALPs should they be the reason for the transparency hint or the stellar evolution anomalies, exceeding the sensitivity of the progenitor LSW experiments by more than three orders of magnitude in  $g_{a\gamma}$.
{\bf JURA} is the tentative follow up project based on the ALPS II optics technology using future dipole magnets under development at CERN. The ALPs II sensitivity could potentially be improved by roughly an order of magnitude for $m_{\rm ALP} \lesssim 0.1\milieV$. Recently, also RF-cavity LSW experiments have been proposed that  may exceed the sensitivity of ALPS\,II \cite{Ferretti:2016aut, Janish:2019dpr}.

Another alternative laboratory experiments is to search for \textbf{long range forces} induced by the ALP or axion field. Potentially, these can be exploited if additional CP violation occurs in the standard model. The USA-led {\bf ARIADNE} project is focusing on this idea for the search of axions in the meV range \cite{Geraci:2017bmq}.

\subsection{Vacuum Polarisation and Bi-refringence Experiments}

As the presence of an external magnetic field could induce photon to ALP conversion, consequently the equation of motions of the photon field are altered. Explicitly, this could lead to a measurable ALP induced bi-refringence of the vacuum. This method has been pioneered by the {\bf PVLAS} experiment \cite{DellaValle:2015xxa}, setting limits on ALPs that are, however, weaker than the ones from LSW experiments.
Similarly, it has recently been proposed to use light by light scattering in a cavity with high power lasers in the presence of a non uniform magnetic field. This could lead to resonant ALP/axion production due to momentum exchange with the inhomogeneous $B$-field, thus changing the polarisation behaviour of photons. Inside an 8\,\tesla{} $B$-field using a cavity pumped with squeezed light this concept may have the sensitivity to axions in the mass range between $\sim\!\!10^{-3}\eV$ -- $1\eV$~\cite{Zarei:2019sva}. As for LSW experiments, these approaches are laboratory experiments and do, hence, not suffer from cosmological or astrophysical uncertainties.

Also the presence of an oscillating DM axion or ALP field would alter the properties of the QED vacuum due to non vanishing axion photon mixing without the need of an external $B$-field.
For circularly polarised photons this would lead to slightly different phase velocities depending on the polarisation direction. Using an optical cavity the phase shift can be enhanced accordingly. The {\bf DANCE} project at Tokyo University uses a optical ring cavity \cite{Obata:2018vvr}, while also Fabry-Perot cavities from gravitational wave detectors have been proposed \cite{Nagano:2019rbw}. The sensitivity of these projects is presently not enough to reach the axion band but exceeds the existing limits at a mass $\lesssim 10^{-10}\ev$.  

A similar approach using loop oscillators can also be used for the microwave  regime as recently demonstrated by the {\bf Upload/Download} project at University of Western Australia. Here, axion-field induced instabilities of the up- and/or down-converted phase noise are exploited. This may give the possibility of a broad band measurement in the \nev{} and tens of \mueV{} mass regions, potentially with sensitivity to QCD axion models \cite{Thomson:2019aht}. 

\subsection{Low-Background Experiments} \label{subsubsec:low_background}
Experiments designed to search for very rare events like WIMP scattering off nuclei or neutrinoless double beta decay typically have extremely low backgrounds  and energy threshold. The background data of these detectors can be analysed for possible ALP or hidden photon signals as a very complementary by-product of these searches. Different  detection processes can be used to search for ALPs or hidden photons from different sources.

The signature of the axio-electric effect~\cite{Derevianko:2010kz} would be an energy deposit in the detector corresponding to the energy transferred to the electron given by the rest mass of the axion plus the kinetic energy of the incident ALP minus the electrons binding energy~\cite{Arisaka:2012pb}.

The axio-electric effect can be used for searches with low background detectors for solar ALPs (solar ALPs energies up to $\sim\!\!10\kev$). Here, the detection threshold needs to be low enough for one to be able to detect the ~keV electron that absorbed the thermally produced ALP.  Limits can then be obtained for different couplings corresponding to the different ALP production mechanisms in the Sun for the mass range below the energy threshold of the detector.
Limits on solar ALPs have been published as by products of WIMP searches like 
XENON100~\cite{PhysRevD.90.062009}, PandaX~\cite{PhysRevLett.119.181806}, LUX~\cite{Akerib:2017uem}, XENON1T~\cite{Aprile:2019xxb} and CDEX \cite{PhysRevD.101.052003}. 
Recently also competitive results from a cryogenic bolometer have been released~\cite{Abdelhameed:2020hys}. Presently, the most stringent limits give $g_{{\rm ALP}e} \lesssim 3\times10^{-12}\gev^{-1}$ in the mass range below 10\kev.  Note that astrophysical limits are roughly an order of magnitude stronger. Very recently the XENON1T collaboration has reported a 3.5\,$\sigma$ Poissonian upward fluctuation in their low-energy electron recoil spectrum which  can be interpreted as possibly due to an excess caused by solar axions~\cite{Aprile:2020tmw}. This, however, is in strong tension with astrophysical arguments~\cite{DiLuzio:2020jjp}.

Solar ALPs can also be searched for using their coherent Bragg conversion into photons in the electric field of crystalline detectors. The conversion rates inside the crystals would depend on their orientation towards the Sun and, hence, have a distinct time modulation that can be searched for. The  DAMA \cite{Bernabei:2001ny}, CDMS~\cite{Ahmed:2009ht} and EDELWEISS~\cite{Armengaud:2018cuy, Armengaud:2013rta} experiments have analysed their data for this modulation and set limits on  $g_{{\rm ALP}\gamma} < 2\times10^{-9}\geV^{-1}$. 

The axio-electric effect can also be used to look for cold DM ALPs  with a mass in the range  between $10\kev$ and $\sim\!\! 500\kev$, that is a parameter range not 
excluded theoretically and experimentally.
Assuming all DM being made up by these kind of ALPs, low background experiments can
set limits on $g_{{\rm ALP}e} \lesssim 10^{-13}\gev^{-1}$ in this mass range.  
Irrespective of mass ranges excluded by astrophysical arguments,  there
 are a number of results on ALPs in the mass range between 
 $\sim\!\!1\ev$ and $500\kev$. The most sensitive ones are presently coming from EDELWEISS \cite{Arnaud:2020svb}, PandaX-II \cite{PhysRevLett.119.181806}, LUX \cite{Akerib:2017uem}, XENON100 \cite{Aprile:2014eoa} and SuperCDMS \cite{Amaral:2020ryn}, depending on the mass range considered. For  $m_{\rm ALP} \lesssim 10\keV$ limits from astrophysical arguments are stricter by about an order of magnitude or more, depending on the mass range.

Also hidden (dark) photons as DM can be searched for by utilising their kinetic mixing~$\epsilon$ with photons. The signature would be a peak at the mass corresponding to the DM particle.  SENSEI \cite{sensei2}, DAMIC \cite{damic2}, CDEX~\cite{PhysRevLett.124.111301},  SuperCDMS \cite{Aralis:2019nfa}, EDELWEISS~\cite{Arnaud:2020svb},  XENON1T\cite{Aprile:2019xxb},  XMASS \cite{Abe:2018owy}, Majorana \cite{Abgrall:2016tnn} and {\bf GERDA} \cite{Agostini:2020xta} have published results in the energy range between a few \eV\ and $\sim\!\!1 \mev$.
These limits could be improved by roughly one order of magnitude with the next generation DARWIN WIMP search experiment for the mass range above $\sim\!\!\kev$~\cite{Aalbers:2016jon}.

\subsection{Current Status, Limits and Projections
}\label{sec:axionstatus}

\begin{figure}
    \centering
    \includegraphics[width=\textwidth]{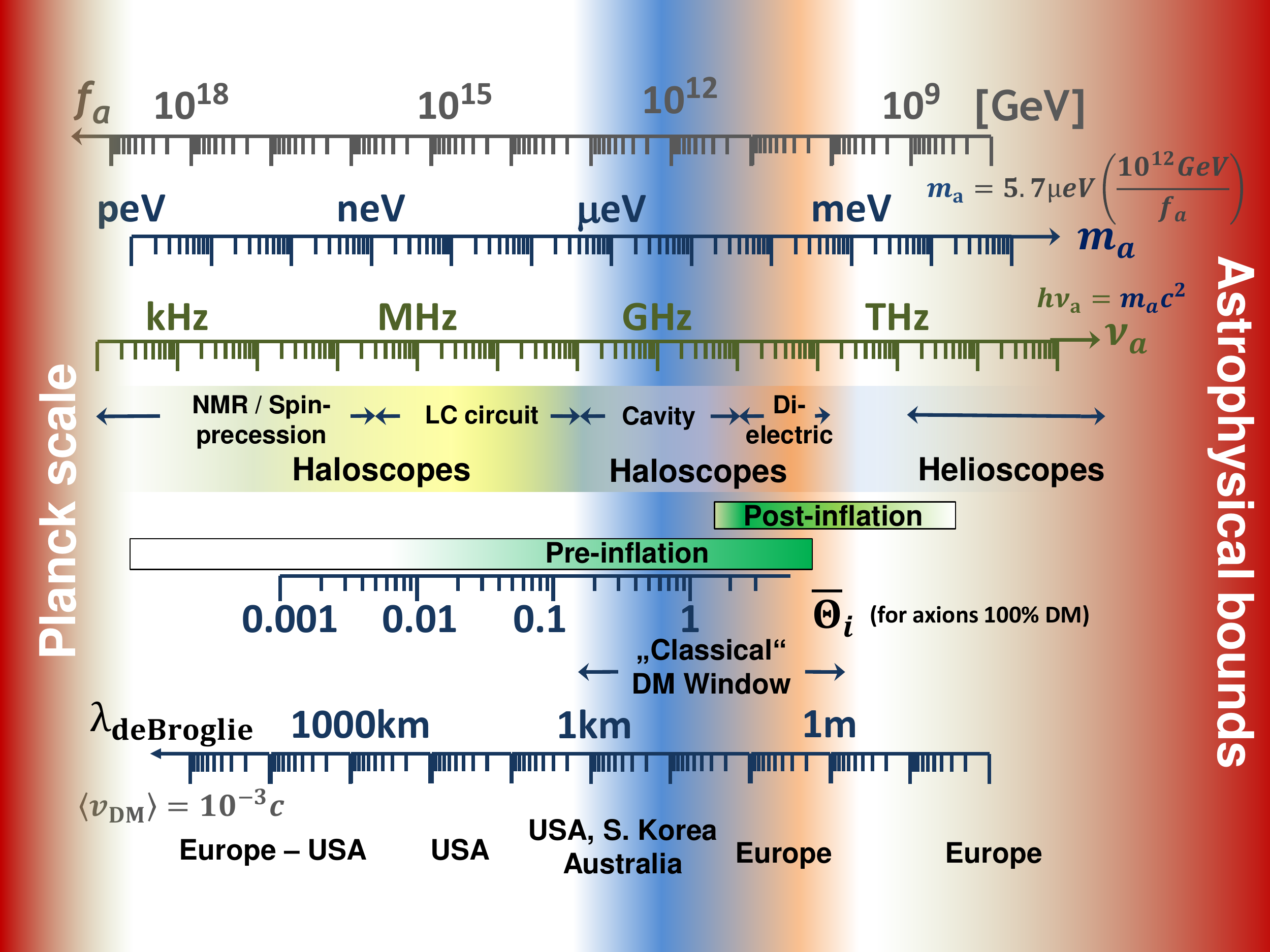}
    \caption{Mass ranges over which different technologies could be sensitive to DM and solar axions. The different QCD axion properties $f_{a}$, $m_{a}$, $\nu_{a}$ are related to each other. For the pre-inflationary PQ-symmetry breaking scenario the corresponding initial $\theta_i$ value is displayed for which axions would explain all DM. Also, the corresponding de Broglie wavelength is shown for the assumption of a the mean axion velocity $10^{-3} c$. }
    \label{fig:axion_mass_range}
\end{figure}

In Fig.~\ref{fig:axion_mass_range} the different experimental concepts with prospect to have sensitivity to the QCD axion line are displayed and are put into context. It is clearly seen that the different concepts perfectly complement each other to close the axion mass range with the prospect for the European community to take over a leading role in covering the mass ranges below 1\nev\ and the ones compatible with post inflationary PQ symmetry breaking above $\sim\!\!25\muev$.

\begin{figure}
    \centering
    \includegraphics[width=1.0\textwidth]{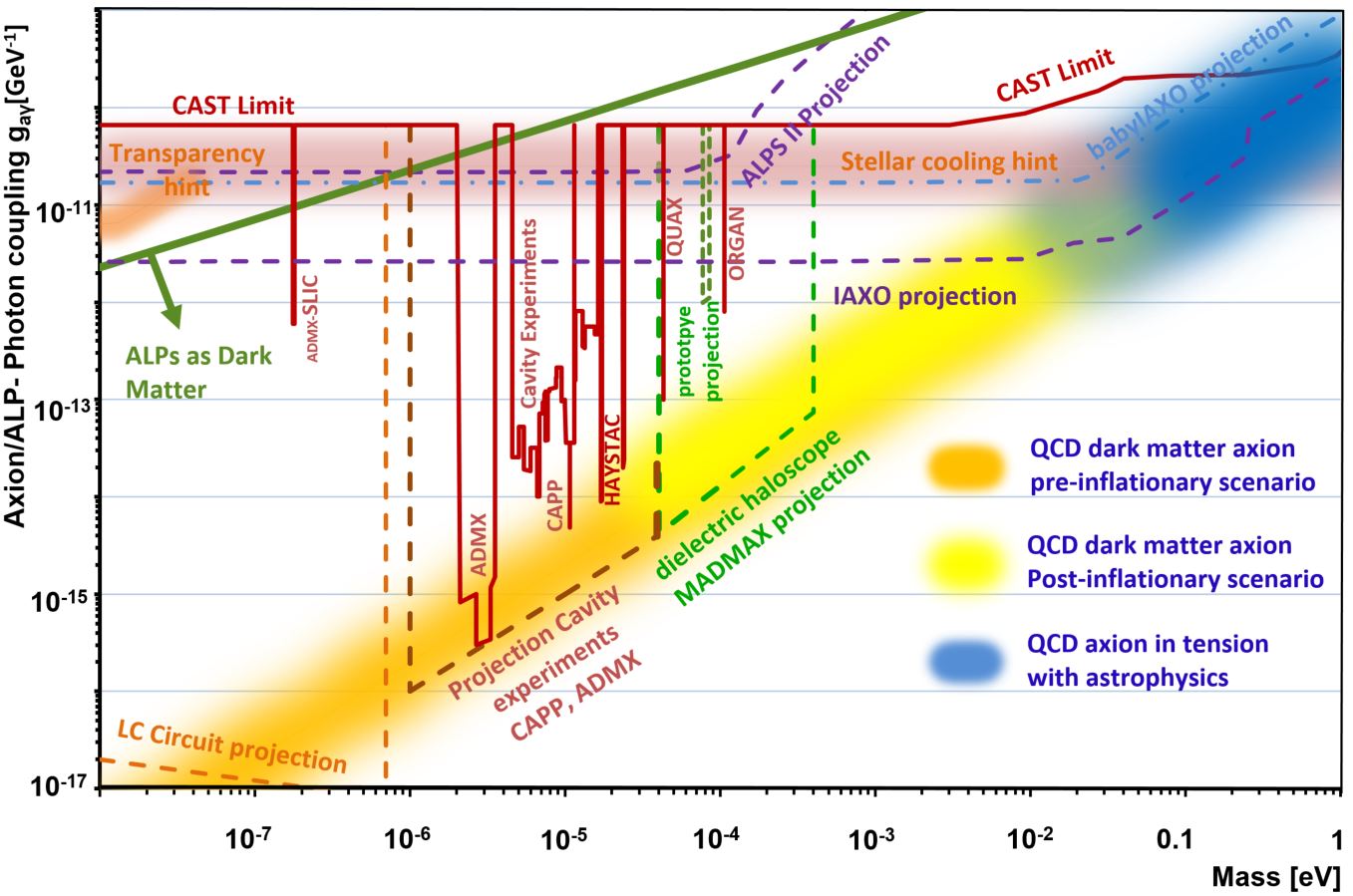}
    \caption{Current limits (red line) and projections (dashed lines) of searches for axions and ALPs via the $g_{a\gamma}$ coupling. Also shown are the different areas in the QCD axion parameter range corresponding to pre- (orange) and post-inflationary (yellow) Peccei Quinn breaking scenario. Note that the two scenarios partly overlap. Shown as shaded orange and violet regions are the ranges consistent with ALPs explaining the transparency and cooling hints, respectively. For clarity, only the highest projected sensitivities are shown.}
    \label{fig:axions_status}
\end{figure}

Fig.~\ref{fig:axions_status} shows the present limits and prospect sensitivities to the coupling constant~$g_{a\gamma}$ as a function of the mass for experiments and proposals sensitive in the "classical" DM axion mass range. 
Earlier cavity experiments have scanned a sizeable part of the open ALPs parameter range~\cite{PhysRevD.40.3153, PhysRevD.42.1297}. However, up to now
ADMX is the only experiment sensitive to the benchmark DM axion models. The next generation cavity experiments like ADMX Gen2, together with other initiatives like CULTASK at CAPP or HAYSTAC could cover the mass range between~1\mueV and 40\mueV within the next decade. 

There are quite some developments to reach the sensitivity to cover the remaining fraction of the QCD axion mass range.
Within the last few years many new approaches for DM axion detection in different mass ranges have been proposed. 
Some of these have lead to collaborations that are on a good way to achieve a sensitivity within a decade to being able to detect DM and/or solar axions in the mass range between $\sim\!\!40\mueV$ and $400\mueV$ and between $\sim\!\! 5\miliev$ and $100\miliev$, like the European based MADMAX or (baby)IAXO collaborations, respectively. For the intermediate mass range (500\mueV\, to few \miliev) some ideas, like plasma haloscopes or topological insulator exist, for which, however, still some R\&D efforts are needed before being able to evaluate their feasibility. 
The mass range between $\sim\!\!1\neV$ and $1\mueV$ could be exploited by the LC-circuit technique.
For axions with a mass below $\sim\!\! 1\neV$, NMR techniques as exploited by the CASPER experiments might be a viable option to reach the required sensitivity but require significant investment.

By covering the DM axion mass regime, the ALPs parameter regions above the axion band in the coupling vs. mass parameter range will automatically be covered, hence exploiting a large fraction of feasible DM ALP candidates. 
Already before reaching this sensitivity  the ALPS II experiment at DESY will cover a parameter range that is consistent with ALPs that could explain the transparency hint and the stellar cooling anomaly (see Sect.\ref{sec:axionbroader}).

\begin{table}[]
\small
    \begin{tabular}{| m{2.0cm}| m{0.5cm}| m{1.1cm}| m{0.75cm}| m{2.85cm}| m{1.25cm}| m{0.8cm}| m{1.4cm}|m{1.35cm}|}    
\hline\hline
    Experiment & Type & Techn.& $g\_$  & Mass range  & Status & Limits &  Location & Timescale\\
\hline\hline
\multicolumn{9}{| l |}{\bf Experiments with expected sensitivity to DM axion benchmark models}\\
\hline
{\bf CASPEr-e}$^{\bf a}$  &$\varnothing$ & NMR & $_{a{\rm N}}$ & $10^{-13}$\eV$-$\,1\neV &   R\&D & {\small ALP}& BU &\\
DM Radio$^{\bf b}$ & $\varnothing$ & LC & $_{a\gamma}$ & 20\neV$-\,0.8$\mueV &     R\&D &  HP  & Stanford & 2025-30 \\
ADMX$^{\bf c}$& $\varnothing$ & C  & $_{a\gamma}$ & 2\muev$-$\,40\mueV &  running & \textbf{axion}$^{\bf \dagger}$ & UW & 2017-30 \\
HAYSTAC & $\varnothing$ & CS & $_{a\gamma}$ &  15\mueV$-$\,35\mueV & running & {\small axion}$^{\bf \ddagger}$ & Yale & 2015-25 \\
CULTASK & $\varnothing$ & SC/MC & $_{a\gamma}$ & 3\mueV$-$\,70\mueV & running & {\small axion}$^{\bf \star}$ & CAPP & 2021-30 \\
{\bf QUAX}$^{\bf d}$ & $\varnothing$& SC/DC & $_{a\gamma}$ & 30\mueV$-\,50$\mueV& in prep. &  {\small ALP}$^{\bf \ast}$ & INFN & 2021-25 \\
{\bf MADMAX}$^{\bf e}$ & $\varnothing$ & DH & $_{a\gamma}$ & 40\muev$-\,400$\mueV & prototype &  & DESY & 2025-35$^{\bf f}$ \\
ORGAN$^{\bf d}$ & $\varnothing$ & DC/CS  & $_{a\gamma}$ & 60\mueV$-\,210$\mueV &  prototype & {\small ALP} & UWA & 2025-35$^{\bf f}$ \\
{\bf IAXO}$^{\bf g}$ & $\odot$ &XR & $_{a\gamma, ae}$ & 1\miliev$-$\,10\eV & in prep. & & DESY & 2023-35 \\
\hline\hline
\multicolumn{9}{| l |}{\bf ALP experiments} \\
\hline
{\bf CASPEr{\tiny -W}}$^{\bf a}$  & $\varnothing$ & NMR & $_{\rm ALP N}$ & $10^{-22}$\eV$-$\,1\mueV &   running & {\small ALP}& HIM/UCB &\\
{\bf GNOME} & $\varnothing$ &  NMR  & $_{\rm ALP N}$ & $10^{-21}$\eV$-\,10^{-10}$\eV & running & ALP & global &2017-24\\
DANCE & $\varnothing$ & OC &  $_{{\rm ALP}\gamma}$ & $\lesssim 10^{-10}$\eV &  R\&D & {\small ALP} & Tokyo &\\
Up/Download & $\varnothing$ & MO &  $_{{\rm ALP}\gamma}$ & $10^{-10}$\eV$-\,10^{-7}$\eV &  prototype & {\small ALP} & UWA &\\
ABRA$^{b}$ & $\varnothing$ & LC & $_{{\rm ALP}\gamma}$ & 1\neV$-$\mueV &  in prep. & {\small ALP} & MIT &\\
SHAFT & $\varnothing$ & LC &  $_{{\rm ALP}\gamma}$ & $\lesssim\,$10\,neV &  R\&D & {\small ALP} & BU &\\
{ADMX-}SLIC & $\varnothing$ & LC &  $_{{\rm ALP}\gamma}$ & $\lesssim\,$0.2\mueV &  R\&D & {\small ALP} & UFL &\\
{\bf ALPS II} & $\mathcal{L}$ & LSW & $_{\rm ALP\gamma}$ & $\lesssim$\,0.1\milieV & constr. & & DESY &2021 \\ 
{\bf RADES}  & $\varnothing$ & MC & $_{\rm ALP\gamma}$ & $\sim\!\! 30\,-\,50$\mueV & R\&D & & CERN &\\ 
{\bf QUAX} & $\varnothing$ & $e^{-}$S & $_{\rm ALPe}$ & 30\mueV$-$\,80\mueV& R\&D & {\small ALP} & INFN & 2021-25\\
{\bf BRASS} & $\varnothing$ & DA & $_{\rm ALP\gamma}$ & 1\mueV$-$\,1000\mueV & in prep. & & UH &2022-23\\
{\bf IAXO}$^{\bf g}$ &  $\odot$ &XR & $_{\rm ALP\gamma}$ & $\lesssim$\,1\,eV & in prep. & & DESY &2025-35\\
\hline\hline
\multicolumn{9}{| l |}{\bf Hidden photon experiments (no axion or ALP coupling)}\\
\hline
{\bf SHUKET} & $\varnothing$ &  DA & $\epsilon$ & 20\mueV$-$\,30\mueV &  in prep. & HP & CEA &2024\\ 
{\bf FUNK} & $\varnothing$ &  DA  & $\epsilon$ & 2\eV$-$\,8\eV & upgrade & HP & KIT&\\
\hline\hline
    \end{tabular}
    \caption{Dedicated DM axion, ALP and hidden photon searches. The type of experiment ($\varnothing$: haloscope, $\odot$: helioscope, $\mathcal{L}$: laboratory experiment), the technology used (NMR: nuclear resonance methods, LC: LC-circuit, C: cavity, SC: superconducting cavity, CS: Cavity using sub quantum limited receiver (squeezed states), MC: Multi cavity, DC: dielectric cavity, DH: dielectric haloscope, XR: X-ray detection, OC: optical cavity, MO: Microwave cavity, $e^{-}$S: Electron Spin cavity, DA: dish antenna, see text in sec \ref{sec:axiondetection} for details), the type of coupling 
    as well as the approximate mass range for maximum sensitivity are given. Also, the current status, type of published limits, current location (for abbreviations see text) and estimated period of measurement times are given. 
    Axion experiments are not listed in the ALP section unless they additionally cover ALP mass range in which they are not sensitive as axion search. For limit sensitivities of ALP and hidden photon searches see text.  Projects with strong European participation are marked in boldface. }
    \label{tab:axion_experiments}
{\footnotesize 
$^{\bf \dagger}${Limits reaching DFSZ sensitivity}

$^{\bf \ddagger}${Limits at 2\,$\times$\,KSVZ sensitivity, running with $1.4\times$ KSVZ sensitivity}

$^{\bf \star}${Limits at 4\,$\times$\,KSVZ sensitivity, running with $1.5\times$ KSVZ sensitivity,  aiming at DFSZ sensitivity within next 2 years}

$^{\bf \ast}${Presently running with $\sim\!\!4\times$ KSVZ sensitivity without scanning}

$^{\bf a}${CASPER electric: to reach axion sensitivity requires significant improvement in hyperpolarisation techniques and availability of sample materials free of paramagnetic impurities. CASPER wind is running with ALPs sensitivity}

$^{\bf b}${DM Radio and ABRACADABRA collaborations merging, toroidal magnet with $1\meter^3$ magnetic volume with $B=5\tesla$  needed}

$^{\bf c}${Will use SC/MC in upcoming runs. UW and/or Fermilab will be the location of future phases}

$^{\bf d}${requires operation in $\gtrsim 14\tesla$ magnet and (sub) quantum limited receiver to reach sensitivity to KSVZ benchmark model}

$^{\bf e}${$9\tesla\, 1.3\meter$ aperture dipole magnet in design phase. For m$_a \gtrsim 120\muev$ development of detection techniques necessary}

$^{\bf f}${starting with prototype measurements with ALP sensitivity running 2021-2024}

$^{\bf g}${First measurement will be done with babyIAXO as an intermediate step towards the full IAXO experiment}

}

\end{table}

A list of dedicated axion, ALP and hidden-photon experiments is given in \ref{tab:axion_experiments}.
Comprehensive and detailed descriptions of the status of axion and ALP theory as well as of search experiments are given in  \cite{Sikivie:2020zpn, Irastorza:2018dyq}.

\subsection{Axion Research: R\&D Efforts and Synergies with Other Physics 
Areas}
\label{sec:axionrecom}

The success of axion searches will depend on some of the key technologies still to be developed. These show much synergy with other areas of fundamental or applied physics. 
Especially CERN, DESY, INFN, MPG and CEA, but also other European and national laboratories play leadership role in many technologies of particle physics. Many of these are essential to the axion community. 
Sharing of this expertise  in terms of magnet development, high quality factor cavities detector knowledge and by their available infrastructure (workshop, beams, etc.) and available resources would be very beneficial.

\subsubsection{High field large aperture magnets}
The experimental efforts that are the most promising to reach the sensitivity for detection of DM axions depend on the availability of an external strong magnetic field. In order to make possible to cover the complete range in parameter space for axions it is therefore essential to support the development of large aperture magnets based on knowledge from particle-physics detector and accelerator projects. This is a task that should be done in cooperation especially with the particle physics   community.
There are other fields that rely on high field magnets like medical physics, fusion experiments or some aspects of solid state physics.
While usually magnets are custom developed for the individual application, the community nevertheless does strongly benefit from the development of the new and innovative technologies, facilitating straight forward production of custom made strong magnets for other experiments.

\subsubsection{Development of (sub-)quantum limited and single photon detectors: } 
Even with the strongest magnets available or planned and even if assuming that axions make up 100\% of DM, the expected signal of experiments utilising axion to photon conversion, is extremely small, making it difficult to exploit the whole band of QCD axion models in parameter space. 
In order to facilitate scanning of the whole experimental mass range within reasonable time scale it is essential to minimise the required measurement time. This depends crucially on the noise level of the used detector. Especially for the mass range between $\sim\!\!10\GHz$ and $\sim\!\!1\THz$ significant improvements are still necessary in order to either reach (or even surpass) the quantum noise limit  or to develop single photon detectors capable of detecting individual photons with these low energies (sub--\milieV). 

For the frequency range below 10\GHz{} (axion mass below 40\mueV) detector R\&D on (sub-)quantum limited detectors should be supported.
Many other fields of fundamental physics research have strong interest in development of (sub) quantum limited sensors:
gravitational wave detectors, radio astronomy and cosmology. But also applications like quantum information technology,  extreme telecommunications (space program, submarines) have common interests that are worthwhile to exploit for synergies on detector R\&D.

For higher mass axions (mass above 100\mueV) single photon detectors for the frequency range above 30\GHz{} need to be developed.
This also is a big topic for the quantum computing technology as well. Many groups working on the development of quantum computing are striving to find applications other than their core business. This opportunity should be exploited. Specifically, it should be evaluated whether the huge commercial interest, hence huge amount of funds being put into this type of R\&D, could be  at least partly utilised for axion research. Note that this research may also be highly beneficial for other topics in astroparticle physics (like development of low-energy threshold detectors for coherent neutrino scattering or low mass WIMP searches)
in quantum computing.
 
\subsubsection{Low loss RF technology and cavities}
Many applications in quantum computing, microwave engineering for telecommunication and radio astronomy depend on careful RF engineering and development of new low loss dielectric materials.
The current R\&D efforts in these fields should be studied and potential common interests and synergies should be evaluated. This could be especially relevant for:
\begin{itemize}
    \item the development and characterisation of low loss dielectric materials by crystal growth (like for example needed for MADMAX dielectric haloscope or the dielectric cavity approach QUAX and ORGAN) 
    \item development of meta-materials for the plasma haloscope and topological insulators 
\end{itemize}
Developments of cavities could be synergetic with activities on accelerator R\&D. Again, the big European labs have a leadership role here that might be exploited. 

\subsubsection{Optical cavities and lasers}
For LSW experiments but maybe also for future light scattering experiments high power lasers and optical cavities with high finesse are essential. 
There already is an ongoing exploitation of the synergies between groups from gravitational wave detectors and LSW experiments. 
Development of extremely stable high power laser is a field with many applications
like solid-state physics, plasma acceleration, gravitational wave detectors, fusion experiments and industry. Development of such devices could easily lead to spin offs.

\subsubsection{Nuclear magnetic resonance technologies and spin coupling}
Some axion experiments, like the CASPER projects, focus their research on the development of magnetic resonance techniques. The key issue here is to measure spin frequencies influenced by the axion field with unprecedented precision. This, potentially, could have an important  impact also for medical physics applications, like improvement of position resolution, and could lead to important improvements and possible spin offs in the field.

\subsubsection{X-ray detectors }
The IAXO solar axion experiment requires a low background X-ray detector for detecting the weak signal.
Here, X-ray optics like the ones developed for X-ray astronomy missions are needed. On the other hand, much of the knowledge gained in low background experiments for WIMP searches but also from neutrinoless double beta decay experiments is used in developing detectors with the required specifications. 

\subsubsection{Cryogenic engineering}
Most efforts require minimisation of thermal noise. Operation of the experiments at cryogenic temperatures is therefore paramount. The experiments that need tuning of the setup, hence, rely on cryogenic rated technologies that partly need to be custom developed.
As an example, the technology for using Piezo motors at cryogenic temperatures is presently being developed. Axion experiments (ADMX and MADMAX) have triggered these developments with companies like Attocube or JPE.\footnote{Communication between JPE and the MADMAX collaboration was established at an APPEC technology forum.}


\subsection{SWOT Tables for Axion/ALPs Experiments}\label{sec:axionswot}
The advantages (strengths~{\bf S}) and limitations (weaknesses~{\bf W}) as well as opportunities~{\bf O} and general risks (threats~{\bf T}) of the individual experimental axion search approaches described above are summarised in the following pages in SWOT tables: haloscope experiments, 
helioscopes, 
laboratory experiments, 
and low background experiments. 

\begin{table}[h!]
  \begin{tabular}{ |p{7.6cm}|p{7.6cm}|}

  \hline\hline
    \multicolumn{2}{|c|}{\bf SWOT Analysis: Haloscope Experiments}    \\ \hline \hline
    Strengths & Weaknesses  \\ \hline
\begin{itemize}
\item cavity technology already shown to have sensitivity to QCD axions
\item dielectric haloscopes quasi broad-band
\item cavities and dielectric haloscopes have the potential to scan the whole ``classical" DM axion mass range
\item precise mass measurement possible
\item NMR methods could cover the very low axion mass range, while LC circuits can fill the gap up to the ``classical" mass window
\item NMR experiments quasi ``table top" experiments
\item measuring frequency allows extremely good resolution with which axion astronomy could be preformed
\item determination of spin and parity possible
\end{itemize}  
&
\begin{itemize}
\item thermal background difficult to overcome
\item cavities difficult to scale for mass range above $\sim\!\!40\mueV$ 
\item very narrow frequency band measurements: need scanning strategy, in case of axion: clustering could miss signal
\item high field large aperture dipole magnets for dielectric haloscope are expensive
\item no direct measurement of coupling strength and DM density possible
\end{itemize}     
\\ \hline\hline
    Opportunities & Threats \\ \hline
\begin{itemize}
\item axion-photon coupling could be significantly larger than for benchmark models
\item could be extended by multiple matched cavities, plasma haloscopes, topological insulators
\item profit from R\&D towards quantum sensing performed in quantum computing community
\item cavities and other alternative concepts could quickly adapt to frequency once mass is known to measure velocity dispersion 
\end{itemize}     
&       
\begin{itemize}
\item axion-photon coupling could be on the lower side leading to haloscopes escaping signal detection if search strategy is not adapted to reach lower band couplings in axion band
\item clustering could reduce the overall axion density by up to $\sim\!\!50$\%  for the post inflationary scenario PQ symmetry breaking
\item for dielectric haloscope: novel large aperture dipole magnet needed
\end{itemize}     
      \\
    \hline\hline
  \end{tabular}
\caption{SWOT analysis for axion and ALP cavity and dielectric haloscope searches.}\label{swot:cavities}
\end{table}
\newpage

\begin{table}
\begin{tabular}{ |p{7.6cm}|p{7.6cm}|}
\hline\hline
    \multicolumn{2}{|c|}{\bf SWOT Analysis: Helioscope Experiments}    \\ \hline \hline
    Strengths & Weaknesses  \\ \hline
\begin{itemize}
\item approach less dependent on cosmological models  and independent of DM halo density
\item required technologies available
\item signal can be easily switched off by pointing away from Sun
\item in certain mass range no need to tune to axion mass
\item sensitivity to $g_{ae}$ and $g_{aN}$ 
\end{itemize}  
& 
\begin{itemize}
\item for part of the mass range: needs step-by-step He-pressure scanning to enter QCD axion band
\item cannot confirm that a possible detection is related to Galactic DM
\end{itemize}     
\\ \hline\hline
    Opportunities & Threats \\ \hline
\begin{itemize}
\item use existing technology developed for other fundamental science physics searches (magnets, X-ray detectors, etc.)
\item could cover QCD axion mass range compatible with DM and stellar cooling anomaly 
\item possibility to carry a ``payload experiment" such as RADES
\item determination of solar axion/ALP mass could guide the path for long range force experiments
\end{itemize}     
&

\begin{itemize}
\item in complex hidden sector models axion production in stellar plasmas could be suppressed
\item detection of solar axions or ALPs that are not, or are just a subdominant component of,  DM  could mislead DM search community
\end{itemize}     

    \\
  \hline\hline
\end{tabular}
\caption{SWOT analysis for axion and ALP helioscope searches.}\label{swot:helioscopes}
\end{table}

\newpage

\begin{table}
\begin{tabular}{ |p{7.6cm}|p{7.6cm}|}
\hline\hline
    \multicolumn{2}{|c|}{\bf SWOT Analysis: Laboratory Experiments}    \\ \hline \hline
     Strengths & Weaknesses  \\ \hline
\begin{itemize}
\item production mechanism independent of astrophysical or cosmological models
\item  re-use and further develop existing technologies (accelerator dipole magnets, long-baseline optical resonators)
\item spin/parity measurement possible
\item potentially table top experiment
\item for some optical cavity DM ALP experiments no external $B$-field needed
\end{itemize}  
 & 
\begin{itemize}
\item benchmark QCD axion models out of reach
\item cannot confirm that a possible detection is related to Galactic DM
 \end{itemize}     
\\ \hline\hline
    Opportunities & Threats \\ \hline
\begin{itemize}
\item LSW experiments can cover parameter ranges compatible with transparency hint and partly parameter range compatible with stellar cooling anomaly
 \item exploitation of synergies with other fundamental physics searches
 \item relatively ``cheap" experiments re-using existing magnets
\end{itemize}     
 &       
\begin{itemize}
\item ALPs that are not, or are just a subdominant component of,  DM could mislead DM search community
\end{itemize}     
 \\
     \hline\hline
\end{tabular}
\caption{SWOT analysis for LSW searches and vacuum polarisation experiments.}\label{swot:LSW_VB}
\end{table}

\newpage

\begin{table}
  \begin{tabular}{|p{7.6cm}|p{7.6cm}|}
    \hline\hline
    \multicolumn{2}{|c|}{\bf SWOT Analysis: Low Background Experiments}    \\ \hline \hline
    Strengths & Weaknesses  \\ \hline
\begin{itemize}
\item by-product of other fundamental physics searches - nearly for free
\item crystal detectors can can probe $g_{a\gamma}$ and  $g_{ae}$
\end{itemize}  
& 
\begin{itemize}
\item not sensitive to 
QCD axion or ALP models with f$_a\gtrsim10^{-8}\gev^{-1}$ that are not excluded by astrophysical arguments
\item solar searches: cannot confirm that a possible detection is related to Galactic DM
\end{itemize}     
\\
\hline \hline
    Opportunities & Threats \\ \hline
\begin{itemize}
\item can exploit some so far inaccessible ranges 
\item analysis of data of low background experiments can lead to better understanding of experimentally obtained spectra/background 

\end{itemize}     
&       

      \\
    \hline\hline
  \end{tabular}
 \caption{SWOT analysis for low background experiments.}\label{swot:alps_low_bkg}
\end{table}

\newpage

\newpage
\section{Broader Context}\label{sec:broadercontext}

The searches for DM in form of WIMPs, axions and ALPs described in detail in the chapters above are embedded into a large international experimental and theoretical effort to identify DM by using tools from (astro)particle physics, astrophysics as well as cosmology. Each of these dark matter search strategies has its unique advantages and challenges, making them highly complementary. Their combined power is leveraged in particular in the context of global fits, see~\cite{AbdusSalam:2020rdj} for an overview of recent activity in this field.
Different ways to detect WIMPs and axion/ALPs are summarised in Sects.~\ref{sec:wimpbroader} and~\ref{sec:axionbroader}, respectively. The potential of direct DM search experiments to perform studies beyond the primary DM channel is discussed in~Sect.~\ref{sec:beyonddm}. A quick discussion of the laboratories and infrastructure required to perform the searches is presented in Sect.~\ref{sec:labs}.

\subsection{WIMP Searches
}\label{sec:wimpbroader}

DM in the form of WIMPs is searched for not only in direct detection experiments, but also indirectly, by looking for the products of WIMP annihilation/decays, by searching for signatures of ``dark" particles being produced at colliders and in beam-dump experiments, and by using astronomical data, including gravitational waves. These approaches are briefly summarised here.

\subsubsection{Indirect Searches
}\label{sec:wimpid}

As discussed in Sect.~\ref{sec:thermalWIMPs}, the freeze-out mechanism relies on a particular value of the product of WIMP pair-annihilation cross section and their relative velocity, in order to obtain the observed relic density.
In most models WIMP particles can also {\bf annihilate today} 
with the same, or  higher, rate~\footnote{If the annihilation rate is velocity dependent, it can change substantially due to the different kinetic
energy of the DM at freeze-out compared to today. 
In particular in the presence of the Sommerfeld enhancement,
the rate today can be even much larger than the  cross section value at freeze-out.} and produce observable signals, as long as a sufficiently 
large DM density is present.
Overdense regions are expected to arise due to gravitational collapse
in the centre of gravitationally bound objects, like galaxies 
or clusters of galaxies. 
Annihilation processes in those regions produce energetic SM 
particles at different epoch and locations in the Universe. 
Depending on the WIMP, different final states are 
produced, but generically photons, neutrinos, light
leptons and hadrons arise as secondary products from fragmentation
and final state radiation~\cite{Bringmann:2012ez, Cirelli:2015gux, Gaskins:2016cha}.
Note that analogous signatures can appear also for decaying, rather than pair-annihilating, DM, and the resulting spatial distributions
of the excesses are very different~\cite{Bertone:2007aw}. 
A prominent example of a decaying DM candidate is the sterile neutrino.
Different searches are conducted depending on the final state.

\paragraph{\bf Photon flux} 
As neutral particles, photons
    propagate without deflection and therefore one can focus on 
    several targets that can be  
    sources of DM annihilations, like the local dwarf galaxies 
    and the central region of the Milky Way. Sources that are too far away 
    or too weak to be  detected, contribute instead to 
    the Galactic and extragalactic gamma-ray backgrounds. 
    The annihilation is proportional to the square of the DM density, so the strength and spatial distribution of the signal is very sensitive 
    to the DM profile.
    Regarding the spectrum, photons from internal bremsstrahlung,
    fragmentation or final state radiation, give a continuous broad  
    spectrum up to energies equal to the DM mass, 
    while the direct annihilation into two photons via one loop diagrams
    generates a monochromatic line exactly at the DM mass and provides
    a more easily identifiable signal~\cite{Bringmann:2012ez}.
    Some photon excesses over the expected astrophysical
    flux have been reported, \eg the Fermi excess from the Galactic centre~\cite{Hooper:2010mq}, but so far they remain unconfirmed 
     and are in tension with the dwarf
    galaxies constraints~\cite{Keeley:2017fbz}. Both Fermi LAT and at larger energies the H.E.S.S. experiment have approached the target of the ``thermal value" of $\sigma_{\textrm{ann}}v \approx 3\times 10^{-26}\cmeter^3/\second$  (compare Eq.~\ref{eq:sigmavvalue})
    and in some specific channels reached it, but these results depend on assuming some specific profile of the DM halo. A forthcoming experiment Cerenkov Telescope Array (CTA) is expected to improve the reach of H.E.S.S.  by an order of magnitude or so.

   \paragraph{\bf Neutrino flux} 
   Neutrinos are also produced in DM annihilation and propagate undeflected but are more difficult to observe at low energies, as they are overwhelmed by the presence of atmospheric neutrinos. At large energies, they can provide limits comparable to the photon flux especially for monochromatic signals~\cite{Aisati:2015vma}. 
   The Sun and the Earth are two very promising sources of neutrinos from WIMP DM annihilation. 
   In the presence of a substantial elastic scattering of DM particles with normal matter, they are slowed down when they pass through an astrophysical object and can remain trapped there. If the equilibrium between capture and annihilation rate is reached~\cite{Silk:1985ax, Peter:2009mk}, as is expected for the Sun, the detection of an annihilation signal could give direct information on the elastic scattering cross section and provide complementary information to direct detection as the capture rate in the Sun includes all DM energies, even those below the threshold for direct detection~\cite{Arina:2013jya, Ibarra:2018yxq}.  Moreover, since the main component of the Sun is  hydrogen, the spin-independent and the spin-dependent scattering cross section strengths are comparable, the neutrino channel gives very competitive limits with respect to direct detection searches for the spin-dependent WIMP-proton elastic cross section.
   Neutrino 
   experiments like SuperKamiokande and IceCube have already set constraints on the DM elastic cross section and on the proton and DM annihilation cross section from upper limits on the flux of neutrinos from the Sun and the centre of the Galaxy~\cite{Choi:2015ara, Aartsen:2015xej, Aartsen:2016zhm}.

 \paragraph{\bf Charged cosmic rays} 
Other annihilation products are
charged SM particles that are deflected by the magnetic field 
in the galaxy and can interact with the Galactic medium. 
For these channels the full, complex, propagation of the charged 
particles has to be taken into account and generically only 
particles produced within a radius of a few \kpc \ around the Earth 
can reach us.
To identify the DM annihilation signal in cosmic rays the astrophysical
backgrounds have to be subtracted and this leads to large uncertainties~\cite{Giesen:2015ufa, Cirelli:2015gux}.
Nevertheless, excesses in the fluxes of antiparticles like positrons, antiprotons 
and even antideuterons, which are not expected to be present in significant amounts in primary cosmic rays, have been interpreted as DM signals~\cite{Gaskins:2016cha}, even though no fully convincing 
signal has been found yet.  
Extracting DM properties from the charged particle spectra is difficult, 
as they are reprocessed by the propagation, but nevertheless information 
on the scale of the DM mass and its main annihilation channel can in principle be extracted.

\paragraph{\bf Complementarity with direct detection}
Indirect DM detection (ID) can provide complementary information to direct detection (DD) and collider searches~\cite{Bergstrom:2010gh, Roszkowski:2017nbc}. Indeed it has usually a better reach at large mass, as the signals become more evident above the backgrounds, and can provide a direct measure of
the DM mass if the monochromatic signal into two photons is measured.
Analysing the spectra 
can give information 
also on the dominant DM annihilation channels and on its spatial distribution
in the Galaxy.  
On the other hand, indirect searches usually suffer from significant and often hard to quantify backgrounds. In addition, the interpretation of observed signal excesses has proven to be very hard in practice as 
astrophysical processes can often lead to similar signatures (e.g., nearby pulsars can lead to an increase in the positron flux~\cite{Profumo:2008ms}). 
In complete models, like those rooted in supersymmetry, the correlation between the
ID and DD signals can be looked for once a detection in any search is made
 and the combination of both gives a much better coverage of the parameter space~\cite{Bergstrom:2010gh, Roszkowski:2017nbc}. 
 With the exception of the constraints from neutrino telescopes a direct model independent comparison of results from indirect and direct detection is not possible as the searches probe the different processes of WIMP annihilation and scattering. In general ID and DD searches probe different couplings in
the dark sector.

\subsubsection{Searches at the Large Hadron Collider
}\label{sec:wimplhc}

As discussed in \ref{sec:thermalWIMPs}, the ``WIMP miracle'' suggests that
particle DM is related to new physics at the electroweak scale.  The LHC is
therefore a prime additional tool for probing WIMP DM physics.  In particular,
we can hope to directly produce DM particles in high-energy proton--proton
collisions through the process inverse to DM annihilation.  Of course, 
even when DM particles are produced, they typically do not leave signatures
in the LHC detectors (see, however, Ref.~\cite{Bauer:2020nld} for a recently developed
model that violates this statement).  Therefore, the only way to establish
their existence is through ``missing momentum'' signatures: the DM particle carries
away transverse momentum, leading to apparent violation of energy-momentum
conservation among the visible particles in the event. There are several obvious
difficulties with missing momentum signatures, though:
\begin{itemize}
  \item 
    Standard Model neutrinos, which are copiously produced for instance in
    the decays of $W$ or $Z$ bosons, lead to very similar signatures.  A detailed
    study of event topologies and kinematic distributions is required to disentangle
    these contributions from real DM signatures. Nevertheless, neutrino-induced
    backgrounds often remain large and must be estimated with high
    precision.

  \item 
    While discovering a new particle species that is invisible to the LHC detectors
    would be a gigantic leap forward for particle physics, this observation alone
    would not prove that the new particle indeed contributes to the DM in the Universe.
    In particular, it could be unstable on cosmological time scales, or its production
    in the early Universe could be suppressed.
\end{itemize}

A large number of LHC searches for DM exist, falling broadly into two main categories:

\textbf{Searches based on top-down and phenomenological models.}~~
    In top-down models, such as supersymmetric extensions of the Standard Model or models with extra dimensions, the DM particle is assumed to be the lightest among a larger number of new states, some of which carry electromagnetic or colour charges.

    If these charged states are not too heavy
    (\mbox{$\lesssim \text{several}\ 100\gev$} for electromagnetically charged states,
    $\lesssim \text{few TeV}$ for coloured particles), they can be produced at the LHC
    in large numbers. They quickly decay to a DM particle and Standard Model particles,
    possibly via multiple steps (cascade decays). The Standard Model particles emitted
    in these decays, together with the missing transverse momentum,
    often constitute striking signatures that give such searches superb sensitivity.
    
    Backgrounds to searches for cascade decays arise from SM processes that
    yield similar final states. For longer cascades, the likelihood for this to
    happen is smaller than for single-step decays.  Careful analysis of kinematic
    distributions can be used to significantly suppress backgrounds and to identify
    the unstable intermediate particles appearing in the cascade.

    In top-down models, \eg models based on (grand) unification,  the dynamics at collider scales depends on only a small number of parameters defined at some higher energy scale.  Such models make specific predictions, which makes their analysis relatively easily manageable. However, they usually do not encircle all observational possibilities. Therefore, it is also useful to consider purely phenomenological models, such as the general MSSM or its subsets, \eg the various versions of the pMSSM with 19 or less free  parameters which are more directly related to collider observables, for instance particle masses. The large number of free parameters and available search channels is, however, also the main complication in searches within such phenomenological models.

\textbf{Searches for generic signatures using simplified models.}~~
    There exist also relatively generic LHC signatures of DM, realised in most
    WIMP models.  These signatures are based on the unavoidable initial state
    radiation in high energy collision: the initial state quarks in a
    proton--proton collision leading to DM production have a high probability
    of radiating off hard gluons, photons, or electroweak bosons. Energetic
    initial state radiation is easily detectable, and together with the
    missing momentum footprint of DM
    particles, can be used to search for DM productions in a fairly
    model-independent way.

    On the other hand, such ``mono-$X$ + missing
    momentum'' signals can arise also in the Standard Model, in particular in
    events containing neutrinos.  This implies large backgrounds and large
    associated systematic uncertainties, ultimately limiting the sensitivity of
    mono-$X$ searches.  Typically, mono-$X$ events associated with the production
    of heavy DM are somewhat harder than those arising from SM processes. While
    this feature is typically exploited in mono-$X$ searches, it leads to only a
    moderate suppression of backgrounds. An additional problem are the large
    theoretical uncertainties associated with the background prediction:
    the rates of processes such as the production of QCD jets (a background to
    mono-jet searches), photons (relevant to mono-photon searches)
    and electroweak gauge bosons (relevant to mono-$Z$ and mono-$W$ searches)
    are very difficult to predict accurately, further reducing the sensitivity
    to DM searches in these channels.

{Besides searching for the DM particle directly via its missing
transverse momentum signature, the LHC collaborations are also carrying out
a vast number of searches that target potential mediator particles, that is
particles that interact with both the SM sector and the DM.  Examples
include (i) heavy vector bosons ($Z^\prime$) arising from a new $U(1)^\prime$ gauge interaction
under which both the DM and some SM particles are charged; new scalar bosons
which couple to the DM via a Yukawa coupling and mix with the SM Higgs boson;
(iii) new charged scalars (such as squarks and sleptons in supersymmetry)
coupled to SM fermions and to fermionic DM.  Thanks to their couplings to
SM particles, such mediators may be observable in processes involving only
SM final states and no missing transverse momentum.  For instance, in the
case of a $Z^\prime$ mediator, processes like $\bar{q} q \to Z^\prime \to \bar{q} q$
offer superb sensitivity up to mass of multiple TeV through a simple bump hunt.
While a discovery of a new mediator particles does not immediately solve the
DM mystery, it would nevertheless inform other, more targeted, searches
by setting a mass scale. For instance, if a $Z^\prime$ boson were found at a mass
$M_{Z^\prime}$ and its couplings to SM particles were measured, we would be able to
calculate its couplings to DM (as a function of the DM mass) based on the
observed relic abundance. We would thus have a clear idea where in parameter
space we may hope to find the DM particle itself.

Comparing LHC results to direct detection limits on dark matter is always
model dependent. For instance, direct detection experiments have a significant
advantage in scenarios in which DM--Standard Model
interactions are mediated by relatively light particles ($\ll 100 \GeV$) and the
DM particle mass is $\gtrsim 10\GeV$.  In such scenarios, the DM production cross section
at a collider decreases rapidly with the collider energy, while low-energy probes
such as direct detection experiments do not suffer from this suppression.
If, on the other hand, DM is very
light ($\ll 10\GeV$), but its interactions depend on heavier new physics, LHC limits
are often stronger. In this regime, recoil energies in direct detection experiments
are very low and thus difficult to detect, while LHC does not 
suffer
from any threshold
effects. This highlights the important complementarity between the
different DM search strategies.}

\subsubsection{Astrophysical Probes of Nature of DM 
}\label{sec:wimpastro}

Current and upcoming gravitational wave detectors offer various opportunities for probing the nature of DM. For instance direct detection of PBHs, compact objects formed from DM (e.g. oscillons, Q-balls), or condensates of sub-eV DM. The nature of DM could also be probed indirectly, via its environmental effects on BH and neutron star mergers, or via the gravitational waves generated by the DM production mechanism. For an overview see Ref.~\cite{Bertone:2019irm}.

Future astronomical surveys, such as Euclid, 
will measure the mass function, density profiles and shapes of DM halos, as well as the offsets between DM, gas and galaxies, with improved precision. This will lead to improved constraints on the DM self-interaction cross section and also the mass of warm and ultra-light DM. For further details see e.g. Ref.~\cite{Buckley:2017ijx,Marsh:2015xka}. The open, asteroid mass, window for PBHs is hard to probe. Possibilities include high cadence, short wavelength lensing observations of sources with small sizes and PBH-driven stellar destruction. For discussion see Ref.~\cite{Montero-Camacho:2019jte}. Astrophysical probes of axions are discussed in Sect.~\ref{sec:axionbroader}.

\subsection{Other Physics at Direct DM Detection Experiments
}\label{sec:beyonddm}

Modern DM detectors are no longer single-purpose experiments, but
rather observatories with a rich physics program beyond their core mission of
searching for DM scattering on nuclei or electrons \cite{Harnik:2012ni,Dutta:2019oaj}.

\subsubsection{Neutrino Physics}
\label{sec:neutrino}

While being an obstacle to the search for DM, the unavoidable background
due to coherent neutral-current neutrino--nucleus scattering (the neutrino floor
discussed in Sect.~\ref{sec:wimpbackgrounds})
offers interesting physics opportunities in itself:

\textbf{Solar neutrinos.} The $pp$~chain of nuclear reactions is
    the dominant energy production mechanism in the Sun, but the associated
    $pp$~neutrinos are extremely difficult to detect due to their low energy
    $\lesssim 400keV$. In DM detectors, on the other hand,
    scattering of solar $pp$ neutrinos on electrons is the dominant
    neutrino-induced background at energies below $\sim\!\!200\keV$~\cite{Harnik:2012ni} and a sub-percent flux measurement is possible~\cite{Aalbers:2020gsn}. (In the case of nuclear recoils, only
    ${}^{8}\text{B}$
    and hep neutrinos are relevant at recoil energies $\gtrsim 
    0.1\keV$~\cite{Gutlein:2010tq}.)    At somewhat higher energies
    ($\lesssim 1.5\MeV$), DM detectors are sensitive
    to neutrinos from the CNO cycle~\cite{Franco:2016ex,Dutta:2019oaj}, with up to several thousand
    neutrino--electron scattering events expected in currently
    planned  multi ton-scale DM detectors.  DM experiments will also provide an independent measurement of the ${}^{8}\text{B}$ neutrino flux, and they may be able to measure for the first time neutrinos from the hep cycle~\cite{Dutta:2019oaj}.
    
    A future precision measurement
    of $pp$ and CNO neutrinos will offer a unique opportunity to
    probe the interior of the Sun much more directly than with neutrinos
    from higher-energy reactions. Such a measurement will be interesting in
    the context of the solar metallicity problem, a persistent tension between
    measured element abundances at the Sun's surface and predictions from
    solar models fitted to data from helioseismology (see~\cite{Serenelli:2016dgz}
    for a review and for further references). Some models
    addressing this tension predict substantially modified neutrino
    fluxes (for CNO neutrinos, the discrepancy between the two leading
    metallicity models \cite{Grevesse:1998cy} and \cite{Asplund:2009eu} is
    up to 38\%), which could be tested in DM detectors,
    see e.g.,~\cite{Vincent:2014jia,Aalbers:2020gsn}.
    
    Important backgrounds to solar neutrino measurements in DM
    detectors will be non-neutrino-induced electron recoils,  in particular $^{222}$Rn, as well as atmospheric neutrino interactions.

 \textbf{Atmospheric neutrinos.} At higher energies, the
    neutrino floor of DM detectors is dominated by atmospheric
    neutrinos. A measurement of the atmospheric neutrino flux at
    energies $<100\MeV$ -- below the energies typically studied
    in neutrino experiments like SuperKamiokande -- is therefore
    feasible~\cite{Dutta:2019oaj,Newstead:2020fie}.

  \textbf{Supernova neutrinos.} In case of a Galactic supernova,
    multi-ton scale DM detectors will record hundreds of
    coherent neutrino--nucleus scattering events~\cite{Chakraborty:2013zua,Lang:2016zhv}.
    This measurement is unique as it is sensitive to neutrinos of all
    flavors, thus providing a clean measurement of the total explosion
    energy and the flavor-averaged neutrino spectrum. Together with
    flavor-sensitive measurements in large-volume detector like DUNE
    and HyperKamiokande, it will help us gain a comprehensive understanding
    of the explosion dynamics.
    
    Even before the actual supernova explosion, DM experiments
    may observe a flux of pre-supernova neutrinos, which are emitted
    during the very final stages of stellar evolution and may thus
    serve as an early warning for an imminent explosion~\cite{Raj:2019wpy}.

\subsubsection{Neutrinoless Double Beta Decay and Other Rare Nuclear Decays}
\label{sec:nudbd}

Many of the experimental challenges which direct DM searches face
are similar to those relevant to searches for
neutrinoless double beta decay. For instance, both types of experiments require
extremely low background levels, large target volumes, and long exposures.
Since several target materials considered for DM detection
(in particular germanium and xenon) have isotopes that are unstable to
double beta decay, the two searches can possibly be merged in future experiments. 
In case of dark matter LXe TPCs, first results on the search for the neutrinoless double beta decay of $^{136}$Xe have been published~\cite{Ni:2019kms} and it has been shown that DARWIN-size detectors 
{may} reach a sensitivity competitive to dedicated double beta experiments -- even without expensive isotopic enrichment~\cite{Agostini:2020adk}.

The potential of DM detectors to search for rare nuclear decays
has been demonstrated spectacularly when XENON1T observed for the first
time double electron capture on ${}^{124}$Xe~\cite{XENON:2019dti}. The sensitivity of multi-ton LXe TPCs to the corresponding neutrinoless decay is discussed in~\cite{Wittweg:2020fak}. The double beta decay of $^{134}$Xe has a rather low $Q$-value but could be targeted in a LXe TPC depleted in $^{136}$Xe~\cite{Barros:2014exa}.  LAr detectors search for the double electron capture of $^{36}$Ar~\cite{Dunford:2018xar}.

\subsubsection{Other Aspects of Physics Beyond the Standard Model}
\label{sec:nubsm}

A large number of well-motivated extensions of the Standard Model
feature signatures that are accessible to DM experiments.
Perhaps the most widely-discussed ones are:

\textbf{Neutrino magnetic moments.} The magnetic moment of neutrinos
    in the Standard Model, extended with Dirac neutrino masses, is $\mu_\nu =
    3.2 \times 10^{-19} \mu_B \times(m_\nu / \eV)$ \cite{Fujikawa:1980yx}, which
    is beyond experimental reach. The value of $\mu_\nu$ can, however, be much
    larger in extensions of the Standard Model, for instance in
    radiative neutrino mass models and in models with flavor symmetries
    \cite{Lindner:2017uvt}. Current experimental
    limits on $\mu_\nu$ are of order $10^{-12} \mu_B$ from astrophysical
    observations, while the best terrestrial limits are of order
    $10^{-11} \mu_B$~\cite{Tanabashi:2018oca}.
    In the current context, it is particularly noteworthy that the
    differential cross section for magnetic moment-induced 
    neutrino--nucleus scattering scales as $d\sigma/dE_R \propto 1/E_R$
    with the recoil energy $E_R$. This implies that low-threshold experiments
    like DM detectors are ideally suited for probing $\mu_\nu$
    beyond the current limits \cite{Harnik:2012ni}.
    With a DARWIN-class experiment, terrestrial
    limits would for the first time rival astrophysical ones~\cite{Huang:2018nxj}.
    
 \textbf{New neutrino interactions.} Recent years have seen
    a surge of interest in new forces mediated by $\lesssim
    \mathcal{O}(\GeV)$ particles,
    motivated by the realisation that a multitude of viable and
    well-motivated DM models can be constructed at this mass scale.
    Scattering processes of neutrinos mediated by such light force mediators
    are enhanced at low recoil energies, making DM detectors
    ideally suited to search for them~\cite{Harnik:2012ni}.
    The largest enhancements are typically achieved
    in models involving also sterile neutrinos, which are independently
    motivated by several anomalous results from short-baseline
    neutrino oscillation experiments \cite{Dentler:2018sju,Diaz:2019fwt,Vincent:2014jia,Gariazzo:2017fdh}.
    
 \textbf{Dark Photons.} Many experiments searching for axions exploit
    the mixing of the latter with photons in magnetic fields. Therefore,
    they are automatically sensitive also to dark photons,
    that is new gauge bosons from a $U(1)^\prime$ gauge group that
    couple to the Standard Model through a kinetic
    mixing term of the form $\epsilon F^\prime _{\mu\nu} F^{\mu\nu}$. Here $\epsilon$
    is a dimensionless coupling constant and $F^{\mu\nu}$, $F^{\prime\mu\nu}$
    are the electromagnetic and $U(1)^\prime$ field strength tensors, respectively.

\subsection{Other Constraints on Axions and ALPs
}\label{sec:axionbroader}
Axions and ALPs as DM candidates are constrained by several theoretical considerations and astrophysical observations as discussed above. 

The mass generation mechanism gives a direct correlation between the axion mass and the energy scale of Peccei-Quinn symmetry breaking described by the axion decay constant $f_a$. Through this, also a linear relationship between the axion mass and the axion coupling strengths exists.
Therefore, in the $m_a$ vs.~$g_{a\gamma}$ parameter range, the plausible regions for axions are situated within a band: the higher the mass, the higher the coupling strength. The uncertainties on the coupling strength itself, i.e. the width of the band depend on the type of coupling and are model dependent \cite{DiLuzio:2016sbl} (see Fig.~\ref{fig:axions_status}). The axion to photon coupling $g_{a\gamma}$ utilised in haloscope, helioscope and LSW experiments is constrained within roughly an order of magnitude. Other couplings, like  the one for axions to electrons $g_{ae}$ can be strongly suppressed for some models. 

The most stringent lower limits for the axion decay constant~$f_a$ come from astrophysical constraints: stellar evolution and the observation of the time spread of neutrinos from the neutron star progenitor of SN87a: for $f_a \lesssim 10^8\GeV$ the axion coupling to ordinary matter would be too high to explain the length of the neutrino burst from SN87a. This constrains the axion mass to a value $\lesssim 15\miliev$~\cite{Raffelt:1990yz, Raffelt:2006cw}. Constraints can also be extracted from evolution of stars in the horizontal branch of the Hertzsprung Russel diagram. Note that these constraints are slightly model dependent (SN evolution, stellar evolution).

The upper limit on~$f_a$ is generally given by the Planck scale for consistency reasons with quantum gravity $f_a \gtrsim 10^{18}\GeV$, which corresponds to a lower limit of the axion mass of $\gtrsim 10^{-13}\eV$.
In the scenario in which Peccei-Quinn symmetry breaking happened after inflation the lower limit for the axion mass is generally believed to be around $m_a \gtrsim 25\mueV$, as otherwise there would be too much axion DM and the Universe would be over-closed. No such bound exists for the scenario in which Peccei-Quinn symmetry breaking happened before inflation. In the latter case it could be argued that a lower axion mass $m_a \lesssim 10^{-7}\eV$, corresponding to a very small initial $\theta_i$ may require fine tuning. This argument can, however, be overcome by considering that a misalignment mechanism (explicit symmetry breaking) may have been at work before the inflationary epoch already, thus driving $\theta$ to a very small value.

For ALPs and hidden photons, no correlation between the energy scale of breaking of the symmetry and ALP mass exist. Hence, the allowed range in the coupling strength vs.~mass parameter space is  much less constrained.
This is explicitly also the case for constraints on the viable mass consistent with ALPs as dark matter.

It is interesting to note that there are astrophysical inconsistencies that could be explained by ALPs: 
The opacity of the Universe for photons with energies around few $\tev$ is increased by the electromagnetic background light of known objects in the Universe like stars, galaxies, etc. Measurements of the $\tev \gamma$-ray flux from objects at cosmological distances with Cerenkov telescopes seem to indicate that the Universe is more transparent for photons of this energy than expected. This could be explained if the emitted photons are converted to ALPs in an external magnetic field (either in the source itself or in the non-vanishing intergalactic magnetic field) and back-converted to photons in the vicinity of our solar system before they are detected~\cite{Horns:2012fx, Brun_2013}. 
The range in the mass versus coupling parameter space compatible with ALPs consistent with this explanation is depicted as "transparency hint" in Fig.~\ref{fig:axions_status}.

In addition there are some inconsistencies in evolution for several stellar systems: The absence of stars in the tip of the horizontal branch of the Hertzsprung-Russell diagram seems to indicate that red giants cool faster than anticipated. Likely, the energy loss in white dwarfs and some neutron stars is higher than expected from standard stellar evolution models. 
While the individual anomalies have only low significance, it should be noted that all anomalies point towards extra energy loss for all the investigated systems and could be explained by ALPs with a mass $\lesssim$ few $\keV$ and $g_{{\rm ALP}\gamma}$ of a few $10^{-11}\,\GeV^{-1}$ \cite{Giannotti:2015kwo, Giannotti:2017hny}, compatible with ALPs consistent with the transparency hint.
The parameter range consistent with the stellar evolution anomalies is shown as a violet band in Fig.~\ref{fig:axions_status}.

ALPs consistent with the transparency hint and the stellar evolution anomalies could be detected with ALPs-II and IAXO. Note, however, that ALPs consistent with the transparency hint would not be good DM candidates as those require mass~$\gtrsim\!\!5\muev$. 
On the other hand, an axion with $m_a\sim\!\!20\milieV$ could explain the stellar evolution anomalies, while being a feasible DM candidate and solving the strong CP problem. For this case the IAXO experiment would have the sensitivity to detect the axion. 

Observations of pulsars and other astrophysical sources with strong magnetic fields are presently under discussion to provide possibilities to observe signatures of axions or ALPs in a very complementary way to Earth bound experiments. DM axions could in this surrounding be resonantly converted to photons if the effective mass of the photon in the plasma of the neutron star magnetosphere coincides with the axion mass. This method could - under optimistic conditions - be sensitive to axions and ALPs with a mass around $10\,\mueV$ to $100\,\mueV$~\cite{PhysRevLett.121.241102}.

Moreover, ALPs and non-minimal axion scenarios could lead to measurable gravitational wave signatures in future detectors such as LISA. Axions and ALPs are originated by spontaneous breaking of approximate symmetries, hence undergoing a phase transition. If this phase transition is first-order enough, the gravitational imprint of the release of energy during the phase transition may be observable. In particular, ALP theories with $\TeV$-scale breaking are good candidates for producing such measurable gravitational wave effects.

In the post inflationary scenario the decay of topological defects contributes to the formation of mini-clusters. 
These could have a mass that allow for their detection by femto- and picolensing~\cite{Kolb:1995bu} of astrophysical objects.
For example, observation of lensing of gamma ray bursts could lead to important constraints on the mass distribution of mini-cluster~\cite{Katz:2018zrn}.

For very high ALP and hidden photon mass $\gtrsim\!\!10\,\MeV$ there are some unconstrained and unexplored regions in the parameter space that are consistent with models for particles resulting from additional $U(1)$~symmetry groups that could also explain the nature of DM. 
These can partly be tested by colliders and beam dump experiments. These parameter regions are the target of large scale 
fixed-target or beam dump experiments, like NA62~\cite{NA62:2017rwk} and SHiP~\cite{Alekhin:2015byh} at CERN, but can also be explored by the LHC. At the LHC sub-$\GeV$ ALPs could be searched for in non-resonant searchers for new physics, as imprints in the high-momentum tails of Standard Model particles. ALPs with higher mass (around the electro-weak scale) can be searched by looking at resonant features in the missing energy distributions. A review on these particle candidates and the experimental efforts focusing on their detection is given elsewhere~\cite{Beacham:2019nyx}.

\subsection{Infrastructure and Deep Underground Laboratories
}\label{sec:labs}

The experimental approaches to directly search for WIMPs and axions are based on very different technologies and have thus very different requirements regarding laboratories and infrastructure.

{\bf Direct dark matter detection experiments} searching for WIMPs rely on environments with lowest backgrounds (see also Sect.~\ref{sec:wimpbackgrounds}). The large rock overburden of {\bf underground laboratories} provides shielding against cosmogenic backgrounds, i.e., backgrounds produced by cosmic ray muons, the only charged particles which are able to survive a few meters of shielding material. These laboratories offer to the experiments sufficient space to erect additional shielding against radiogenic backgrounds from the environment and the required auxiliary buildings to run the experiment (cleanroom, counting house etc.) as well as the relevant infrastructure to prepare (material analysis etc.), build (workshops, construction halls, etc.) and operate the experiments (power, liquefied gases, network, etc.).  Europe currently hosts fourteen different DM projects at four laboratories. Some key parameters of these laboratories are listed in Table~\ref{Experiments_labs:table}. A more detailed description can be found in the APPEC Double Beta Decay report~\cite{Giuliani:2019uno}.

\begin{table}[b!]
\begin{center}
\begin{tabular}{|m{3.5cm}|m{2.5cm}|m{2.5cm}|m{2.5cm}|m{2.5cm}|}
    \hline \hline
Laboratory    & LNGS & LSC & LSM & Boulby \\ \hline
Country & Italy & Spain & France & UK \\
Depth (m.w.e) & 3600 & 2450 & 4800 & 2820 \\ 
Muon Flux ($\mu/\meter^{2}/\second$) & $3\times10^{-4}$ & $3\times10^{-3}$ &  $5\times10^{-5}$ & $4\times10^{-4}$ \\
Volume ($\meter^3$) & 180000 & 8250 & 3500 & 4000 \\ 
Access & Road & Road & Road & Shaft \\
Personnel & {\cal O}(100) & {\cal O}(10) & {\cal O}(10) & {\cal O}(5) \\ 
DM Experiments$^*$ & 8 & 2 & 3 & 1 \\ \hline \hline 
\end{tabular}
\caption{Main features of the large European underground laboratories hosting DM experiments: Laboratori Nazionali del Gran Sasso (LNGS), Laboratorio Subterraneo de Canfranc (LSC), Laboratoire Subterrain de Modane (LSM), and Boulby Underground Laboratory (Boulby). $^*$Only projects running or under commissioning. 
} 
\label{Experiments_labs:table}
\end{center}
\end{table}

The Italian {\bf Laboratori Nazionali del Gran Sasso (LNGS)} is still the largest underground laboratory world-wide. It features three large experimental halls underground, a large surface campus and is sufficiently deep for a DARWIN-like LXe dark matter TPC. LNGS is rather conveniently located less than 2~hours away from Rome's airports. Due to environmental concerns by the local authorities, operations at LNGS became more complicated in the recent years. One direct consequence impacting on astroparticle physics and direct DM searches is the future ban of organic liquid scintillators from the laboratory. The {\bf Laboratorio Subterraneo de Canfranc (LSC)} in the Spanish Pyrenees and the {\bf Laboratoire Subterrain de Modane (LSM)} in the French Alps are significantly smaller, however, the muon flux at the deeper LSM is almost 7~times lower compared to LNGS. The {\bf Boulby Laboratory}, hosted in a working potash and salt mine in the North East of England, provides a very low level of $<3\Bq/\meter^2$ of Rn-activity in the air. 
Several other, smaller underground laboratories with less rock overburden exist in Europe which are usually used and operated by individual Universities. The most important large laboratories for DM searches outside Europe are SNOLAB in northern Ontario (Canada), Jinping (CJPL) in southern Sichuan (China), SURF in South Dakota (USA), Kamioka in Japan and Yangyang in South Korea. SNOLAB ($\phi=3 \times 10^{-6}$\,muons/m$^2$/s) and CJPL ($\phi=2 \times 10^{-6}$\,muons/m$^2$/s) are the laboratories with the lowest muon fluxes in the world. In South America a longer-term opportunity is being explored by the ANDES project~\cite{Aihara:2020LA}.
In general, the European DM community would benefit from having a deeper laboratory in Europe to further reduce cosmogenic backgrounds; a possible realisation could be the extension of the Modane laboratory which has already been under discussion. However, it is important to stress that any laboratory needs to provide the auxiliary infrastructure (surface space etc.) and that the existing facilities are of utmost importance for the continuation of the ongoing and upcoming projects. 

As pointed out in Sect.~\ref{sec:wimpbackgrounds}, the strict control of the radiopurity of materials used in DM experiments is essential to achieve the background requirements.  Axion helioscopes also need low-background X-ray detectors and benefit from the direct detection expertise. Underground laboratories provide the environment for material analysis which is becoming more and more demanding. All laboratories have facilities to perform germanium spectroscopy~\cite{facilitiesGe} but only some of them have also ICP-MS equipment~\cite{facilitiesICPMS} and radon detection and mitigation systems~\cite{facilitiesRn}. Specific detectors such as BiPo-3~\cite{facilitiesBiPo}, the world-leading GeMPI HPGe facilities~\cite{Heusser2006495} and DArT~\cite{facilitiesDart} or dedicated analysis techniques~\cite{Lindemann:2013kna} have been developed by individual research groups or within international collaborations. Copper electroforming units, 3D printing facilities, clean underground workshops as well as extended underground storage facilities to avoid cosmogenic activation are highly needed for the upcoming generation of DM experiments but not yet standard in all laboratories. 
Such central support facilities must be adequately staffed with personnel. More generally, low-background  facilities are costly in construction and operation.

The upcoming projects will have to acquire data in stable conditions for several years in order to obtain the required exposures. The host laboratories must thus provide redundant safety systems for power, network, gas supply etc.~in order to cope with external problems such as power losses.

Apart from the already mentioned common requirements such as underground space, low muon flux, (Rn-free) cleanrooms, underground production/storage to avoid activation, etc.~the different experimental technologies used for direct DM searches often require dedicated infrastructure systems: examples include crystal growth facilities (crystal detectors), underground test platforms and large-volume mK-cryostats (bolometers), radiopure SiPM packaging facilities (LAr), underground storage of large amounts of (cryogenic) noble gases (LAr, LXe) as well as facilities to obtain large quantities of low-background underground argon: within the GADMC program these are plants for UAr extraction (Urania), purification (Aria) as well as storage (Argus), see Sect.~\ref{sec:LArTechnologies}.

Coordinating and leveraging synergies in the sector may be crucial to the further development of the underground science infrastructures. The DM community and all experiments demanding ultra-low background conditions would greatly benefit from the implementation of a transnational network across the 
underground laboratories,  facilitating the implementation of common regulations, operational standards and procedures (security, safety, management of resources and materials) as well as open access policies, sustainable collaboration and shared infrastructure and support facilities.
Several attempts to create such a network have been made in the past at European level and are also being pursued on a global scale, \eg the initiative led by  LNGS and SNOLAB to form an Underground Global Research Infrastructure~\cite{GSOreport2019}.
A further step towards this goal would be the establishment and operation of an European Research Infrastructure Consortium (ERIC) --  the {\em European Laboratory of Underground Science}. The low-background community, including the projects searching for the neutrinoless double beta decay, would greatly benefit from such infrastructure. 
{In fact, one of the recommendations of the APPEC Double Beta Decay report~\cite{Giuliani:2019uno} states: ``The European underground laboratories should provide the required space and infrastructures for next generation double beta decay experiments and coordinate efforts in screening and prototyping."
}

Increased cooperation with CERN would be highly beneficial, especially where scientific and technological synergies can be profitably exploited. The case of \LAr\ is a prominent example of a significant overlap between the detector technologies developed for DM searches and those developed towards the large-scale LAr-based neutrino experiments. It would be desirable  that CERN becomes more open to the DM community, as this would enhance synergies that may result from the development of common technologies.

The dedicated experimental searches for {\bf axions and ALPs} have very different requirements on infrastructure and laboratory. 
While many of the efforts are still at a smaller R\&D scale, those projects reaching out for sensitivity to detect DM axions require good infrastructure. Their boundary conditions are defined by the the experimental requirements: 
\begin{itemize}
    \item operation of large aperture superconducting magnets,
    \item minimisation of electromagnetic (thermal) noise in the detector surrounding, 
    \item operation of ultra-sensitive quantum detectors.
\end{itemize}

High field superconducting magnets with the needed aperture require enough {\bf space in well-equipped large experimental halls}. In particular, their operation usually is coupled to {\bf availability of a good cryogenic infrastructure}, i.e., sufficient liquid helium and liquid nitrogen supply as well as the necessary electric installations.
A {\bf stable and low background in terms of electromagnetic radiation} in the frequency range of interest should also be granted. 
Additionally, any next generation axion experiment will largely benefit from the vicinity to labs with world-leading knowledge on accelerator  technology, superconducting magnet R\&D,  detector research, low loss RF-technologies and cryogenic engineering.
Many of these requirements are  fulfilled by the big European particle physics laboratories like CERN, DESY, CEA-IRFU, INFN, but also at other locations as neutron-, solid-state-, quantum computing or astronomical  and gravitational wave research facilities.

Indeed, DESY at Hamburg (Germany) is the host institution for the ALPS LSW project and has additionally been selected as the site for MADMAX and  (baby)IAXO. 
At DESY the experimental halls formerly used by the HERA experiments,   coming with the necessary infrastructural conditions, are presently being re-commissioned for alternative use, also for the mentioned axion/ALP experiments.
This will make DESY the largest axion/ALP hub world-wide. 
CERN, on the other hand, was host of the 
CAST experiment and is strongly involved in the magnet 
design for babyIAXO and IAXO, as well as in the RADES R\&D effort. 

{Finally, there is growing synergy in computing software and hardware. 
Processing and interpretation of complex data streams collected with direct dark matter detection experiments, be it WIMP, axion/ALP, or other searches, increasingly relies on sophisticated signal and background recognition tools and statistical methods, as well as on the availability of modern computing infrastructure. Similar requirements apply to other branches of astroparticle physics, as well as to neighbouring fields of particle physics, cosmology or astrophysics, and beyond. Examples include the use by the WIMP DM experiments of the computing Grid developed for LHC or of tools developed initially for the HEP community, such as Geant4, Garfield, \etc, which are expanded according to the needs of the DM community. Interdisciplinary efforts among different communities, both theory and experiment (often in fact co-operating in various working groups), are becoming increasingly important and should be strongly supported. }

\newpage

\section{List of Recommendations}\label{sec:finalrecom}
In this section we summarise the recommendations of the Report.

{Recommendation~1.\ \ \ 
The search for dark matter with the aim of detecting a direct signal  of DM particle   interactions with a detector should be given top priority in astroparticle physics, 
and in all particle physics, and beyond, as a positive measurement will provide the most unambiguous confirmation of the particle nature of dark matter in the Universe.}\\[-0.1cm]

{Recommendation~2.\ \ \ 
The diversified approach to probe the broadest experimentally accessible ranges of  particle mass and interactions  is needed to ensure the most conservative and least assumption-dependent exploration of hypothetical candidates for cosmological dark matter or subdominant relics. 
}\\[-0.1cm]

{Recommendation~3.\ \ \ 
The experimental underground programmes with the best sensitivity to detect signals induced by dark matter WIMPs scattering off the target should receive enhanced support to continue efforts  to  reach down to the so-called neutrino floor on the shortest possible timescale. }\\[-0.1cm]

{Recommendation~4.\ \ \ 
European participation in DM search programmes and associated, often novel, R\&D efforts, that currently do not offer the biggest improvement in sensitivity should continue and be encouraged with view of a long-term investment in the field and the promise of potential interdisciplinary benefits. We recommend that coordinated programmes are established for dark matter detector development. 
}\\[-0.1cm]

{Recommendation~5.\ \ \ The long-term future of underground science in Europe would strongly benefit from creating 
a distributed but integrated structure of underground laboratories for the needs of the forthcoming generation of new experiments, and beyond.
This strategic initiative would be most efficiently implemented by forming the {\em European Laboratory of Underground Science}.}\\[-0.1cm]

{Recommendation~6.\ \ \
European-led efforts should focus on  axion and ALPs mass ranges that are
complementary to the established cavity approach and this is where European teams have a  unique opportunity to secure the pioneering role in achieving sensitivities in axion/ALP mass ranges not yet explored by experiments conducted elsewhere. In parallel,  R\&D efforts to improve experimental sensitivity and to extend the accessible mass ranges should be supported.
}\\[-0.1cm]

{Recommendation~7.\ \ \ 
Continuing dedicated
and diverse theoretical activity should be encouraged not only in its own right but also as it provides some highly stimulating, and mutually  beneficial, interdisciplinary environment for DM and new physics searches.
}\\[-0.1cm]


\newpage

\appendix
\section{ Appendix: The Mandate of the Report }\label{ref:appa}

The mandate from the APPEC Scientific Advisory Committee defining the scope of the Report is provided below.

To aid in the discussions and to devise concrete recommendations for the next steps in direct DM detection in the next decade, the DM direct detection committee should provide an assessment of the current and future scientific opportunities in non-accelerator DM searches over the next 10-year period, in particular delivering:
\begin{itemize}
    \item 
The {\em global context} of DM particle searches, including the existing hints or evidence for DM particles, an inventory of alternatives for the particle nature of DM, and an inventory of present and best estimates of future sensitivities from measurements or observations from methods other than direct detection.
    \item An inventory of {\em existing DM experiments} and the technologies adopted by these, with current most competitive results.
    \item A {\em comparative SWOT analysis} of existing, planned and proposed technologies for DM direct detection with the potential to surpass current sensitivities in the next decade with the eventual goal of reaching or surpassing the so-called neutrino floor.\footnote{The cross section where the background from coherent neutrino-nucleus scattering becomes relevant.} This SWOT analysis should include: (a) an inventory of technology challenges, R\&D paths to address them, required resources and schedule to achieve staged goals and ensure scientific advancement and discovery potential; (b) a list of fundamental limits of the various technologies, including best estimates of evolution of systematic uncertainties (including those on nuclear physics and how they are related to this field and e.g. neutrinoless double beta decay), background levels in ROI; (c) a discussion of the importance and the limitations of additional signatures such as directionality and annual modulation.

    \item An assessment of the {\em required infrastructure} in Europe, including maintenance and upgrades of existing facilities.
    \item A list of (possible) {\em technological and scientific synergies} between the different direct detection technologies and with research and R\&D outside of this field.
    \item An inventory of physics, astronomy or {\em other research} that can be done in addition to DM direct detection with the various technologies. In addition it would be important to discuss if such other research can be done even within the specifically proposed DM experiments. Synergies with other experiments of indirect, accelerator and cosmology DM searches should also be considered, including possible technical and R\&D synergies, e.g with CERN, other laboratories and industry.
    \item 
    Any other recommendations within the scope of DM direct searches that the committee deems relevant.
\end{itemize}


\glsaddall



\printglossary[title=Glossary and Acronyms, type=\acronymtype]

\newpage
\addcontentsline{toc}{section}{References}
\bibliographystyle{JHEP}
\bibliography{refs}

\end{document}